\documentclass[fleqn,usenatbib]{mnras}

\usepackage{newtxtext,newtxmath}
\usepackage[T1]{fontenc}
\usepackage{graphicx}
\usepackage{amsmath}
\usepackage[usenames,dvipsnames]{xcolor}
\usepackage[normalem]{ulem}
\usepackage{listings}
\usepackage{macros}
\usepackage{tikz}
\usetikzlibrary{matrix, positioning, fit}
\usetikzlibrary{calc, shapes, arrows, positioning}
\tikzstyle{arrow} = [thick,->,>=stealth]

\newcommand{\eee}{\equiv}
\newcommand{\eref}[1]{Eq.~(\ref{#1})}
\newcommand{\pCAeref}[1]{Eqs.~(\ref{#1}}
\newcommand{\pCBeref}[1]{, \ref{#1}}
\newcommand{\pCCeref}[1]{, \ref{#1})}
\newcommand{\betac}{{\beta_\text{c}}}
\newcommand{\bigB}{\mathcal{B}}
\newcommand{\bigE}{\mathcal{E}}
\newcommand{\bigK}{\mathcal{K}}
\newcommand{\bigKt}{\mathcal{K}_\text{t}}
\newcommand{\bigL}{\mathcal{L}}
\newcommand{\bigM}{\mathcal{M}}
\newcommand{\bigQ}{\mathcal{Q}}
\newcommand{\betasigma}{{\beta_\sigma}}
\newcommand{\hatF}{\hat{F}}
\newcommand{\hatP}{\hat{P}}
\newcommand{\hatvarrho}{\hat{\varrho}}
\newcommand{\hatZ}{\hat{Z}}
\newcommand{\Mz}{M_0}
\newcommand{\Ra}{R_\text{a}}
\newcommand{\ra}{r_\text{a}}
\newcommand{\rz}{r_0}
\newcommand{\rhoz}{\rho_0}
\newcommand{\rhow}{\rho_\text{w}}
\newcommand{\gamerNS}{\textsc{gamer-2}}
\newcommand{\gamer}{{\textsc{gamer-2}\;}}
\newcommand{\PIANISTpy}{\texttt{PIANISTpy}\;}
\newcommand{\PIANISTpyNS}{\texttt{PIANISTpy}}
\newcommand{\agama}{\textsc{Agama}\;}
\newcommand{\agamaNS}{\textsc{Agama}}
\newcommand{\galpy}{\texttt{galpy}\;}
\newcommand{\galpyNS}{\texttt{galpy}}
\newcommand{\PYTHON}{\texttt{python}\;}
\newcommand{\JAX}{\texttt{JAX}\;}
\newcommand{\JAXNS}{\texttt{JAX}}
\newcommand{\Npar}{N_\text{par}}
\newcommand{\Vt}{V_\text{t}}
\newcommand{\Vr}{V_\text{r}}
\newcommand{\Zt}{Z_\text{t}}
\newcommand{\sigmar}{\sigma_\text{r}}
\newcommand{\sigmat}{\sigma_\text{t}}

\newcommand{\aref}[1]{Appendix~\ref{#1}}
\newcommand{\fref}[1]{Fig.~\ref{#1}}

\hypersetup{urlcolor=cyan}
\definecolor{codegreen}{rgb}{0,0.6,0}
\definecolor{codegray}{rgb}{0.5,0.5,0.5}
\definecolor{codepurple}{rgb}{0.58,0,0.82}
\definecolor{backcolor}{rgb}{0.95,0.95,0.92}

\lstdefinestyle{mystyle}{
        language=Python,
        identifierstyle=\color{codegreen},
	backgroundcolor= \color[rgb]{0.97,0.97,0.97},
	commentstyle=\color{blue},
	keywordstyle=\color{magenta},
	numberstyle=\tiny\color{codegray},
	stringstyle=\color{codepurple},
        moredelim=**[is][\color{red}]{~}{~},
	basicstyle=\ttfamily\footnotesize,,
	breakatwhitespace=false,
	breaklines=true,
	captionpos=b,
	keepspaces=true,
	numbers=left,
	numbersep=5pt,
	showspaces=false,
	showstringspaces=false,
	showtabs=false,
	tabsize=1,
}

\providecommand{\rme}{\mathrm{e}}
\title[\PIANISTpy and Anisotropic Equilibrium Models]
      {\PIANISTpyNS: A \JAXNS-based \PYTHON package for velocity-anisotropic astrophysical equilibrium models}

\author[B. T. Chiang, F. C. van den Bosch and H.-Y. Schive]
{Barry T. Chiang$^{1,2}$\thanks{E-mail: barry.chiang@yale.edu},
Frank C. van den Bosch$^{1}$, and Hsi-Yu Schive$^{2,3,4,5}$
\vspace*{8pt}
\\
$^{1}$Department of Astronomy, Yale University, New Haven, CT 06511, USA.\\
$^{2}$Institute of Astrophysics, National Taiwan University, Taipei 10617, Taiwan.\\
$^{3}$Department of Physics, National Taiwan University, Taipei 10617, Taiwan.\\
$^{4}$Center for Theoretical Physics, National Taiwan University, Taipei 10617, Taiwan.\\
$^{5}$Physics Division, National Center for Theoretical Sciences, Taipei 10617, Taiwan.}

\date{Accepted 2026 November 22. Received 2026 November 22; in original form 2026 June 30}

\pubyear{2026}
\begin{document}

\label{firstpage}
\pagerange{\pageref{firstpage}--\pageref{lastpage}}
\maketitle

\begin{abstract}
  The hierarchical assembly of dark matter haloes and galaxies via mergers and anisotropic mass accretion naturally imprints non-trivial velocity anisotropy onto haloes and galaxies, an empirical fact emerging from state-of-the-art cosmological simulations. However, dynamical studies based on idealised numerical simulations typically assume isotropic velocity distributions for the systems studied. This is mainly because no tools exist to construct anisotropic initial conditions with fully self-consistent (i.e., physical and instability-free) truncation boundary conditions for finite spatial extents. Here we present the first such analytical framework that performs truncation in phase space, with two distinct approaches to guarantee either the monotonicity of output distribution functions or identical truncated density profiles. We additionally present the new open-source \JAXNS-based GPU-accelerated \PYTHON package \PIANISTpy (\textbf{P}article \textbf{I}nitial conditions with \textbf{ANIS}o\textbf{T}ropy) for constructing spherical, non-rotating initial conditions, supporting arbitrary double-power-law analytical radial density profiles and a library of analytical anisotropy profiles. This framework further permits the construction of systems with non-analytical anisotropy profiles via an arbitrary linear superposition of all implemented analytical anisotropy models as basis functions. \PIANISTpy also outputs various 3D velocity profiles and line-of-sight-integrated observables with end-to-end differentiability for rapid gradient-based Monte Carlo inference, and self-consistent local dark matter velocity distributions of anisotropic haloes beyond the unphysical Maxwell--Boltzmann simplification adopted in terrestrial detection experiments. We discuss possible use cases and demonstrate \PIANISTpyNS's orders-of-magnitude higher numerical solver accuracy and superior dynamical stability of particle initial conditions relative to existing tools, including \galpy and \agamaNS. The public source code is available at \url{https://github.com/barrytchiang/PIANISTpy} with extensive documentation at \url{https://pianistpy.com}.
\end{abstract}


\section{Introduction}\label{sec:Introduction}

It is common practice to model galaxies \citep[e.g.,][]{Penarrubia2008, SanchezAlmeida2024}, star clusters \citep[e.g.,][]{McLaughlin2005, Baumgardt2017}, dark matter (DM) haloes \citep[e.g.,][]{Freundlich2020MNRAS4914523F, Dattathri2025}, and subhaloes \citep[e.g.,][]{vandenBosch2018MNRAS4754066V, Errani2020MNRAS491} as spherical and isotropic systems due to the relative ease of theoretical modelling and constructing particle initial conditions. However, typically these structures are aspherical and exhibit non-trivial velocity anisotropy, as predicted by both theory and numerical simulations, and supported by observational inferences.

On cosmological scales, DM haloes and galaxies form and evolve preferentially along large-scale cosmic web filaments \citep[e.g.,][]{Dekel2009, Snedden2016MNRAS4552804S, Kraljic2018, OKane2024MNRAS5341682O, Medlock2026}. The resulting mass growth via anisotropic accretion \citep[e.g.,][]{VeraCiro2011, Libeskind2014, Kang2015ApJ8136K, Shi2015, Orkney2023MNRAS5253516O, Mezini2025} and mergers \citep[e.g.,][]{Wu2014, Wang2020, Wang2025ApJ979223W} naturally gives rise to velocity anisotropy. This natal anisotropy can subsequently be modified by a variety of physical processes, such as baryonic feedback \citep[e.g.,][]{El-Badry2017ApJ835193E}, dynamical friction \citep[e.g.,][]{Debattista1998, Debattista2000}, core scouring \citep[e.g.,][]{Milosavljevic2001, Rantala2019}, and tidal evolution \citep[e.g.,][]{Bianchini2017, Chiang2025}.

Quantitatively, the velocity anisotropy of spherical systems is customarily defined as 
\begin{equation}\label{eqn:betadef}
\betasigma(r) \equiv 1 - \frac{\sigma^2_\rmt(r)}{2\sigma^2_\rmr(r)} \in(-\infty,1]\,,
\end{equation}
relating the tangential $\sigma_\rmt(r)$  and radial $\sigma_\rmr(r)$ velocity dispersion profiles. Isotropy corresponds to $\betasigma = 0$, while systems with $\betasigma > 0$ ($< 0$) are radially (tangentially) anisotropic. High-resolution cosmological simulations consistently show that dark matter haloes exhibit velocity structures deviating from isotropy \citep[e.g.,][]{Diemand:2007qr, Ascasibar2008, Navarro2010MNRAS40221N, Ludlow2011MNRAS4153895L, Lemze2012, Klypin2016MNRAS4574340K, vdBosch.etal.19, He2024ApJ976187H}, with typical values ranging between $\betasigma \sim 0.1$\textendash$0.4$. Similarly, simulated galaxies also often display significant radial velocity anisotropy \citep[e.g.,][]{El-Badry2017ApJ835193E, Gonzalez-Samaniego2017MNRAS4724786G, Wang2020MNRAS4915188W, Orkney2023MNRAS5253516O}, especially in their stellar haloes where $\betasigma$ can reach values as high as $0.9$ \citep{Mondal2024MNRAS5333426M, Mondal2026}. Observationally, it has long been known that elliptical galaxies have anisotropic velocity distributions \citep[e.g.,][]{Davies.etal.83, Bender.88, vdMarel.91, Cretton.vdBosch.99, Cappellari.etal.07}, and the same applies for stars in the Solar neighbourhood \citep[][]{Wielen1977AA60, Dehnen.Binney.98, Tian.etal.15}, Milky Way dwarf galaxies \citep{GeringerSameth2015, Goldstein2022, Vitral2024, Vitral2026, Nguyen2026}, and other disc galaxies \citep[][]{Gerssen.etal.97, Kregel.etal.05, Gerssen.etal.12}. Deviations from velocity isotropy have also been found in globular clusters \citep[e.g.,][]{Jindal2019, Vasiliev2021}, the Galactic stellar halo (\citealt{Bird2019}; see also references in \citealt{Hattori2017}), and the orbital kinematics of satellite galaxies around hosts ranging from sub-Milky Way to massive cluster scales \citep{Wojtak2013, Mitra2024, Chiang2026a}; in galaxy clusters, the orbital anisotropy of member galaxies also varies across different galaxy types \citep{Mamon2019, Lotz2019}.

Yet, it is still extremely common to assume isotropic velocity distributions when modelling star clusters, galaxies or dark matter haloes \citep[e.g.,][]{McLaughlin2005, Hilker2007A&A463119H, Miocchi2013, Sonnenfeld2013, Tortora2019}, especially when this is done using idealised numerical simulations \citep[e.g.,][]{Hayashi2003ApJ584541H, Penarrubia2010MNRAS4061290P, vandenBosch2018MNRAS4754066V, Delos2019PhRvD100f3505D, Errani2021MNRAS50518E, Benson2022MNRAS5171398B, Ogiya2022MNRAS5102724O, Du2024PhRvD110b3019D}. This persistent reliance on the isotropic assumption, despite overwhelming evidence to the contrary, stems from a lack of frameworks and tools to generate anisotropic initial conditions that maintain dynamical stability within truncated, finite spatial extents.

On separate physical grounds, truncated models are also required not only for a finite total mass but for a finite spatial extent. The stellar distributions of spheroidal galaxies, of brightest cluster galaxies, or even of the intracluster light \citep[e.g.,][]{2009ApJS..182..216K, 2013ApJ...764L..31K, 2005ApJ...618..195G} occupy a finite region within their host dark matter haloes. Satellite galaxies, subhaloes and star clusters are similarly tidally limited by their hosts \citep[e.g.,][]{King1966AJ7164K, Penarrubia2008, Errani2021MNRAS50518E}. Idealised $N$-body studies of such systems need equilibrium initial conditions of finite extent, so that the further dynamical evolution reflects the physics under study rather than the relaxation of an improper boundary condition \citep[e.g.,][]{Chiang2025, Chiang2026}.

Even within the simplest isotropic case, where the distribution function (DF) $f(E)$ depends only on energy $E$ and can be trivially computed from an input density profile $\rho(r)$ via the Eddington inversion \citep{Eddington1916MNRAS76572E}, ensuring the dynamical stability of truncated boundaries is still non-trivial. Such truncation is, however, required for density models with infinite spatial extents, such as the NFW profile \citep{Navarro:1995iw}, which otherwise yields a divergent total mass. The earliest and still most prevalent approach is to `directly truncate' or `modify' the density profile. Various choices, such as the standard Heaviside-like hard truncation \citep[e.g.,][]{Teyssier2013MNRAS4293068T, vandenBosch2018MNRAS4754066V}, exponential truncation \citep[e.g.,][]{Springel1999MNRAS307162S, Errani2021MNRAS50518E}, or more elaborate parametrisations \citep[e.g.,][]{Baltz2009JCAP01015B}, modify the density profile $\rho(r)$ directly, with little or no regard for the corresponding DF; existing tools like \galpy \citep{Bovy2015ApJS21629B} and \textsc{nbopy} \citep{Errani2020MNRAS491} fall into this category. Even when the DFs are re-computed from the truncated density profiles \citep[e.g.,][]{Choi2007MNRAS381987C}, as is done for example in \agama \citep{Vasiliev2019MNRAS4821525V}, truncation choices that are `improper' in phase space, which are not known a priori, manifest as truncation-induced inflection features in the DFs that directly drive instabilities in the initial conditions (\citealp{Weinberg2023MNRAS5254962W}; see also \citealp{Dattathri2025}).

The mathematically fully self-consistent truncation approach that avoids such boundary artefacts is to directly truncate the DF $f(E)$ via energy $E$ re-parametrisation. This method, pioneered by \citet{Drakos2017MNRAS4682345D}, is conceptually similar to the energy-lowering procedure of the isothermal King model \citep{King1966AJ7164K}. Such an approach offers several key advantages: (1) if the pre-truncated DFs are everywhere non-negative (physical) and monotonic (dynamically stable), the truncated DFs automatically inherit these properties; (2) the resulting density and velocity profiles are inherently truncated at a finite radius while remaining fully self-consistent with the DF; (3) the boundaries of such constructed particle initial conditions are in perfect dynamical equilibrium and require no further relaxation; and (4) the truncated DF and velocity distributions are exact solutions to the collisionless Boltzmann equation, free of numerical `imperfections' inherent to approximate iterative schemes like the method of \citet{Schwarzschild1979ApJ232236S}.

This problem becomes significantly more complex for systems with velocity anisotropy. Dynamically, while truncation-induced perturbations in isotropic systems are typically confined to the outer radii near the truncation boundary \citep[e.g.,][]{vandenBosch2018MNRAS4754066V}, they pose a far more fundamental problem for anisotropic models. In radially anisotropic systems, such destabilising perturbations can swiftly trigger radial orbit instability \citep[e.g.,][]{Antonov1973, Henon1973, Trenti2006}, destroying spherical symmetry and distorting the density distribution across a wide range of radii. This creates a critical ambiguity in dynamical studies, making it difficult to cleanly disentangle innate physical instabilities from artefacts induced by the truncation method itself. Extending the energy-based re-parametrisation framework to these systems is therefore essential.

In phase space, the primary challenge lies in the fact that anisotropy introduces an explicit dependence on angular momentum $L$ into the DF, $f(E, L)$, which no longer maps one-to-one to a given density profile $\rho(r)$ \citep[e.g.,][]{Dejonghe1986PhR133217D, Dejonghe1989ApJ343113D}. Analytical $f(E, L)$--$\rho(r)$ conversions exist only for specific families\footnote{Other anisotropic frameworks exist that differ fundamentally from the mainstream $E$--$L$ and $\rho \to f$ inversion approach described in this work. \citet{Binney2026} exemplifies one such forward-modelling framework, which constructs DFs directly in action space by exploiting the adiabatic invariance of constituent orbits.}, such as constant-anisotropy models \citep[e.g., \S4.3.2 of][]{Binney&Tremaine2008}, Cuddeford--Osipkov--Merritt models \citep{Osipkov1979PAZh577O, Merritt1985AJ901027M, Cuddeford1991MNRAS253414C}, or quasi-separable models \citep{Gerhard1991MNRAS250812G, Baes2004MNRAS35118B, Baes2007AA471419B}, and even in these cases, the physical requirement that $f(E, L) \geq 0$ is not guaranteed and must be verified numerically. Consequently, any such phase-space-based `recipe' for self-consistent truncation must address the following technical complexities: (1) identifying a suitable truncation hyper-surface in the 2D $E$--$L$ phase space is non-trivial; (2) it is unclear if standard $f$--$\rho$ conversion formulae remain valid under such truncation; and (3) anisotropic DFs are subject to stricter physical constraints \citep[e.g.,][]{An2006ApJ642752A, Ciotti2010MNRAS4011091C}, where an otherwise well-behaved density profile may yield unphysical or unstable distribution functions. It is thus not obvious how to determine an optimal truncation parametrisation that preserves the input anisotropy profile without introducing unphysical or non-monotonic features in the DF.

In this work, we present the first phase-space-truncation framework for anisotropic spherical systems that addresses the complexities outlined above. In particular, we introduce two distinct truncation methodologies designed to meet different modelling requirements: the first preserves the `well-behaved' functional forms of the truncated DFs, while the second guarantees identical output truncated density profiles across varying anisotropy configurations. Furthermore, we demonstrate that by exploiting the linear nature of distribution functions in phase space, one can perform a self-consistent superposition of truncated DFs. This allows for the construction of flexible, radially varying anisotropy profiles that extend beyond the limitations of known closed-form analytical families. As we will show, this framework provides a robust tool for generating dynamically stable, anisotropic initial conditions and performing self-consistent kinematic modelling for finite-extent systems.

Based on this novel algorithm, here we also introduce and describe the new open-source \JAXNS\footnote{\url{https://docs.jax.dev/}}-accelerated \citep{jax2018github} \PYTHON package \PIANISTpy (\textbf{P}article \textbf{I}nitial conditions with \textbf{ANIS}o\textbf{T}ropy) that constructs self-consistently phase-space-truncated (an)isotropic spherical systems with non-divergent central gravitational potential, such that the stability of truncation boundaries is ensured and requires no further dynamical relaxation. This new tool enables systematic exploration of anisotropic spherical systems, as we will demonstrate with a few examples. The public \texttt{GitHub} source code repository is available at \url{https://github.com/barrytchiang/PIANISTpy}. The more detailed user guide and extensive documentation can be found at \url{https://pianistpy.com}.

The paper is organised as follows. \S\ref{sec:Overview} describes the overall code design of \PIANISTpyNS. \S\ref{ssec:DF_Truncation} presents the phase-space truncation framework and contrasts the two complementary truncation methodologies with concrete examples. \S\ref{ssec:DF_Superposition} demonstrates the construction of flexible, non-analytical anisotropy profiles via linear superposition of truncated DFs. \S\ref{ssec:DF_Physicalness} discusses physical constraints on anisotropic DFs. \S\ref{sec:DynamicalStability} benchmarks numerical solver accuracy and the dynamical stability of \PIANISTpyNS-generated initial conditions against existing tools. \S\ref{sec:OP_Vel} explores the velocity distributions and projected kinematic profiles of anisotropic systems. \S\ref{sec:User_Case} discusses possible application cases of \PIANISTpyNS. Lastly, \S\ref{sec:Summary} concludes. For technical details, \aref{app:Double_Power_Law_Rho} derives the potential normalisation constants of analytical density profiles; \aref{app:DF} provides analytical derivations of the truncation framework and DF-derived quantities; and \aref{app:JAX_imp} documents the \JAX implementation and numerical optimisation.

\section{General Overview of \PIANISTpy}\label{sec:Overview}

\subsection{Workflow and units}\label{ssec:Units}

In \PIANISTpyNS, the entire DF truncation pipeline and all phase-space-related calculations are handled under the `natural' dimensionless normalisation scheme of \mbox{$G\eee \rhoz/C \eee \rz \eee 1$} first introduced by \citet{Widrow2000ApJS13139W}, henceforth the \textit{internal} units which all \PIANISTpy outputs default to. For further unit conversions as often required for general physical quantities (e.g., particle initial conditions) and observables (e.g., velocity dispersion profiles), \PIANISTpy also natively supports physical units via \texttt{astropy} \citep{Astropy2022ApJ935167A}, and model units $G\eee \Mz \eee \rz \eee 1$ \citep[e.g.,][]{Henon1973, Heggie.Mathieu.86}. All \PIANISTpyNS-generated outputs can be converted to in between these different unit conventions. Here, $G$ denotes the gravitational constant, while $\Mz$, $\rz$, and $\rhoz$ are user-specified characteristic mass, length, and density scales, as discussed below. The density-profile-dependent numerical prefactor $C$ is defined below.

Consider a generic system generated by some physical phase-space DF $f(\bx,\bv) \geq 0$. The local mass density $\rho(\bx)$, the mass $M$ enclosed within a given volume, and the associated gravitational potential $\Phi(\bx)$ given by the Poisson equation are 
\begin{align}\label{eqn:rho_DF_Poisson}
	\begin{aligned}
		&\rho(\bx)= \int \rmd^3\bv \, f(\bx,\bv),\\
		&M = \int\int \rmd^3\bv \, \rmd^3x \, f(\bx,\bv),\\
        &\nabla^2 \Phi(\bx) = 4 \pi G \rho(\bx).
	\end{aligned}
\end{align}
For spherical systems with no net rotation with respect to the Centre of Mass (CM), the corresponding DFs are fully parameterised by the relative specific energy $\bigE$ and specific angular momentum $L = |\bL| \eee |\bx\times \bv|$ \citep[e.g.,][]{Baes2007AA471419B}, where
\begin{align}\label{eqn:unit_nor_4}
	\begin{aligned}
 &\bigE(r,v) = \Psi(r) - \frac{1}{2}v^2 && \;\;\;\:\:\text{(relative specific energy)},\\
 &\Psi(r) = -\Phi(r) + \Phi_0 && \;\;\;\:\:\text{(relative potential energy)},\\
 &Q_{\pm}\eee \bigE \pm \frac{L^2}{2 \ra^2} && \;\;\;\:\:\text{(Osipkov\textendash Merritt $Q$ parameter)}.
	\end{aligned}
\end{align}
Here, $r\eee|\bx|$ denotes the 3D radius with respect to the CM, $v = |\bv|$ is the instantaneous 3D velocity of individual member particles, and $\ra$ is the anisotropy radius. The constant $\Phi_0$ sets the zero point of the relative potential and is fixed by the unit normalisation scheme below; $Q_\pm$ is the combination of $\bigE$ and $L$ that replaces $\bigE$ as the energy-like argument of the DFs of the (Cuddeford--)Osipkov--Merritt models, the sign distinguishing the radially (Type~I) and tangentially (Type~II) biased variants (\S\ref{ssec:DF_Truncation}).

\PIANISTpy builds upon the unit-normalisation scheme of \citet{Widrow2000ApJS13139W} that is both natural and numerically advantageous for phase-space-based calculations of such systems. In particular, the dimensionless radius $R$, 3D velocity vector $\bV$, angular momentum $\bigL$, density profile $\varrho$, DF $F$, relative potential $P$, relative specific energy $Z$, and other derived parameters are defined as 
\begin{align}\label{eqn:relative_potential_energy}
	\begin{aligned}
        &R \eee r/\rz ,\\
        &\bV \eee \bv\times\big(4\pi G \rhow \rz^2\big)^{-1/2},\\
        &\bigL \eee L\times\big(4\pi G \rhow \rz^4\big)^{-1/2},\\
		&\varrho \eee \rho/ \rhow,\\
        &F \eee f\times\big(4\pi G\big)^{3/2}\rhow^{1/2}\rz^3,\\
        &P \eee \Psi\times(4\pi G \rhow \rz^2\big)^{-1},\\
        &Z \eee \bigE\times(4\pi G \rhow \rz^2\big)^{-1},\\
        &\bigQ_\pm \eee Q_\pm\times(4\pi G \rhow \rz^2\big)^{-1} = Z\pm \frac{\bigL^2}{2\Ra^2},\\
        &\Ra \eee \ra/\rz,
	\end{aligned}
\end{align}
where $\rhow\eee\rhoz/C$ and $\rz$ are the characteristic density and radius (length scale) of the system in question. As such, the relative potential is always \textit{unit-normalised} with the boundary conditions \mbox{$P(R=0)=1$} at the centre and $P(\Rt) = 0$ at the truncation radius $\Rt$ where the density profile first vanishes\footnote{For systems with non-vanishing density distribution out to spatial infinity, the outer boundary condition is simply $P(R\rightarrow\infty) = 0$.}, $\varrho(\Rt) = 0$. This is enforced by a suitable choice for the density normalisation prefactor $C \in \mathbb{R}^{+}$ (as discussed below) and for the reference value of the gravitational potential $\Phi_0$ as defined in \eref{eqn:unit_nor_4}. With this normalisation scheme, $Z(P) \in [0,1]$ and $\bigQ_\pm(P,R) \in [0,1]$. Also, by explicitly anchoring the unit-normalisation to the \textit{untruncated} density profile, the numerical value of $C$ remains unchanged even after truncation described in \S\ref{sec:DF_Truncation}.

As a concrete example, consider the family of double-power-law analytical density profiles \citep{Zhao1996MNRAS278488Z}
\begin{align}\label{eqn:double_power_law_MT}
	\begin{aligned}
		&\rho(r) = \rhoz\times \Big(\frac{r}{\rz}\Big)^{-\gamma} \Big[1+\big(\frac{r}{\rz}\big)^\alpha\Big]^{(\gamma-\beta)/\alpha},\\
		&\varrho(R)= C\times R^{-\gamma} (1+R^\alpha)^{(\gamma-\beta)/\alpha},
	\end{aligned}
\end{align}
where $\alpha$ determines the sharpness of transition from the inner density slope $\varrho(R\ll1)\propto R^{-\gamma}$ to outer slope $\varrho(R\gg1)\propto R^{-\beta}$. Many commonly adopted density profiles are special cases $\{\alpha,\beta, \gamma\}$ of \eref{eqn:double_power_law_MT}, including NFW $\{1, 3, 1\}$, Dehnen $\{1, 4, \gamma\}$, Hernquist $\{1, 4, 1\}$, modified isothermal sphere $\{2, 2, 0\}$, and Plummer $\{2, 5, 0\}$ \citep[e.g., \S5.4.7 of][]{Mo2010gfebookM}. By computing the corresponding relative potential, the numerical value of $C(\alpha, \beta, \gamma)$ is uniquely determined by imposing the condition $P(R=0)\eee1$ and can be expressed entirely in terms of the Gamma function (see \aref{app:Double_Power_Law_Rho}).  For the double-power-law profiles to be physical, we are restricted to $\alpha \geq 1/2$ (physical consistency in phase space; e.g., \citealp{Baes2021}), $\beta >2$ (finite potential at infinity) and $0\leq \gamma < 2$ (non-decreasing density with decreasing radius and finite central potential, respectively).

Separately, given the scale-free nature of purely gravitational systems, model units $G\eee \Mz \eee \rz \eee 1$ are often adopted in idealised $N$-body simulations \citep[e.g.,][]{Penarrubia2008, vandenBosch2018MNRAS4754066V, Chiang2025}. In \PIANISTpyNS, the model-unit convention always anchors $\rz$ to the scale radius as defined in the double-power-law density profiles, \eref{eqn:double_power_law_MT}. The characteristic mass scale $\Mz$ is the total mass integrated out to a multiple of the scale radius. Namely, $\Mz$ is set by the user-input parameter $c_\rho =\texttt{concentration}$ such that
\begin{align}\label{eqn:M_nor_1}
	\Mz(c_\rho; \alpha,\beta,\gamma) &\eee \int_0^{c_\rho \rz} \rmd r \, 4\pi r^2\rho(r; \alpha,\beta,\gamma)  \nonumber\\
    &= 4\pi \rz^3\rhoz \times q(c_\rho; \alpha,\beta,\gamma).
\end{align}
This constant $q(c_\rho; \alpha,\beta,\gamma)$, which can be numerically evaluated via \eref{eqn:q_beta_closed}, provides a direct unit mapping from the internal units defined in \eref{eqn:relative_potential_energy}. Taking the NFW density profile as an example; if one adopts $c_\rho$ as the overdensity-based concentration parameter, then \eref{eqn:M_nor_1} yields $q(c_\rho; 1, 3,1) = \ln(1+c_\rho)-c_\rho/(1+c_\rho)$ and $\Mz(c_\rho)$ now corresponds to the virial halo.

For data outputs, the default internal units are generally useful only for the DF and other direct phase-space quantities as those defined in \eref{eqn:relative_potential_energy}. All the other practical applications (e.g., generating particle initial conditions for simulations or line-of-sight velocity dispersion profiles to be compared with observations) will generally use the model or the physical units.

\subsection{Package structure}\label{ssec:Package_structure}

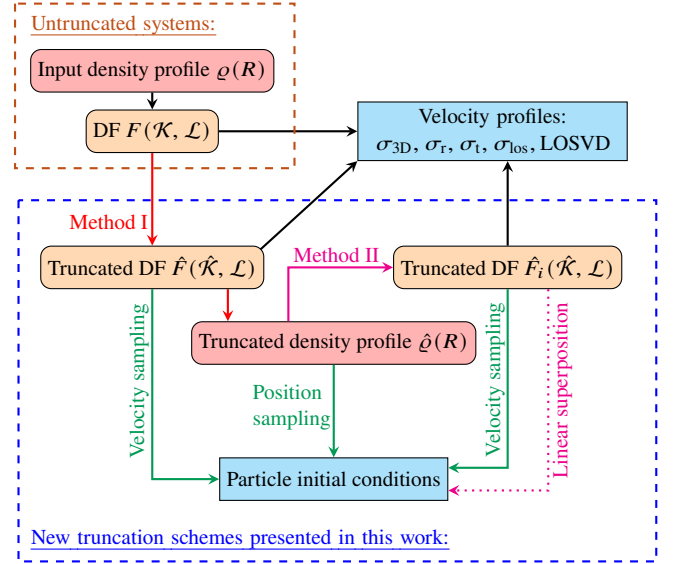
\begin{figure}
	\centering
	\begin{tikzpicture}[node distance=1cm]
		\node (Ini_Den)  [rectangle, rounded corners, minimum width=2em, minimum height=2em,text centered, draw=black, fill=red!30] {\mbox{Input density profile $\varrho(R)$}};
		\node (Ini_DF)   [rectangle, rounded corners, minimum width=2em, minimum height=2em, text centered, draw=black, fill=orange!30, below of=Ini_Den, node distance=0.8cm] {DF $F(\bigK, \bigL)$};
		\node (sigma)  [rectangle, text width=12em, minimum width=2em, minimum height=2em, text centered, draw=black, fill=cyan!30, right of=Ini_DF, xshift=3.5cm] {Velocity profiles:\\ $\sigma_\text{3D}, \sigmar, \sigmat, \sigma_\text{los}, \text{LOSVD}$};
		\node (MethodI_DF) [rectangle, rounded corners, minimum width=2em, minimum height=2em, text centered, draw=black, fill=orange!30, below of=Ini_DF, node distance=1.8cm] {Truncated DF $\hatF(\hat{\bigK}, \bigL)$};
		\node (MethodI_Den) [rectangle, rounded corners, minimum width=2em, minimum height=2em,text centered, draw=black, fill=red!30, right of=MethodI_DF, node distance=1.2cm, xshift=1.2cm, yshift = -1.0cm] {Truncated density profile $\hatvarrho(R)$};
		\node (MethodII_DF)  [rectangle, rounded corners, minimum width=2em, minimum height=2em, text centered, draw=black, fill=orange!30, right of=MethodI_DF, xshift=3.7cm] {Truncated DF $\hatF_i(\hat{\bigK}, \bigL)$};
		\node (PAR_IC)   [rectangle, minimum width=2em, minimum height=2em, text centered, draw=black, fill=cyan!30, below of=MethodI_Den, node distance=1.8cm] {Particle initial conditions};
		\coordinate [left=1cm of MethodI_DF.north] (tr) ;
		\draw[arrow] (Ini_Den) -- (Ini_DF);
		\draw[arrow, red] (Ini_DF) -- (MethodI_DF)node[xshift=-0.575cm, yshift=0.6cm] {Method I};
		\draw[arrow, red] ([xshift=1.0cm]MethodI_DF.south) -- ([xshift=1.0cm]MethodI_DF.south |- MethodI_Den.north);
		\draw[arrow, Magenta] ([xshift=-0.6cm]MethodI_Den.north) |- (MethodII_DF)node[xshift=-2.265cm, yshift=0.17cm]{Method II};
		\draw[arrow, ForestGreen] (MethodI_DF) |- (PAR_IC)node[rotate=90, xshift=1.3cm, yshift=2.57cm] {Velocity sampling};
		\draw[arrow, ForestGreen] (MethodII_DF) |- ([yshift = 0.15cm]PAR_IC.east)node[rotate=90, xshift=1.17cm, yshift=-0.62cm] {Velocity sampling};
		\draw[arrow, dotted, Magenta] ([xshift = 5.5cm]MethodII_DF) |- ([yshift = -0.15cm]PAR_IC.east)node[rotate=90, xshift=1.33cm, yshift=-1.51cm] {Linear superposition};
		\draw[arrow, ForestGreen] (MethodI_Den.south) -- ++(0,-0.25cm) -| (PAR_IC)node[xshift=-0.32cm, yshift=0.95cm, text width=1.5cm] {Position sampling};
		\draw[arrow] (Ini_DF) -- (sigma);
		\draw[arrow] ([xshift=-0.05cm, yshift=-0.05cm]MethodI_DF.north east) -- (sigma.south west);
		\draw[arrow] (MethodII_DF.north) -- (MethodII_DF.north |- sigma.south);
		\draw[Bittersweet,thick,dashed] ($(Ini_Den.north west)+(-0.2,0.6)$)  rectangle ($(Ini_DF.south east)+(1.0,-0.2)$);
		\draw[blue,thick,dashed] ($(MethodI_DF.north west)+(-0.287,0.6)$)  rectangle ($(PAR_IC.south east)+(2.75,-0.8)$);
		\node[Bittersweet, text width=3cm] at (-0.1, 0.6) {\uline{Untruncated systems:}};
		\node[blue, text width=10cm] at (3.4, -6.2) {\uline{New truncation schemes presented in this work:}};
	\end{tikzpicture}
	\caption{Schematic design of \PIANISTpyNS. For an arbitrary input density profile $\varrho(R)$ that extends to spatial infinity, the associated (an)isotropic DF $F(\bigK, \bigL)$ is first computed, where the `energy-like' variable $\bigK = Z$ or $\bigQ_\pm$ depends on the adopted anisotropy model and $\bigL$ denotes the dimensionless angular momentum  defined in Eqs.~(\ref{eqn:relative_potential_energy},~\ref{eqn:DF_Functional_form_MT}). If everywhere physical and monotonic in $\bigK$ for dynamical stability, the DF can be self-consistently truncated in phase space $\hatF(\hat\bigK, \bigL)$ (Method I) while preserving the anisotropy profile $\beta_\sigma(R)$. The resulting truncated density profile $\hatvarrho(R)$ with finite spatial extent can be used to construct other anisotropic DFs $\hatF_i(\hat\bigK, \bigL)$ (Method II). The linear combination of a collection $\hatF_i(\hat\bigK, \bigL)$ for $i =2, 3, ...$ constructed from the same $\hatvarrho(R)$--potential pair can create more complex $\beta_\sigma(R)$, as discussed in \S\ref{ssec:DF_Superposition}. From these truncated DFs, \PIANISTpy then constructs particle initial conditions (\S\ref{sec:DynamicalStability}) and various velocity profiles (\S\ref{sec:OP_Vel}).}
	\label{fig:flowchart}
\end{figure}

\begin{figure}
        \centering
	\includegraphics[width=0.9\linewidth]{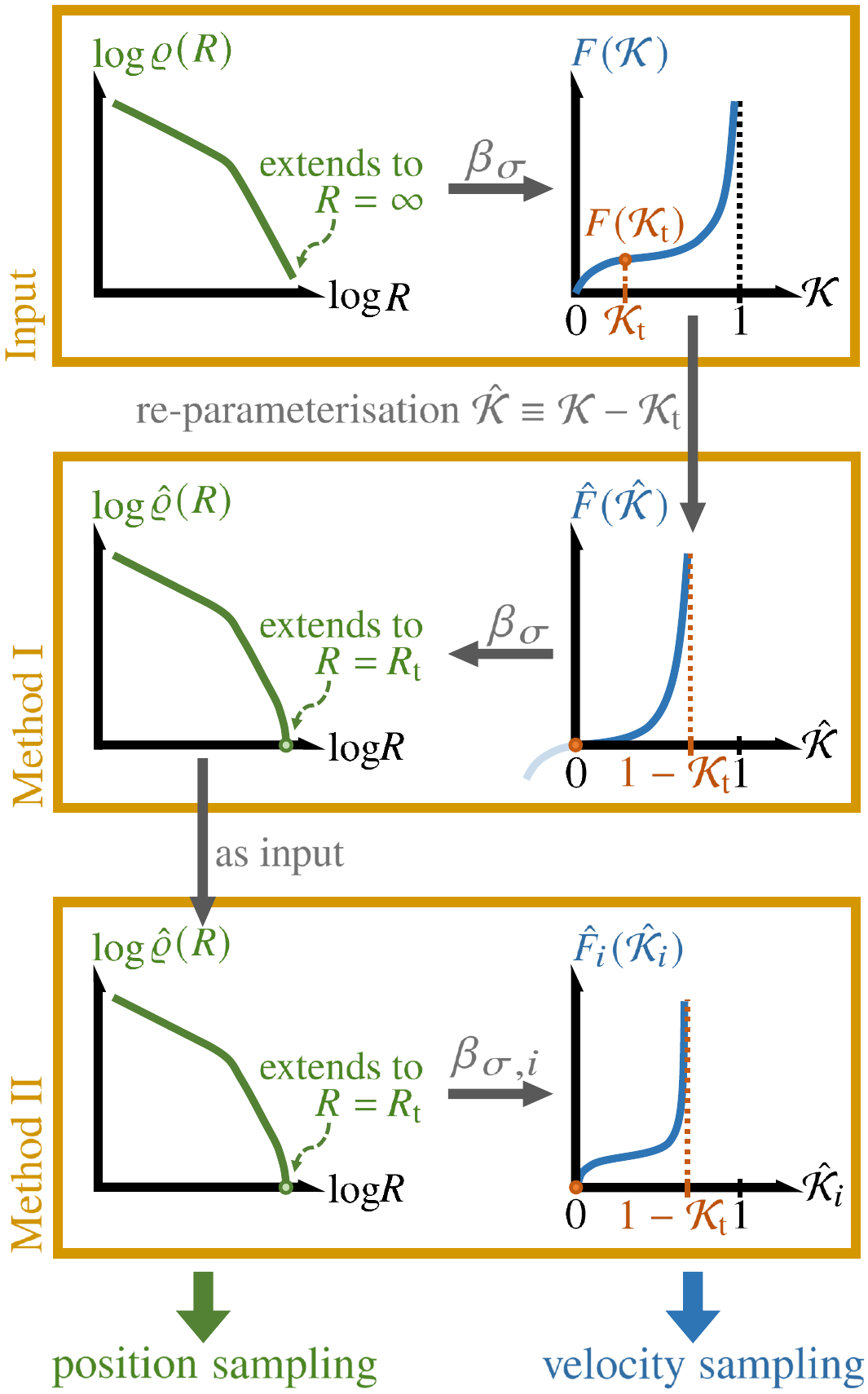}
	\caption{Schematic diagram of the two distinct truncation methods implemented in \PIANISTpy (see \S\ref{ssec:DF_Truncation} for details); the caret symbol denotes truncated quantities. \textbf{Input:} For a user-specified density profile $\varrho$ and anisotropy model $\betasigma$, we first construct the corresponding untruncated DF. \textbf{Method~I:} We `truncate' the DF in phase space by a re-parametrisation of the `energy-like' variable $\bigK$, by shifting $(\bigKt, F(\bigKt)) \rightarrow (0,0)$ and defining $\hat{\bigK} \eee \bigK -\bigKt$. We then integrate over all velocities to obtain self-consistently truncated density profile $\hatvarrho$ with finite truncation radius $\Rt$. Increasing the truncation value $\bigKt \in (0 ,1)$ reduces the total truncated mass. \textbf{Method~II:} We invert from a phase-space-truncated potential--density pair at $\bigKt$ to compute other distinct DFs in the form of \eref{eqn:DF_Functional_form_MT}, separable in both $\bigK$ and $\bigL$, under the same integration range $\hat\bigK \in [0, 1-\bigKt]$. This approach allows the construction of systems with identical truncated density profile $\hatvarrho$ but different anisotropy profiles $\beta_{\sigma,i}$.}
	\label{fig:Trun_Illustration}
\end{figure}

\fref{fig:flowchart} sketches the general workflow of \PIANISTpyNS. Starting from an input (untruncated) density profile such as \eref{eqn:double_power_law_MT}, one first constructs the DF corresponding to the desired velocity anisotropy profile $\betasigma(R)$. As an analytical density profile generally extends to spatial infinity, truncation is necessary to better represent the initial conditions of a physical gravitationally self-bound system with finite spatial extent. \PIANISTpy provides two complementary DF truncation approaches conceptually illustrated in \fref{fig:Trun_Illustration} and detailed in \S\ref{sec:DF_Truncation}. In essence, \textit{Method I} truncates an input density profile in phase space by re-parameterising the untruncated DF by a user-input truncation value $\bigKt$. This method guarantees self-consistency of the truncation boundary, but different anisotropy models produce distinct truncated density profiles. In applications that require identical truncated density profiles but with different anisotropy, the user can resort to \textit{Method II}, which inverts the truncated DF directly from an already Method-I-truncated potential-density pair.

The numerical implementation of \PIANISTpy is organised into three hierarchical layers. The innermost layer comprises stateless numerical primitives, including grid construction, numerical quadrature, cubic Hermite interpolation, and the Abel inversion kernel, which work for all input density profiles and are designed to remain stable and unchanged across all future extensions of the package. The middle layer contains the core physics routines, organised by feature (DF construction and truncation, particle sampling, and observables), each implemented as a self-contained block with a consistent function signature so that Method~I and Method~II solvers remain interchangeable at the call site. In particular, DF construction is shared across truncation methods, while the Poisson ODE system and the root-finding solver that enforces the self-consistent mass condition (see \S\ref{sec:DF_Truncation}) are encapsulated within the truncation routines. The outermost layer handles all workflow orchestration, including the user-facing interface and file I/O, and contains no numerical logic of its own. This layered architecture ensures that future additions require modifications only to the relevant physics layer, leaving the numerical primitives and the user-facing workflow interface unaffected.

\subsection{Classes and outputs}\label{ssec:OP}

\begin{figure*}
	\begin{lstlisting}[frame=tb, escapechar=~]
import ~pianistpy~ as pp
import astropy.units as u

# an untruncated NFW density profile with a constant velocity anisotropy of 0.25
halo = pp.Component(pp.DoublePowerLaw.nfw(), pp.Anisotropy(beta_c=0.25))
# compute the untruncated DF
df_untrun = pp.compute_df(halo)
# DF truncation via Method I, by requiring the final truncated mass equals the virial mass of untruncated NFW halo with concentration of 10 (or enclosed mass out to 10 times the scale radius)
df_trun = pp.truncate(df_untrun, pp.Truncation(mass_t=1.0, concentration=10))
# cast into physical units: a halo of virial mass 1e12 Msun and R_vir = c * r_0 = 200 kpc
df_trun = df_trun.to("physical", mass_0=1e12 * u.Msun, r_0=20 * u.kpc)
# draw 1e7 equal-mass particles in physical units centred at (0, 0, 0), and output as an HDF5 file
particles = pp.Sampler(df_trun).sample(10_000_000, seed=1)
particles.write("NFW_Const0p25_Mt1e12_PARIC.h5")\end{lstlisting}
	\caption{Importing \PIANISTpyNS, followed by an example code block to generate a DF-truncated particle initial condition of an NFW profile with a constant velocity anisotropy of $\betac = 0.25$. All outputs default to the internal units (see \S\ref{ssec:Units}), unless explicitly specified as in this example for a halo of total mass $10^{12}\,{\rm M_{\odot}}$. Code excerpts to reproduce, or to generate particle initial conditions used in, the subsequent figures are provided at \url{https://pianistpy.com}.}\label{fig:python_example}
\end{figure*}

Here, we provide a high-level overview for each of the main classes and outputs in \PIANISTpyNS: potential--density pair (\S\ref{sssec:OP_P_Rho_R}), DFs (\S\ref{ssec:OP_DFs}), various velocity dispersion profiles (\S\ref{sssec:OP_Vel}), and discrete particle sampling and initial conditions (\S\ref{sssec:OP_PARIC}). \fref{fig:python_example} provides a concrete end-to-end example from density profile initialisation to particle initial condition generation, illustrating how the entire \PIANISTpy workflow can be executed in just a few lines of code. Natively, \PIANISTpy supports HDF5 (via the \texttt{h5py} package \citep{Collettecollette2013python, Collette2022zndo6575970C}), plain text, and binary formats for file I/O.

\subsubsection{Potential--density pair}\label{sssec:OP_P_Rho_R}

The \texttt{DoublePowerLaw} class serves two distinct purposes within the code design of \PIANISTpyNS: (1) initialise (un)truncated input radial profiles and (2) output computed (un)truncated radial profiles. This class directly interacts with potential--density pairs, which consist of tabulated arrays of relative potential, density, and enclosed mass profiles on logarithmically spaced radial grids. In particular, \texttt{DoublePowerLaw(rho\_alpha,rho\_beta,rho\_gamma)} initialises untruncated potential--density pair instances of analytical double-power-law density profiles following \eref{eqn:double_power_law_MT}. Each truncated DF also stores the corresponding potential--density pair.

\subsubsection{Distribution functions}\label{ssec:OP_DFs}

The \texttt{df} module contains DF-related features in \PIANISTpyNS, including numerical methods specific to each class of analytical anisotropy models (\S\ref{ssec:DF_Truncation} and \aref{app:DF_Prelim}), Method I and II truncation pipelines (\aref{app:DF_Truncation}), and the linear superposition of Method-II-truncated DFs (\S\ref{ssec:DF_Superposition} and \aref{app:DF_superposition}).

As detailed in \S\ref{ssec:DF_Truncation}, \PIANISTpy supports DF construction derived from the following sets of analytical anisotropy profiles
\begin{align}\label{eqn:betasigma_list}
	\betasigma(R) = 
	\begin{cases}
		\betac & \text{(constant-anisotropy)},\\
		\dfrac{1}{1+(\Ra/R)^2} & \text{(OM Type~I: $\bigQ_{-}$)},\\
		\dfrac{1}{1-(\Ra/R)^2} & \text{(OM Type~II: $\bigQ_{+}$)},\\
		\dfrac{1+\betac(\Ra/R)^2}{1+(\Ra/R)^2} & \text{(COM Type~I: $\bigQ_{-}$)},\\
		\dfrac{1-\betac(\Ra/R)^2}{1-(\Ra/R)^2} & \text{(COM Type~II: $\bigQ_{+}$)}.
	\end{cases}
\end{align}
The constant-anisotropy model features radius-independent velocity anisotropy across the whole system. Osipkov\textendash Merritt (OM) models have an isotropic $\betasigma(R\ll\Ra) \simeq 0$ central region, and become completely radially anisotropic $\betasigma(R\gg\Ra) \simeq 1$ at large radii in Type~I or completely tangentially biased $\betasigma(R\rightarrow\Ra) \simeq -\infty$ near $\Ra$ in Type~II. At $R=\Ra$, the Type~II model comprises purely circular particle orbits and prompts a discontinuity in the velocity space; the DFs become unphysical for $R>\Ra$ \citep{Merritt1985AJ901027M}. Cuddeford--Osipkov--Merritt (COM) models are an extension of the OM parameterisation, featuring constant anisotropy at the centre $\betasigma(R\ll\Ra) \simeq \betac$ but otherwise reducing down to the OM models at large radii. Some example profiles are shown in the top row of \fref{fig:DF_Trun_DFs}.

\texttt{Anisotropy(beta\_c)} deals with DFs with constant (radius-independent) anisotropy profiles of \texttt{beta\_c} $=\betac \in (-\infty,1)$. If \texttt{beta\_c=0.0}, the subclass reduces to isotropy, which is treated with separate numerical routines (given the simplified analytical expressions) for speeding up calculations.

\texttt{Anisotropy(beta\_c,r\_a,kind)} deals with DFs with COM models of both radially (\texttt{kind="I"}) and tangentially (\texttt{kind="II"}) biased types. \texttt{r\_a} is the normalised anisotropy radius $\Ra\in(0,\infty)$ defined in \eref{eqn:relative_potential_energy}. If the central anisotropy is set to zero (\texttt{beta\_c=0.0}), the subclass reduces to OM models and is treated with separate numerical routines (given the simplified analytical expressions) for speeding up calculations.

\texttt{Component(profile,anisotropy)} defines one mass component of a system (e.g., a dark matter halo) by pairing an input density profile with one of the above anisotropy models. \texttt{compute\_df(component)} then returns the corresponding untruncated DF.

For the above two anisotropy subclasses, there are two distinct phase-space truncation methods, corresponding to Method I and Method II illustrated in \fref{fig:Trun_Illustration}. For Method I, \texttt{truncate(df\_untrun,Truncation(k\_t))} truncates an untruncated DF. The function argument \texttt{k\_t} is the truncation value $\bigKt$ defined later in \eref{eqn:DF_Truncation_MT}; a target mass \texttt{mass\_t} or radius \texttt{r\_t} can be specified instead. For Method II, the input arguments will remain identical (where now the input potential--density pair has already been phase-space truncated). The DF output files also encode information like the DF type and $\bigKt$ value in the file header.

\texttt{superpose(\{df\_trun:weight,...\})} performs linear superposition of Method-II-truncated DFs (with analytical anisotropy profiles) to produce flexible and non-analytical anisotropy profiles. The argument accepts a \PYTHON dictionary that maps each DF to its corresponding weight. The weights are automatically normalised if their sum is non-unity. In general, such superposed DFs have no closed-form analytical parameterisation in phase space.

\texttt{compute\_df} numerically verifies that the output DF is everywhere non-negative (physical) and monotonic (dynamically stable). Otherwise, the output records the value(s) where negativity and non-monotonicity are identified in the DF. Every output also carries the checks of its own numerical accuracy against the relative-error goal of $10^{-6}$ (\aref{app:JAX_imp}): a result above $10^{-4}$ is returned with a warning, and above $10^{-2}$ the computation stops unless explicitly overridden.

\subsubsection{Local and projected velocity distributions}\label{sssec:OP_Vel}

The \texttt{observables} module computes various differential and integrated velocity profiles from an input DF, each by its own function: radial $\sigma_\rmr$ (\texttt{sigma\_r}), tangential $\sigma_\rmt$ (\texttt{sigma\_t}), or line-of-sight $\sigma_{\rm los}$ (\texttt{sigma\_los}) velocity dispersions, as well as the overall line-of-sight velocity distribution (\texttt{losvd}) and the local velocity distribution evaluated at a fixed 3D radius from the system's CM (\texttt{pdf\_v}). Additionally, \PIANISTpy can compute these projected kinematic quantities averaged over centred or offset circular and rectangular apertures, as later illustrated in \fref{fig:Projection_Ops}. Several concrete examples are also provided in \S\ref{sec:OP_Vel}.

\subsubsection{Particle initial conditions}\label{sssec:OP_PARIC}

\texttt{Sampler(df\_trun).sample(n)} generates particle initial conditions from an input DF and the associated potential--density pair with \texttt{n} equal-mass particles. The detailed discrete particle sample implementation and related numerical routines are documented in \aref{app:num_par}.

\section{Phase-space Truncation}\label{sec:DF_Truncation}

\begin{figure*}
	\includegraphics[width=0.99\linewidth]{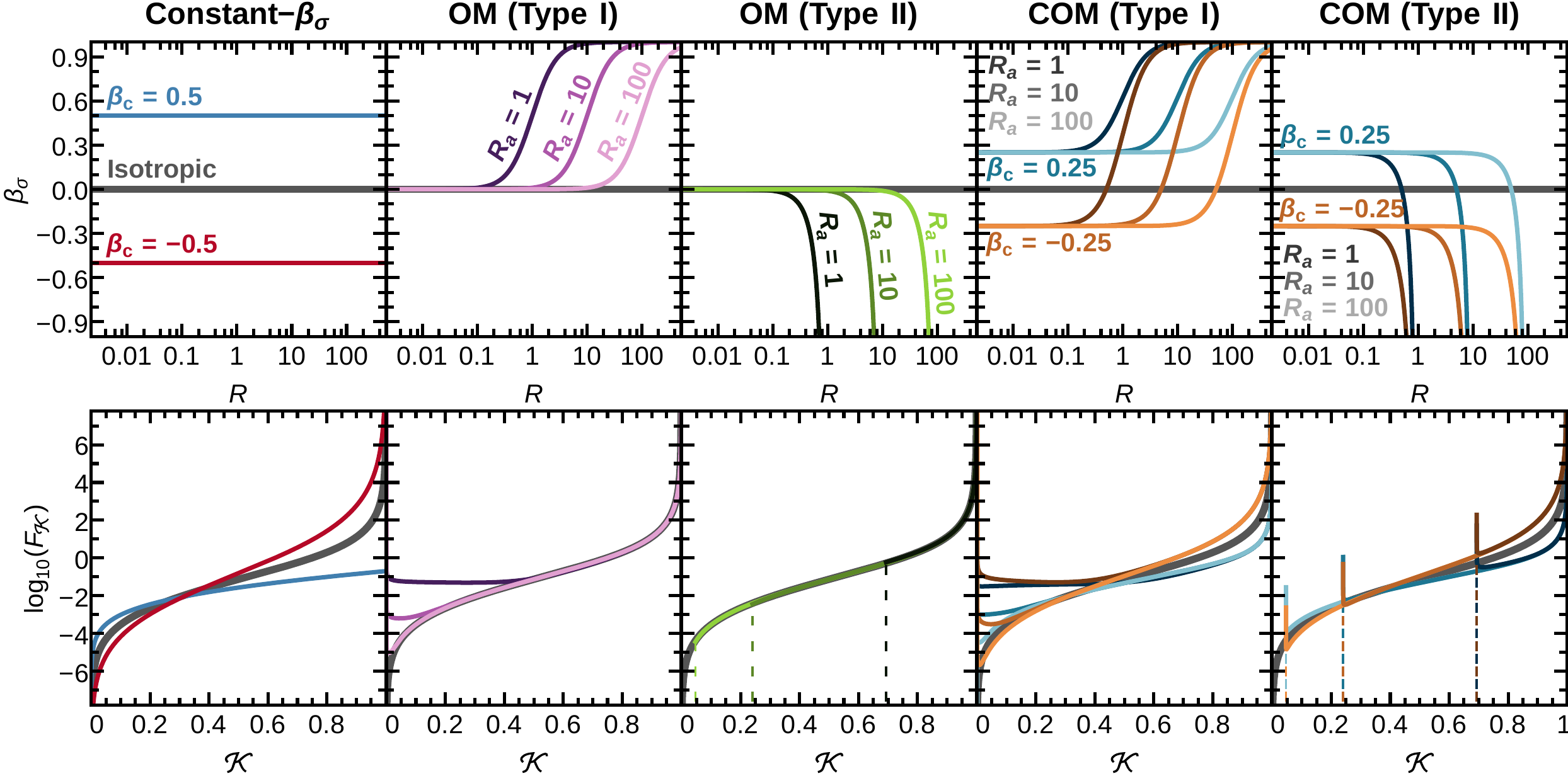}
    \caption{Anisotropy profiles (\textit{top row}) and untruncated DFs (\textit{bottom row}) for an untruncated NFW density profile with concentration $c_\rho = 10$. Different columns show different analytical anisotropy models implemented in \PIANISTpyNS, as indicated at the top. Within each column, different coloured curves correspond to different values of the model parameters, as indicated in each top panel; the isotropic case (dark grey) is reproduced in all columns for reference. Only the $\bigK$-part of each $F(\bigK, \bigL)$, denoted $F_\bigK(\bigK)$ as defined in \eref{eqn:DF_Functional_form_MT}, is shown; here $\bigK = Z$ for the constant-anisotropy model and $\bigK = \bigQ_{\pm}$ for (C)OM models. Note that all (C)OM Type~I models exhibit positively divergent low-$\bigK$ tails, while Type~II models terminate at $\bigQ_+ = P(\Ra)$ (vertical dashed lines), the relative potential at the anisotropy radius beyond which the model is analytically undefined; see \S\ref{ssec:DF_Truncation} and \aref{app:DF_Prelim}. At the scale radius $R = 1$ we have $\bigK = P(1) = \ln 2 \simeq 0.693$.}
	\label{fig:DF_Trun_DFs}
\end{figure*}

\subsection{Truncation of (an)isotropic DFs}\label{ssec:DF_Truncation}

\PIANISTpy implements constant (including isotropic), radial and tangential OM, and COM models, whose respective anisotropy profiles were introduced in \eref{eqn:betasigma_list}. Importantly, the DFs across all these models can be identically parameterised by the separable functional form\footnote{Other spherical equilibrium DFs that are separable in $E$ and $\bigL$ but with a non-power-law angular-momentum factor, such as the form \citet{Wojtak2008} fitted to simulated cluster haloes, are not of this form.}
\begin{align}\label{eqn:DF_Functional_form_MT}
	F(\bigK, \bigL) = F_\bigK(\bigK)\times \bigL^{-2\betac},
\end{align}
where $\bigK = Z$ for constant anisotropy and $\bigK = \bigQ_{\pm}$ for (C)OM models (see \aref{app:DF_Prelim} for details).

\PIANISTpy supports two distinct and complementary methods for truncating the density profiles of systems whose distribution function has the form of \eref{eqn:DF_Functional_form_MT}, as schematically illustrated in \fref{fig:Trun_Illustration}. Unlike truncation approaches that directly modify the density profile, which pose issues of DF self-consistency or dynamical instability around truncation boundaries (see \citealp{Weinberg2023MNRAS5254962W}, \citealp{2026MNRAS.546ag050B} and \S\ref{ssec:OP_Comparison}), both methods presented below perform truncation directly in phase space. Here, we focus on their key features; detailed derivations and specific caveats are presented in \aref{app:DF_Truncation}. In what follows, all truncated quantities are denoted by a caret (e.g., $\hat{F}$).

\medskip

\textbf{Method I:} This method is a generalisation of the energy truncation scheme for isotropic spherical systems formulated by \citet{Drakos2017MNRAS4682345D}. The dimensionless variable $\bigK\in [0,1]$ can be either $Z$ or $\bigQ_\pm$, depending on the anisotropy model of interest. The first step is to pick a truncation value $\bigKt \in (0,1)$. The corresponding \textit{truncated} DF $\hatF$ is parameterised by $\hat{\bigK} \eee \bigK-\bigKt \in [0,1-\bigKt]$, a translation of the energy-like variable such that $\hat{\bigK} = 0$ at the truncation boundary, according to
\begin{align}\label{eqn:DF_Truncation_MT}
	\hatF(\hat{\bigK}, \bigL) = 
	\begin{cases}
		F(\hat{\bigK} + \bigKt, \bigL) - F(\bigKt, \bigL) & \;\;(\hat{\bigK} \geq 0),\\
		0 & \;\;(\hat{\bigK} \leq 0),
	\end{cases}
\end{align}
where $F(\bigK, \bigL)$ is the original \textit{untruncated} DF\footnote{Note that the pure angular momentum dependence $\bigL^{-2\betac}$ is unchanged under truncation due to its separability in \eref{eqn:DF_Functional_form_MT}. In practice, \PIANISTpy directly truncates and outputs $F_\bigK(\bigK)$ following the identical procedure of \eref{eqn:DF_Truncation_MT}, instead of $F(\bigK, \bigL)$; see \aref{app:DF_Truncation}.}. This re-definition $\hatF(\hat{\bigK}, \bigL)$ is always possible by an appropriate shift of the reference constant $\Phi_0$ in \eref{eqn:unit_nor_4}. The truncated relative potential $\hatP$ is obtained by numerically integrating the Poisson equation
\begin{align}\label{eqn:Poissonr_Truncation_MT}
\frac{\rmd^2\hatP}{\rmd R^2} + \frac{2}{R} \frac{\rmd\hatP}{\rmd R} = -\hatvarrho(\hatP),
\end{align}
with the boundary conditions $\hatP(R=0) = 1-\bigKt$ and $\rmd\hatP/\rmd R\big|_{R\rightarrow0}$ set by the input inner density slope $\gamma$ (Eq.~[\ref{eqn:central_slope}])\footnote{This is one fundamental departure from \citet{Drakos2017MNRAS4682345D}, where their stated condition $\rmd\hatP/\rmd R|_{R=0}=0$ is valid strictly for $\gamma<1$ only, and is not fully self-consistent with the input potential--density pair near $R\to 0$ for $\gamma \geq1$. See \aref{app:DF_trun_I} for the technical details.} out to the truncation radius $\Rt$, where $\hatP(\Rt) \eee 0$. The corresponding truncated density profile simply follows from
\begin{align}\label{eqn:densityprof_Truncation_MT}
\hatvarrho(\hatP) = \int \rmd^3 \bV \, \hatF(\hat{\bigK}, \bigL),
\end{align}
where $\hat{\bigK}$ and $\bigL$ carry implicit dependence on $\hatP$.

Aside from preserving the input anisotropy profile $\betasigma(R)$ by construction, the algebraic re-parameterisation of \eref{eqn:DF_Truncation_MT} also leaves the inner density power-law slope of the truncated density profile $\hatvarrho$ invariant and unchanged from the input untruncated profile $\varrho$, regardless of $\bigKt$ or the adopted anisotropy model (see \aref{app:DF_inner_asymp}). In addition, if the untruncated DF is everywhere physical ($F \geq 0$) and monotonic (based on the stability condition $\partial F/\partial \bigK \geq 0$ of \citealp{Antonov1973} and \citealp{Hjorth1994ApJ424106H}), truncation via this phase-space re-parameterisation not only yields self-consistent boundary conditions but also guarantees that these well-behaved properties in $F$ are handed down to $\hatF$. \PIANISTpy can be used to find the value of $\Zt$ for a targeted $\Mt$ or $\Rt$ for any implemented density profile by specifying, in lieu of $\bigKt$, either $\Mt$ in units of $\Mz$ or $\Rt$ in units of $\rz$. Note that the truncation radius $\Rt$, the shape of the truncated density profile near $\Rt$, and the total truncated mass $\Mt$ will all vary across different anisotropy models, even for an identical input density profile at a fixed $\bigKt$. For applications that instead require the truncated density profiles to be identical across different anisotropy models \citep[as for example in][]{Chiang2025, Chiang2026}, this characteristic of Method~I necessitates the complementary Method~II framework discussed next.

\medskip

\textbf{Method II:} This method guarantees identical output density profiles by directly inverting an existing truncated $\hatvarrho(R)$--$\hatP(R)$ pair of one anisotropy model into that of other anisotropy models (see \aref{app:DF_trun_II} for technical details). This comes at the expense of altering the shapes of the output DFs, potentially introducing unphysical behaviour or non-monotonicity into them. For example, unlike Method~I, Method~II admits a DF only if $\betac > 1/2 - \lambda_\rmt$ for an input density that vanishes at the truncation radius as $\hatvarrho \propto \hatP^{\,\lambda_\rmt}$ (see \aref{app:DF_trun_II}); for all fiducial examples shown in this work with Method-I-truncated isotropic NFW model as input, this physical constraint translates to the requirement $\betac > -2$. By construction, these truncated DFs share the same $\bigKt$ and an identical $\hatvarrho(R)$, differing only in their velocity structures.

\begin{figure*}
	\includegraphics[width=0.99\linewidth]{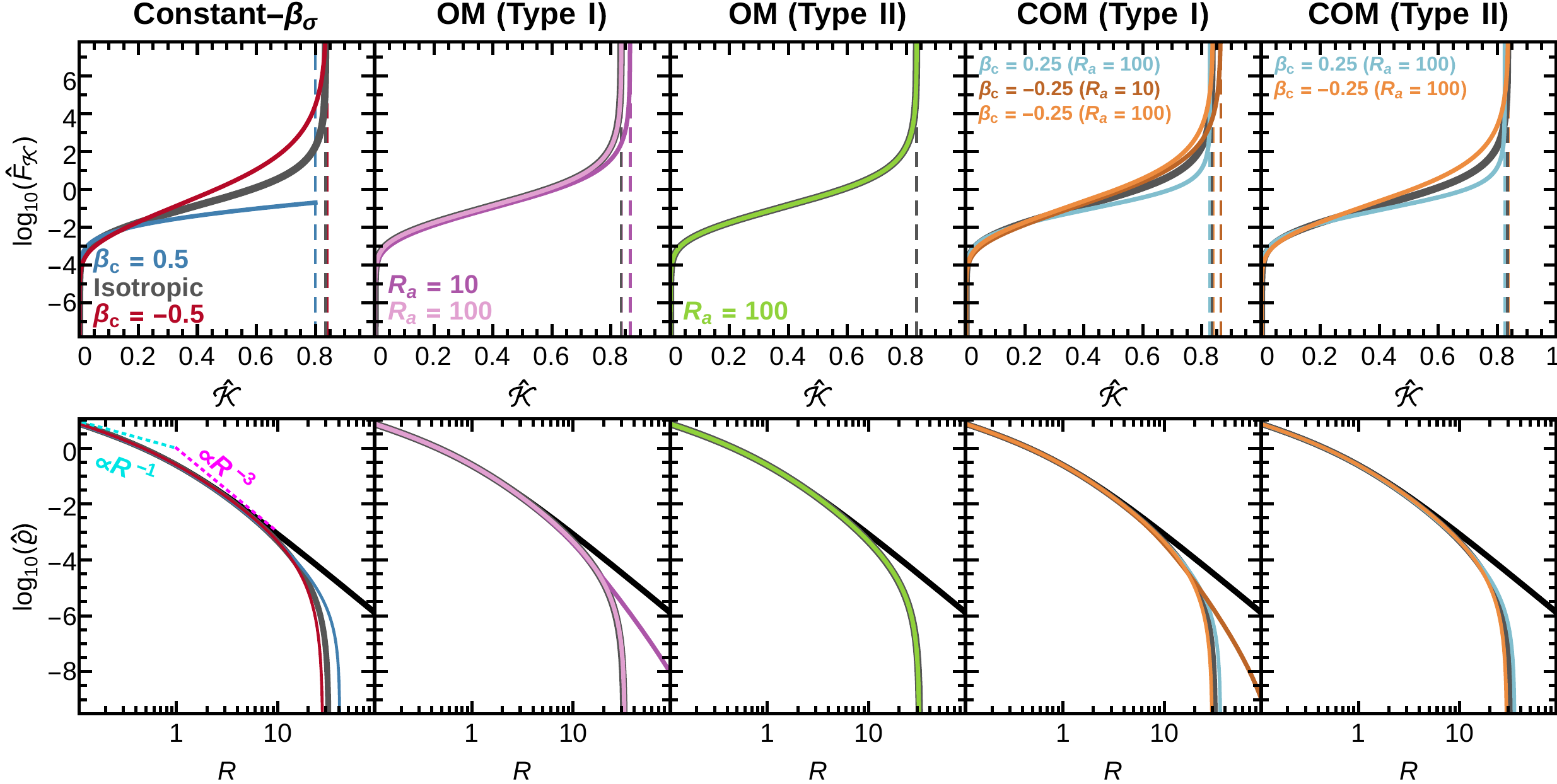}
        \caption{Same as \fref{fig:DF_Trun_DFs}, but now for Method-I-truncated DFs (\textit{top row}; only $\bigK$-part and see Eq.~[\ref{eqn:DF_Truncation_MT}]) and the corresponding density profiles (\textit{bottom row}), constructed from the same input NFW density profile truncated to have total mass equal to the untruncated NFW virial mass (vertical dashed lines in the top row mark $\bigK = 1 - \bigKt$, highlighting the anisotropy-dependent truncation values $\bigKt$). In the bottom row, we also show the untruncated NFW density profile (black reference curve) with the inner $\propto R^{-1}$ and outer $\propto R^{-3}$ slopes annotated. Not all models shown in \fref{fig:DF_Trun_DFs} appear here (see text for details).}
	\label{fig:DF_Trun_varrho}
\end{figure*}

\begin{figure*}
	\includegraphics[width=0.99\linewidth]{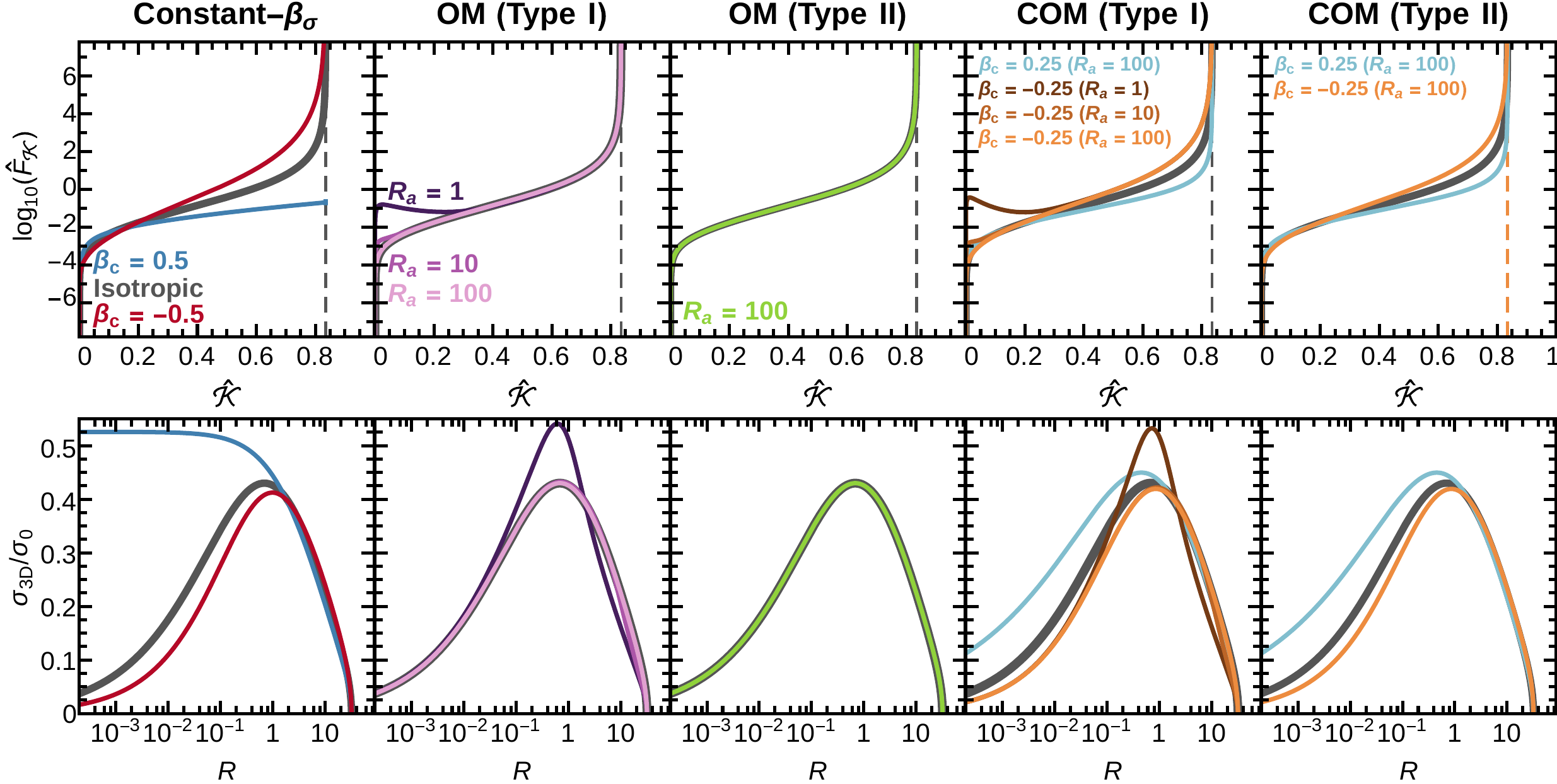}
        \caption{Same as \fref{fig:DF_Trun_DFs}, but now for Method-II-truncated DFs (\textit{top row}; only $\bigK$-part) and the corresponding 3D velocity dispersion profiles $\sigma_\text{3D}(R)/\sigma_0$ (\textit{bottom row}), all constructed by inverting the same Method-I-truncated isotropic potential--density pair shown in the leftmost column of \fref{fig:DF_Trun_varrho}. Here $\sigma_0 \eee \sqrt{G\Mz/\rz}$ is the model-unit velocity scale (\S\ref{ssec:Units}). Cases that do not yield a physical truncated DF under Method~II are not shown. By construction, all models share an identical truncated density profile $\hatvarrho$, with differences appearing exclusively in their velocity structures, a direct illustration of the mass--anisotropy degeneracy.}
	\label{fig:DF_Trun_sigma}
\end{figure*}

Figs.~\ref{fig:DF_Trun_DFs},~\ref{fig:DF_Trun_varrho}, and~\ref{fig:DF_Trun_sigma} together illustrate the two truncation methods across different anisotropy models, using an input NFW density profile with concentration $c_\rho = 10$ throughout. \fref{fig:DF_Trun_DFs} shows the untruncated baseline anisotropy profiles $\betasigma$ (top row) and corresponding DFs (bottom row). From left to right, the columns show the constant-anisotropy, OM Type~I, OM Type~II, COM Type~I, and COM Type~II models. These untruncated DFs serve as the direct input to Method~I, and indirectly to Method~II. Next, \fref{fig:DF_Trun_varrho} shows the corresponding Method-I-truncated DFs (top row) and density profiles $\hat{\varrho}$ (bottom row), following the same column layout and colour annotations as in \fref{fig:DF_Trun_DFs}. The truncation value $\bigKt$ of each curve is numerically determined and chosen such that the total truncated mass equals the untruncated NFW virial mass; this fiducial choice is adopted here for illustration only and is not a physical requirement. Any other target $\Mt$ or $\Rt$ can be specified, provided that a physical solution exists, which \PIANISTpy verifies numerically. Lastly, \fref{fig:DF_Trun_sigma} shows the Method-II-truncated DFs (top row) and the corresponding 3D velocity dispersion profiles $\sigma_\text{3D}$ (bottom row), following the same layout as in \fref{fig:DF_Trun_DFs}. Here, all models are constructed from the same Method-I-truncated isotropic potential--density pair shown in the leftmost column of \fref{fig:DF_Trun_varrho}. Across all three figures, only the $\bigK$-part of each DF, $F_\bigK(\bigK)$ as defined in \eref{eqn:DF_Functional_form_MT}, is shown. In Figs.~\ref{fig:DF_Trun_varrho} and~\ref{fig:DF_Trun_sigma}, we only show those cases that simultaneously obey $\hatF \geq 0$ and the requirement for the truncated mass.

We now discuss each anisotropy model in turn, highlighting the associated caveats of each truncation method and pointing to specific features in Figs.~\ref{fig:DF_Trun_DFs}--\ref{fig:DF_Trun_sigma}. The leftmost column of \fref{fig:DF_Trun_DFs} shows the constant-anisotropy model for $\betac = 0.5$ (blue), $0$ (dark grey), and $-0.5$ (red). These untruncated DFs share the same qualitative shape: a vanishing low-$\bigK$ tail, strict monotonicity throughout, and a positively divergent high-$\bigK$ limit as $\bigK \to 1$. The sole exception is $\betac = 0.5$, for which the central DF remains finite rather than divergent, a direct consequence of the Cusp Slope--Central Anisotropy theorem \citep{An2006ApJ642752A}, discussed in \S\ref{ssec:DF_Physicalness} below. Because all these DFs are everywhere physical and monotonic, Method~I truncation can be applied across the full range $\bigKt \in (0,1)$ for every value of $\betac$. As shown in the leftmost column of \fref{fig:DF_Trun_varrho}, the resulting truncated DFs inherit this well-behaved character, with the truncation boundary condition smoothly attenuating the outer density profile while preserving the inner $\propto R^{-1}$ density slope. The Method-II-truncated DFs for this family (leftmost column of \fref{fig:DF_Trun_sigma}) are equally well-behaved; the $\sigma_\text{3D}$ profiles vary systematically with $\betac$ at fixed truncated density profile, with radially (tangentially) biased models yielding higher (lower) central dispersions relative to the isotropic benchmark.

The second column of Figs.~\ref{fig:DF_Trun_DFs}--\ref{fig:DF_Trun_sigma} shows the OM Type~I (radially biased) model for $\Ra = 1, 10,$ and $100$ (darkest to lightest pink). The most striking feature of the untruncated DFs in \fref{fig:DF_Trun_DFs} is the positively divergent low-$\bigQ_-$ tail, which gives $\rmd F/\rmd\bigQ_- < 0$ and is associated with dynamical instability \citep{Hjorth1994ApJ424106H, Meza1997ApJ490136M}. As $\Ra \rightarrow \infty$, the OM Type~I DF asymptotes to the isotropic case, and the divergent tail recedes toward $\bigQ_- \to 0$. Because of this low-$\bigQ_-$ divergence, Method~I truncation is restricted to $\bigKt \geq \bigQ_-^\text{min}$, where $\bigQ_-^\text{min}$ is the value at which $F$ reaches its global minimum. Furthermore, for models with small $\Ra$ where the outer constituent orbits are nearly radial and hence non-local, sufficiently small $\bigKt$ will still yield a truncated density profile that extends to spatial infinity with divergent total truncated mass. A sufficiently large $\bigKt$ is required for the truncation boundary to become finite, but the total truncated mass then already falls below $\Mvir$ in this particular example. Hence, these models do not have $\bigKt$ that simultaneously yield a physical truncated DF and $\Mt = \Mvir$. This explains why the $\Ra=1$ model is not shown in \fref{fig:DF_Trun_varrho}. For the cases that do admit a viable solution, the truncated density profiles (bottom row of \fref{fig:DF_Trun_varrho}) show that radially biased models naturally yield truncation radii considerably larger than that of the isotropic case, reflecting the non-local nature of nearly radial orbits. Turning to \fref{fig:DF_Trun_sigma}, Method~II successfully removes the divergent low-$\bigQ_-$ tails entirely for all $\Ra$ shown, producing physically well-behaved truncated DFs. The one exception is the $\Ra = 1$ case, which retains a $\rmd\hatF/\rmd\hat{\bigQ}_- < 0$ feature around $\hat{\bigQ}_- \sim 0.2$, indicative of dynamical instability\footnote{Under the same isolation test described in \S\ref{ssec:OP_Comparison}, the corresponding particle initial condition remains stable for $\sim$18~Gyr before relaxing into an aspherical configuration.}.

The third column of Figs.~\ref{fig:DF_Trun_DFs}--\ref{fig:DF_Trun_sigma} shows the OM Type~II (tangentially biased) model for $\Ra = 1, 10,$ and $100$ (darkest to lightest green). Their untruncated DFs terminate exactly at the relative potential at the anisotropy radius $\bigQ_+ = P(\Ra)$, due to the factor $(1 - R^2/\Ra^2)^{-1}$ causing an essential discontinuity at $R = \Ra$ and no physical solutions for $R>\Ra$ (see \aref{app:DF_Prelim} for details). Method~I naturally respects this by restricting truncation to $\bigKt > P(\Ra)$, i.e. $\Rt < \Ra$. The same model-specific constraint also applies to the physical validity of Method-II-constructed solutions, as shown in \fref{fig:DF_Trun_sigma}.

The fourth and fifth columns of Figs.~\ref{fig:DF_Trun_DFs}--\ref{fig:DF_Trun_sigma} show the COM Type~I and Type~II models, respectively, for $\betac = \pm 0.25$ and $\Ra = 1, 10, 100$ (darkest to lightest shading within each $\betac$ group). The COM family generalises the OM framework by allowing a non-zero central anisotropy $\betac$, and this richer parameter space is immediately reflected in the diversity of untruncated DF shapes visible in \fref{fig:DF_Trun_DFs}. Crucially, the low-$\bigK$ tail behaviour of COM models follows OM models of the same Type, irrespective of the value of $\betac$. For example, in \fref{fig:DF_Trun_DFs}, the COM Type~I curves with $\betac = \pm 0.25$ both resemble the OM Type~I family. The COM Type~II curves all end at $\bigQ_+ = P(\Ra)$ as the OM Type~II, but now exhibit positive divergence as $\bigQ_+ \rightarrow P(\Ra)^{+}$, because the factor $(1 - R^2/\Ra^2)^{1-\betac}$ in the model parameterisation (see Eq.~[\ref{eqn:linearity_beta_1}]) no longer carries an integer power, such that the DF is no longer analytic across $R = \Ra$ (cf. this same factor reduces to a polynomial due to integer power in the OM baseline and the DF is regular and simply terminates at $P(\Ra)$ without such divergence, as further discussed in \aref{app:DF_Prelim}). Consequently, all the truncation caveats identified for the OM models carry over to the COM family of the same Type, with the allowed range of $\bigKt$ and the viability of Method~I depending jointly on $\betac$ and $\Ra$. Cases that do not admit a physically viable Method~I solution under the $\Mt = \Mvir$ requirement are not shown in \fref{fig:DF_Trun_varrho}. The Method-II results in \fref{fig:DF_Trun_sigma} demonstrate that inverting from the shared isotropic potential--density pair produces well-behaved truncated DFs across the COM family for those admitting physical DF solutions (the $\betac = -0.25$, $\Ra = 1$ case retains a non-monotonic low-$\bigK$ feature, as for OM Type~I), with the velocity dispersion profiles in the bottom row spanning a wide range of shapes that reflect the combined influence of $\betac$ and $\Ra$ on the velocity structure at fixed density. 

Taken together, Figs.~\ref{fig:DF_Trun_DFs}--\ref{fig:DF_Trun_sigma} demonstrate that the two truncation methods are broadly applicable across all implemented anisotropy families, with Method~II offering a robust alternative whenever Method~I does not admit a viable solution or produces undesirable outer density profile features.

\subsection{Linear superposition of truncated DFs}\label{ssec:DF_Superposition}

Distribution functions $\hat F_i$, which are generated from the same underlying truncated potential--density pair via Method-II, can be linearly superposed to construct a new DF $\hat F=\sum_i c_i\hat F_i$ (with $\sum_i c_i=1$ and $\forall c_i > 0$\footnote{In principle, superposition with $\sum_i c_i=1$ but allowing $c_i < 0$ could still yield physically self-consistent composite DFs, as long as they remain everywhere physical in phase space. \PIANISTpy enforces the constraint $\forall c_i > 0$ to avoid issues with this physical caveat and in numerical implementations.}) that is guaranteed to correspond to the same truncated density profile (see \aref{app:DF_superposition} for technical details). By using the analytical anisotropy models implemented in \PIANISTpy (Eq.~[\ref{eqn:betasigma_list}]) as basis functions, this allows the construction of systems with a wide range of anisotropy profiles that are not captured by any known closed-form analytical model. \PIANISTpy can be used to initialise particle initial conditions (\S\ref{sec:DynamicalStability}) or to forward model various velocity profiles and distributions (\S\ref{sec:OP_Vel}) for any such composite DF.

\begin{figure}
    \centering
    \includegraphics[width=0.99\linewidth]{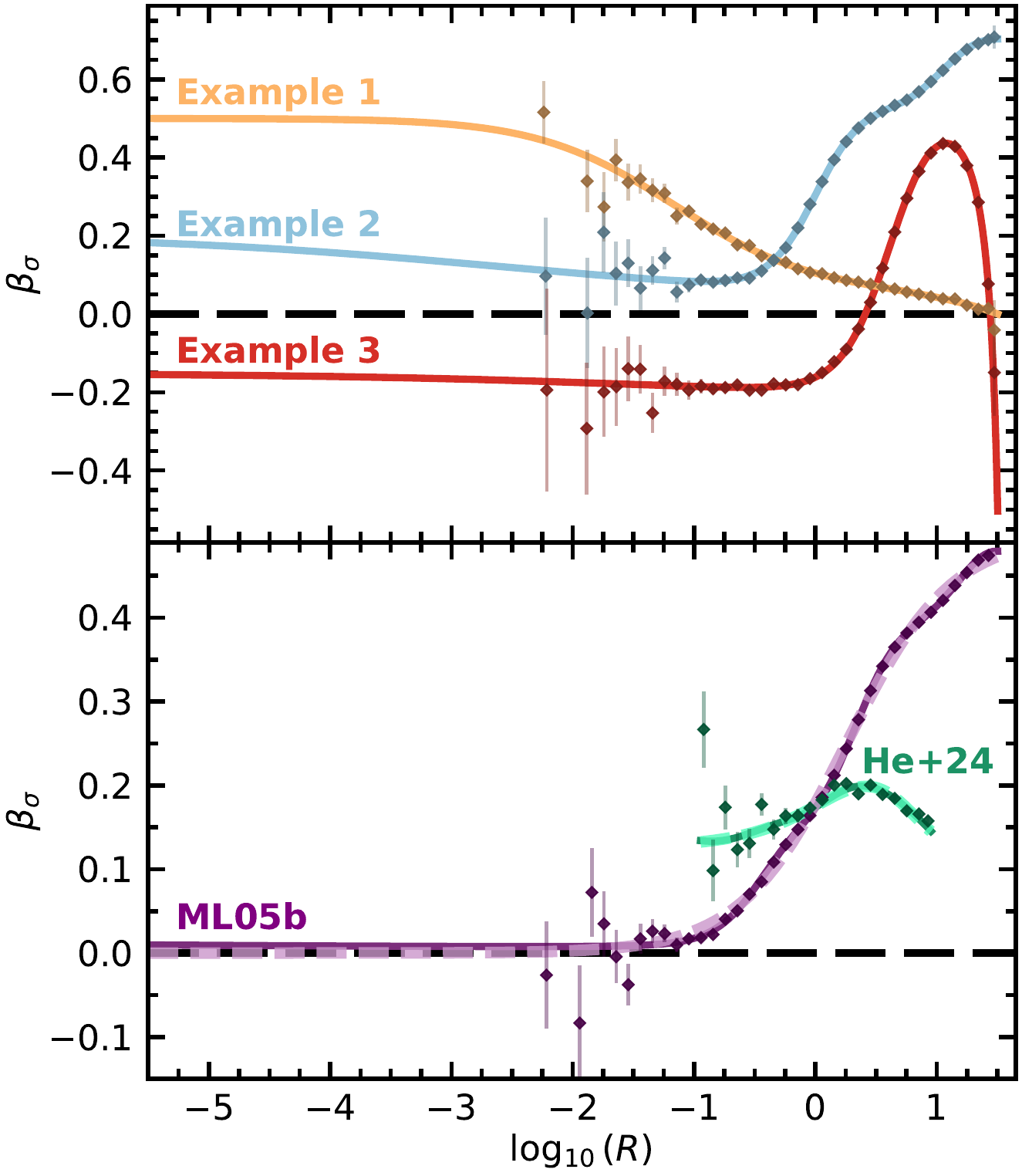}
    \caption{Velocity anisotropy profiles of composite DFs constructed via linear superposition of Method-II-truncated DFs, all inverted from the Method-I-truncated isotropic potential--density pair shown in the leftmost column of \fref{fig:DF_Trun_varrho}. \textit{Top:} Example~1 superposes constant-anisotropy models; Example~2 mixes constant-anisotropy and OM Type-I models; Example~3 superposes COM models (see text for details). Solid curves show the anisotropy profiles computed directly with \PIANISTpyNS, while the diamond symbols show the results inferred from $N$-body realisation (also generated with \PIANISTpyNS) with $10^7$ particles. Error bars indicate $1\sigma$ bootstrap uncertainties. \textit{Bottom:} Target anisotropy profiles of simulated dark matter haloes (dotted curves) from \citet{2005MNRAS.363..705M} (purple) and \citet{He2024ApJ976187H} (green), and the respective \PIANISTpyNS-best-fit models (solid curves) via linear superposition of seven basis DFs. Diamonds and error bars are again derived from $N$-body realisations, now with $10^8$ (purple) and $10^6$ (green, sampled over only the fitted radial extent) particles.}
\label{fig:Linear_Super_betasigma}
\end{figure}

\fref{fig:Linear_Super_betasigma} showcases the anisotropy profiles of three such composite DFs, all constructed from the Method-II-truncated isotropic potential--density pair shown in the leftmost column of \fref{fig:DF_Trun_varrho}. In the top panel, Example~1 is an equal-weight four-component superposition of constant-anisotropy models with $\betac = +0.5, +0.1, -0.2,$ and $-0.4$. Example~2 mixes two OM Type-I models with anisotropy radii $\Ra = 1$ and $\Ra = 10$ with a constant-anisotropy component $\betac = 0.2$, all at equal weights. Example~3 superposes with equal weight a COM Type-I model ($\betac = -0.3$, $\Ra = 4$) with a COM Type-II model ($\betac = -0.1$, $\Ra = 40$). As shown in \fref{fig:Linear_Super_betasigma}, these superpositions produce anisotropy profiles with rich radial variation that are not captured by any known closed-form analytical model. The analytical predictions are validated directly against $N$-body realisations with $\Npar = 10^7$ particles (diamond symbols), with $1\sigma$ bootstrap uncertainties shown as error bars, confirming that \PIANISTpy accurately recovers the intended anisotropy profile across the full radial range.

To further showcase the practical usage of this flexible framework, we consider in the bottom panel of \fref{fig:Linear_Super_betasigma} two example anisotropy profiles measured directly from dark matter haloes at redshift zero in cosmological simulations. The first is the best-fitting profile of \citet{2005MNRAS.363..705M}, $\betasigma(R) = R/[2(R + 1.8)]$ for haloes with $c_\rho = 10$; the second is the best-fitting $\betasigma(R)$ that \citet{He2024ApJ976187H} report for TNG300 haloes of $10^{12.6\text{--}13}\Msun$, defined over the radial range $0.113 \leq R \leq 8.97$. These targets (purple and green dotted curves, respectively) are compared with their \PIANISTpyNS\ best-fit counterparts (solid curves), each a linear superposition of seven basis DFs, which recover the first (second) target to a maximum absolute deviation in $\betasigma$ of $9.7\times10^{-3}$ ($1.7\times10^{-3}$) over its entire range of validity. We additionally show the $N$-body realisations of the two best-fit models, with $10^8$ (purple) and $10^6$ (green, sampled over the fitted radial range only) particles. This proof-of-concept exercise demonstrates, for the first time, the construction of fully self-consistent and exact DF and $N$-body models (as opposed to iterative and approximate alternatives) that are best-fit to non-analytic input anisotropy profiles.

We emphasise that these anisotropy profiles are constructed without explicit computation of any new DF beyond the Method-I-truncated isotropic base case. Rather, they are obtained by direct evaluation of the simplified analytical expression given by \eref{eqn:beta_sigma_sup}. Empirically, each numerical evaluation of a seven-member composite profile in the bottom-panel example of \fref{fig:Linear_Super_betasigma} takes only $\sim10\,\mu$s on a single workstation-grade CPU core (and $\sim15\,\mu$s together with its gradient with respect to the weights), and a full minimax fit over the library of 246 basis DFs about $6$~s. The evaluation cost is overhead-dominated below about a hundred basis DFs and grows linearly with their number beyond a few hundred, thus permitting efficient Markov chain Monte Carlo (MCMC)-based parameter exploration over the space of these composite anisotropy models.

\subsection{Physical constraints on density and anisotropy models}\label{ssec:DF_Physicalness}

For a given physical density profile $\varrho(R)$, not all anisotropy profiles $\betasigma(R)$ produce DFs that are guaranteed to be everywhere physical (i.e., $F\geq0$). Similarly, certain analytical density profiles yield unphysical DFs even under the restrictive assumption of velocity isotropy \citep[e.g.,][]{Baes2020AA634A109B, Dattathri2025}. Alongside direct numerical exploration with \PIANISTpyNS, analytical inspection of the DF in the limit $P\rightarrow1$ offers useful guiding principles for navigating the space of $\varrho(R)$--$\betasigma(R)$ combinations. Specifically, the Cusp Slope\textendash Central Anisotropy theorem \citep{An2006ApJ642752A} states that the central velocity anisotropy of any density profile with a pure power-law inner cusp $\varrho(R\rightarrow0)\propto R^{-\gamma}$ has a physical upper limit,
\begin{align}\label{eqn:CSCA}
	\betasigma(R\rightarrow0) \leq \gamma/2,
\end{align}
above which the DF becomes unphysical, with $F(P\rightarrow1) < 0$. As validated and extended in \citet{Ciotti2010MNRAS4011091C, Ciotti2010MNRAS4081070C}, this is a necessary condition to ensure that resulting DFs remain physical at the centre. For example, a Plummer sphere ($\gamma = 0$) cannot be supported by a radially anisotropic central core.

The boundary case $\betasigma(R\rightarrow0) = \gamma/2$ requires a more careful treatment beyond the original derivation of \citet{An2006ApJ642752A}. At this boundary, a detailed asymptotic analysis of the central DF (see \aref{app:DF_inner_asymp}) shows that its behaviour as $\bigK \rightarrow 1$ is governed by a profile-dependent exponent $\kappa'$. For the family of double-power-law density profiles given by \eref{eqn:double_power_law_MT}, this exponent takes the explicit form
\begin{align}\label{eqn:kappa_prime}
    \kappa' \equiv \frac{\alpha}{2-\gamma} - 
    \left(\frac{3}{2} - \betac\right),
\end{align}
so that, provided $\lambda_\rmc \eee \alpha/(2-\gamma)$ is not a positive integer, the central DF diverges as $(1-\bigK)^{\kappa'}$ for $\kappa' < 0$, diverges logarithmically for $\kappa' = 0$, and is finite for $\kappa' > 0$; for integer $\lambda_\rmc$ it is finite irrespective of $\kappa'$ (see \aref{app:DF_inner_asymp}). Importantly, since phase-space truncation preserves the inner density power-law slope (see \S\ref{ssec:DF_Truncation}), Eqs.~(\ref{eqn:CSCA},~\ref{eqn:kappa_prime}) apply to both untruncated and truncated DFs. For example, an NFW profile ($\alpha = \gamma = 1$) with $\betac = 0.5$ lies precisely on the boundary of \eref{eqn:CSCA} with $\lambda_\rmc = 1$, and thus has a finite central DF, as shown in the leftmost column of Figs.~\ref{fig:DF_Trun_DFs}--\ref{fig:DF_Trun_sigma}.

\section{Dynamical Stability of Anisotropic Systems}\label{sec:DynamicalStability}

\subsection{Comparison against existing initial condition codes}
\label{ssec:OP_Comparison}

In this section, we benchmark \PIANISTpyNS's numerical utilities by testing the accuracy of the numerical solver and the dynamical stability of particle initial conditions. We compare \PIANISTpy against two widely used codes that are publicly available and provide similar functionalities\footnote{Other codes for generating initial conditions are also available, but do not allow the construction of spherical anisotropic systems with a truncated density profile. In particular \href{https://github.com/ndrakos/ICICLE}{\textsc{icicle}} \citep{Drakos2017MNRAS4682345D} is limited to isotropic systems, even though it supports energy-based truncation. \href{https://wwwmpa.mpa-garching.mpg.de/~volker/galic/}{\textsc{galic}} \citep{Yurin:2014pga} only implements the Hernquist profile and offers no truncation utility. \href{https://bitbucket.org/ymiki/magi}{\textsc{Magi}} \citep{Miki2018MNRAS4752269M} supports only isotropic systems and offers no truncation utility. \href{https://github.com/rerrani/nbopy}{\textsc{nbopy}} \citep{Errani2020MNRAS491} supports only isotropic systems with `hard' truncation.}:
\begin{itemize} 
\item \href{https://docs.galpy.org/}{\galpy} \citep{Bovy2015ApJS21629B} is a potential-based galactic dynamics package in which fixed, analytical (or analytically expanded) gravitational potentials serve as the foundational input for all downstream calculations, including the evaluation of DFs. Given its strictly potential-forward code design, \galpy does not possess a native numerical Poisson solver (cf.\ Eq.~[\ref{eqn:Poissonr_Truncation_MT}]) to derive a self-consistent potential--density pair from a DF alone. For truncation, \galpy first initialises the DF $F(Z, \bigL)$ of the \textit{untruncated} density profile (e.g., a pure NFW extending to spatial infinity), and then imposes a hard Heaviside cutoff in energy space,
\begin{align}\label{eqn:galpy_truncation} 
\hat{F}(Z, \bigL) = F(Z, \bigL) \times \Theta\!\left[P_\rmt - Z\right], 
\end{align} 
where $\Theta$ is the Heaviside step function and $P_\rmt = P(\Rt)$ is the dimensionless relative potential evaluated at the truncation radius. This approach differs fundamentally from \PIANISTpyNS's phase-space truncation in two respects. First, whereas \PIANISTpy continuously re-parameterises $\bigK \to \hat{\bigK} = \bigK - \bigKt$ and shifts the DF baseline to enforce the boundary condition $\hatF(\hat{\bigK}=0, \bigL) = 0$ exactly at $\Rt$, \galpy retains the full DF amplitude $F(P_\rmt, \bigL)$ up to $\Rt$ and abruptly zeros it beyond, introducing a finite jump discontinuity of magnitude 
$F(P_\rmt, \bigL)$ at $Z = P_\rmt$. Second, and more critically, $F(Z, \bigL)$ in \eref{eqn:galpy_truncation} is never re-inverted from the truncated density profile (unlike \PIANISTpyNS's Method~I, which explicitly re-derives $\hatvarrho$ from $\hatF$ via Eq.~[\ref{eqn:densityprof_Truncation_MT}]), so the resulting $\hatF$ and the potential--density pair are physically self-inconsistent\footnote{Directly integrating such truncated DF over velocity space yields a density profile that systematically falls below the untruncated profile near $\Rt$. Even if one instead re-inverts the DF directly from the hard-truncated density profile via \eref{eqn:hatF_bigE_constant_beta}, the resulting DF will be unphysical ($\hatF < 0$) and ill-behaved (with inflection points) at the truncation boundary \citep{Weinberg2023MNRAS5254962W}, unless all orbits at $\Rt$ are completely tangential \citep{Baes2023MNRAS5196065B}.}. 

\item \href{https://github.com/GalacticDynamics-Oxford/Agama}{\agama} \citep{Vasiliev2019MNRAS4821525V} is an action-based package whose DFs are constructed natively in the space of three action variables $\mathbf{J} \equiv (J_r, J_\phi, J_z)$. Because actions are adiabatic invariants under slow potential changes, \agama supports a fully self-consistent numerical Poisson solver that iterates between the DF and the gravitational potential until convergence. However, since the relative energy $Z$ is a nonlinear combination of the actions, this architecture does not preserve the classical separability $F(Z, \bigL) = F_Z(Z) \times \bigL^{-2\betac}$ that is crucial to \PIANISTpyNS's phase-space truncation framework (\S\ref{ssec:DF_Truncation}). Instead, \agama truncates the density profile directly in configuration space by applying an exponential suppression, after which the DF is self-consistently re-inverted from the suppressed density profile numerically. While this guarantees internal self-consistency between the truncated $\hat{\varrho}(R)$ and $\hat{P}(R)$, the truncation is imposed in configuration space with no phase-space counterpart, and the resulting DF near $\Rt$ is not guaranteed to be everywhere physical and/or monotonic \citep{Weinberg2023MNRAS5254962W}.
\end{itemize}

\begin{figure}
    \includegraphics[width=\linewidth]{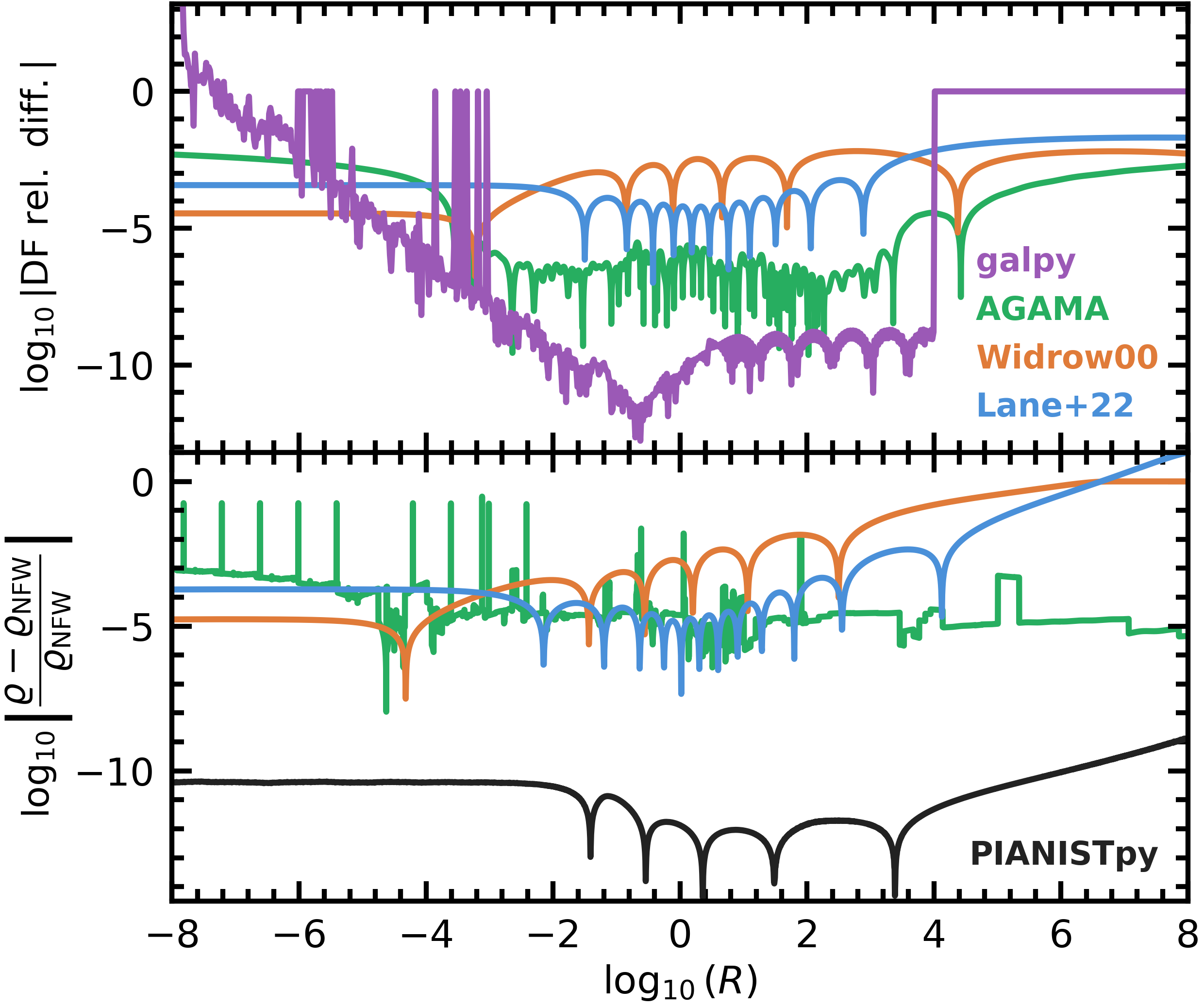}
    \caption{Roundtrip numerical accuracy benchmark for an untruncated isotropic NFW profile, comparing \galpy (purple), \agama (green), \PIANISTpy (black), and the analytical fitting functions of \citet{Widrow2000ApJS13139W} (orange) and \citet{Lane2022MNRAS5105119L} (blue), the latter two passed through the same density integration and Poisson solve as \PIANISTpyNS's DF. \textit{Top:} Relative difference in the isotropic DF $F(Z)$ with respect to \PIANISTpyNS's output, mapped onto radius via $Z = P(R)$ by the analytical NFW potential. \textit{Bottom:} Relative error in the roundtrip-inferred density against the analytical NFW profile; \galpy is omitted as it lacks a native Poisson solver.}
    \label{fig:NFW_Benchmark_Iso}
\end{figure}

\begin{figure*}
	\includegraphics[width=\linewidth]{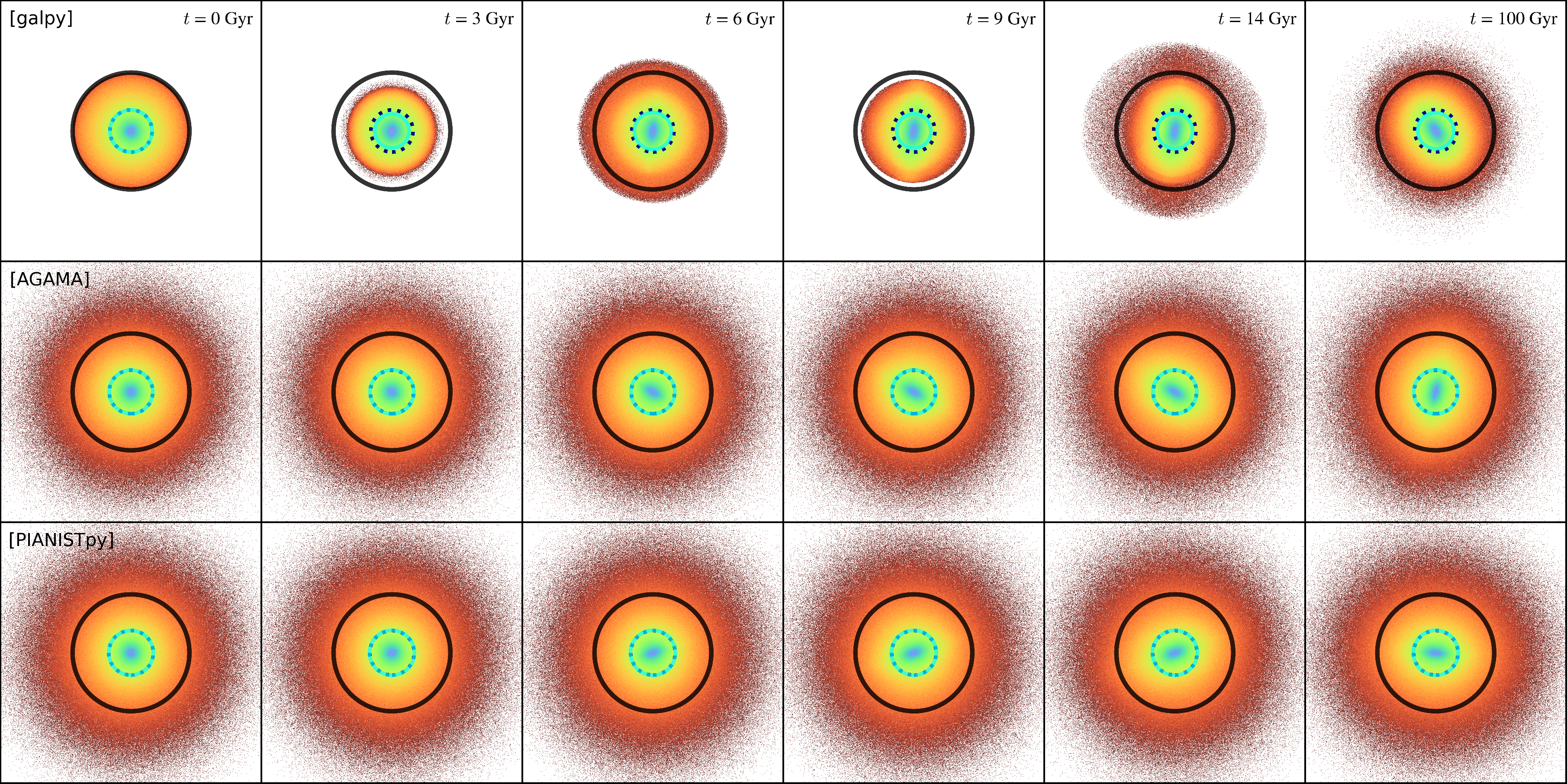}
    \caption{Time evolution of truncated NFW haloes in isolation with initial anisotropy $\betac = 0.5$, resolved with $10^7$ particles and colour-coded by projected surface density. Different rows correspond to initial conditions generated by \galpy (top), \agama (middle), and \PIANISTpy (bottom), all with total mass equal to the untruncated NFW virial mass; see text for details. Columns show different epochs, as indicated. The black solid circle marks the initial virial radius; the dark blue dotted (cyan solid) circle marks the initial (instantaneous) half-mass radius enclosing 50\% of the total mass.}
	\label{fig:Isolation_Comp_beta0p5}
\end{figure*}

We first examine the accuracy of the numerical solvers of \galpyNS, \agamaNS, and \PIANISTpy via the following `roundtrip' test, all in \texttt{float64} (machine precision $\simeq 10^{-16}$). Starting from an analytical untruncated NFW density profile, each code numerically evaluates the corresponding isotropic DF $F(Z)$ at its default numerical settings or finer. To reconstruct the radial density profile, each code integrates the respective numerical DF over all velocities via \eref{app:Den_Def} to obtain the density as a function of potential $\varrho(P)$; the Poisson equation \eref{eqn:Poissonr_Truncation_MT} is then solved numerically to map $\varrho(P)$ back onto radial profile $\varrho(R)$. This reconstructed profile is then compared against the input analytical NFW density profile to quantify the roundtrip accuracy. High roundtrip accuracy is particularly critical in the \PIANISTpyNS\ truncation framework: for example, constructing a Method-II-truncated initial condition requires a multi-step pipeline $\varrho \to F \to \hatF \to \hat{\varrho} \to \hatF_i$ (see \fref{fig:Trun_Illustration}), where numerical errors can rapidly amplify across each inversion step.

\fref{fig:NFW_Benchmark_Iso} shows the results of this roundtrip test. We include \galpy (purple), \agama (green), and \PIANISTpy (black), as well as two analytical fitting functions for the isotropic NFW DF: that of \citet{Widrow2000ApJS13139W} (orange), which has been widely used in the literature, and the more accurate expression of \citet{Lane2022MNRAS5105119L} (blue), adopted as the default isotropic NFW DF in \galpyNS. For both fitting functions, we apply the same density integration and Poisson solve as for \PIANISTpyNS's DF to compute the reconstructed density profile. The top panel shows the relative difference in the isotropic $F(Z)$ with respect to \PIANISTpyNS's output, mapped onto radius via $Z = P(R)$ (i.e., $V = 0$) using the analytical NFW potential for visualisation; \PIANISTpy itself is therefore not shown in this panel. The bottom panel compares the relative error in the roundtrip-inferred density against the analytical NFW profile, now including \PIANISTpy as one of the compared methods; since \galpy lacks a native Poisson solver, it is omitted from this panel.

Quantitatively, the top panel reveals that the DF computed with \galpy agrees with \PIANISTpyNS's output to better than $3\times10^{-8}$ over the radial interval $10^{-3} \lesssim R \lesssim 10^{4}$. \galpy outputs zero for all DF entries at $R \geq 10^{4}$; at $R \lesssim 10^{-3}$, its DF contains non-finite or negative entries. These points correspond to $\log_{10}(|\text{DF rel. diff.}|)$ at or above zero in the figure. The DF computed with \agama agrees with that of \PIANISTpy to better than 0.5\% across the entire range of radii probed. In the bottom panel, \PIANISTpy reconstructs the input density to better than $10^{-8}$ over the full sixteen decades in radius shown, and to better than $10^{-10}$ for $R \lesssim 10^{4}$. \agama achieves a roundtrip density accuracy at the level of $10^{-6}$ to $10^{-2}$, while the fitting functions of \citet{Widrow2000ApJS13139W} and \citet{Lane2022MNRAS5105119L}, accurate to $10^{-5}$--$10^{-2}$ for $R \lesssim 10^{2}$, degrade to order unity for $R \gtrsim 10^{6}$. \PIANISTpy thus outperforms all other methods across the full radial range, typically by five or more orders of magnitude (see \aref{app:num_DF_high_tail} for the numerical treatment of the central DF).

Importantly, the level of numerical accuracy shown in this baseline example is representative of \PIANISTpyNS. An explicit validation on 113 models\footnote{The full table of tested models and respective accuracy quantification can be found at \url{https://pianistpy.com/docs/reference/accuracy}.} spanning 16 input density profiles, the isotropic, constant-anisotropy, and (C)OM Type~I models, and their model parameters gives a relative error of $10^{-7}$ or better in the density recovered from the DF (typical median error of $6\times10^{-12}$), at every radius over $10^{-8} < R < 10^{8}$. In contrast, the measured roundtrip accuracy of the other two codes degrades with the anisotropy due to more nontrivial analytical expressions. Tested on the constant-anisotropy NFW model over $-1/2 \leq \betac \leq 1/2$, the roundtrip density relative error of \agama has $2.3\times 10^{-5}$--$1.1\times10^{-3}$ at median and $0.16$--$2.8$ at maximum over $10^{-3} < R < 10^{4}$. Separately, the \galpyNS-computed DFs agree with the \PIANISTpy baselines to $1.2\times10^{-5}$--$1.8\times10^{-2}$ at median and $4.4\times10^{-4}$--$5.2$ at maximum over $10^{-3} < R < 10^{4}$. These differences in part reflect specific choices in solver design: \galpy and \agama perform numerical differentiation, integration, and interpolation throughout, whereas \PIANISTpy carries exact symbolic and analytical expressions up to the final numerical integration.

Next, we compare the dynamical stability of particle initial conditions generated with \PIANISTpyNS, \galpyNS, and \agamaNS. We construct NFW $N$-body realisations of $10^7$ particles with constant anisotropy $\betac = 0.5$ and concentration $c_\rho = 10$, using each code's native truncation procedure under the condition that the truncated total mass equals the untruncated NFW virial mass. For \PIANISTpyNS, we adopt the Method-II-truncated $\betac = 0.5$ system constructed from the isotropic potential--density pair shown in the leftmost column of \fref{fig:DF_Trun_sigma}. For \galpyNS, the total mass requirement translates to hard truncation precisely at the NFW virial radius, $\Rt = \rvir/\rz = c_\rho = 10$. \agama truncates the density profile using an exponential cutoff of the form $\exp[-(r/r_\text{cut})^\xi]$. By adopting their default value of $\xi = 2$, the total mass constraint yields $r_\text{cut} = 1.277\,\rvir$. In terms of the walltime for $10^7$ particle sampling itself, measured with the build of the submitted version, \PIANISTpy completes within 29~s (17~s) on a single core (32 cores) of a workstation-grade CPU, in comparison to 77~s (47~s) for \agamaNS, and 250~s (serial only) for \galpyNS.

We dynamically evolve each initial condition in isolation for 100~Gyr ($\simeq 1570$ model time units) using the code \gamer \citep{Schive:2017sdo}, which supports Adaptive Mesh Refinement (AMR) and hybrid CPU/GPU parallelisation. We adopt a simulation box of size $800\,\rz = 80\,\rvir$, covered by a $128^3$ root grid and nine refinement levels; a cell is further refined if the total particle number therein exceeds four. This yields a maximum spatial resolution of $0.012\,\rz$ required to properly resolve the innermost density cusp (see \citet{Chiang2026} and Appendix~A of \citet{Chiang2025} for detailed convergence tests). The large box size ensures that all particles always remain sufficiently distant from simulation boundaries, which have outflow boundary conditions.

\begin{figure}
        \centering
	\includegraphics[width=0.99\linewidth]{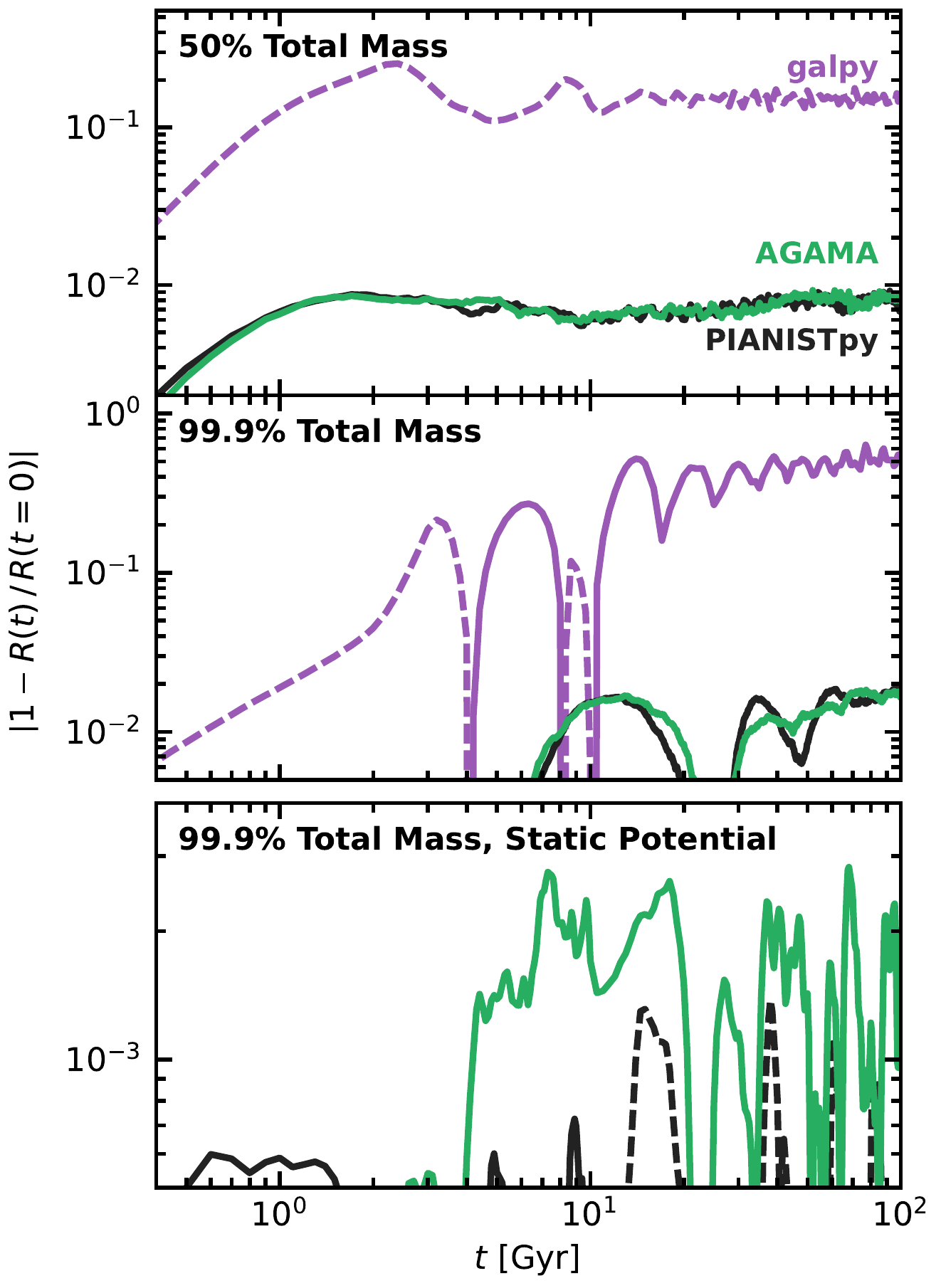}
    \caption{Fractional change $|1 - R(t)/R(t{=}0)|$ in the instantaneous radius enclosing 50\% (top panel) or 99.9\% (middle panel) of the total mass over time, for the $N$-body systems shown in \fref{fig:Isolation_Comp_beta0p5} generated by \galpy (purple), \agama (green), and \PIANISTpy (black) separately evolved in isolation with \gamerNS. The bottom panel again shows the 99.9\%-mass radius for the same \agama and \PIANISTpy initial conditions, but now instead as tracer particles integrated under the respective model's own static potential to test truncation boundary stability with minimal solver discreteness noise (see text for details). Solid and dashed curves denote expansion ($R(t) > R(t{=}0)$) and contraction ($R(t) < R(t{=}0)$), respectively.}
	\label{fig:Isolation_Comp_Renc}
\end{figure}

\fref{fig:Isolation_Comp_beta0p5} shows the time evolution of the projected surface densities for the initial conditions generated with \galpy (top row), \agama (middle row), and \PIANISTpy (bottom row), with the initial virial radius (black solid circle) and initial half-mass radius enclosing 50\% of the total mass (dark blue dotted circle) annotated. The instantaneous half-mass radius $R_{50\%}$ at each subsequent snapshot is marked by a cyan solid circle. \fref{fig:Isolation_Comp_Renc} quantifies this radial evolution through two complementary diagnostics: the half-mass radius (top panel) tracks the bulk dynamical evolution at intermediate radii, and the instantaneous radius enclosing 99.9\% of the total mass $R_{99.9\%}$  (middle panel) specifically targets the truncation boundary stability where the three truncation schemes differ most sharply.

First, \galpyNS's initial condition is violently unstable at all radii. The hard Heaviside energy cutoff leaves the truncation boundary far from dynamical equilibrium, causing large radial oscillations that slowly damp out as the system settles into virial equilibrium. Hence, the approach used by \galpy is unsuitable for constructing dynamically stable initial conditions. \agama achieves substantially better dynamical stability than \galpy at all radii, as expected from its self-consistent DF inversion from the exponentially truncated density profile. Both $R_{50\%}$ and $R_{99.9\%}$ remain within $\simeq 1$--$2\%$ of the respective initial values throughout. Finally, \PIANISTpyNS's self-consistent phase-space truncation also produces a consistently stable initial condition. 

To cleanly distinguish if the similar level of $R_{99.9\%}$ evolution measured in \agama and \PIANISTpy initial conditions is dominated by the discreteness noise of the \gamer gravity solver and particle integration under the fiducial numerical setup, we separately evolve all $10^7$ particles of each initial condition as tracers in the respective model's own static spherical potential for $100$~Gyr, using \agamaNS's native orbit integrator\footnote{Each static potential is represented by \agamaNS's \texttt{Multipole} expansion (spherical monopole) on a radial grid of 200 nodes, and all orbits are integrated with \texttt{agama.orbit} at its default settings.}, as shown in the bottom panel of \fref{fig:Isolation_Comp_Renc}. Overall, \PIANISTpy is more stable by a factor of two in the maximum $R_{99.9\%}$ deviation over $100$~Gyr, and by a factor of five in the median deviation after the first $10$~Gyr. Furthermore, \PIANISTpyNS's phase-space truncation circumvents the possible instability-driving inflection features in DFs induced by a sharp density truncation (\citealp{Weinberg2023MNRAS5254962W}; see also \S\ref{ssec:DF_Truncation}). For example, for the \agama initial condition now exponentially truncated by $\xi = 8$ under the same total mass constraint ($r_\text{cut} = 1.072\,\rvir$), whose DF is no longer everywhere monotonic, the same dynamical evolution in isolation yields $R_{99.9\%}$ deviations of up to $2.3\%$ within the first $5$~Gyr (cf.\ $0.2$--$0.4\%$ for the $\xi = 2$ \agama and \PIANISTpy initial conditions) and of up to $3.1\%$ over $100$~Gyr, above the benchmarked \gamer discreteness noise floor.

\subsection{Collective shape evolution and radial orbit instability}\label{ssec:ROI}

\begin{figure}
	\includegraphics[width=\linewidth]{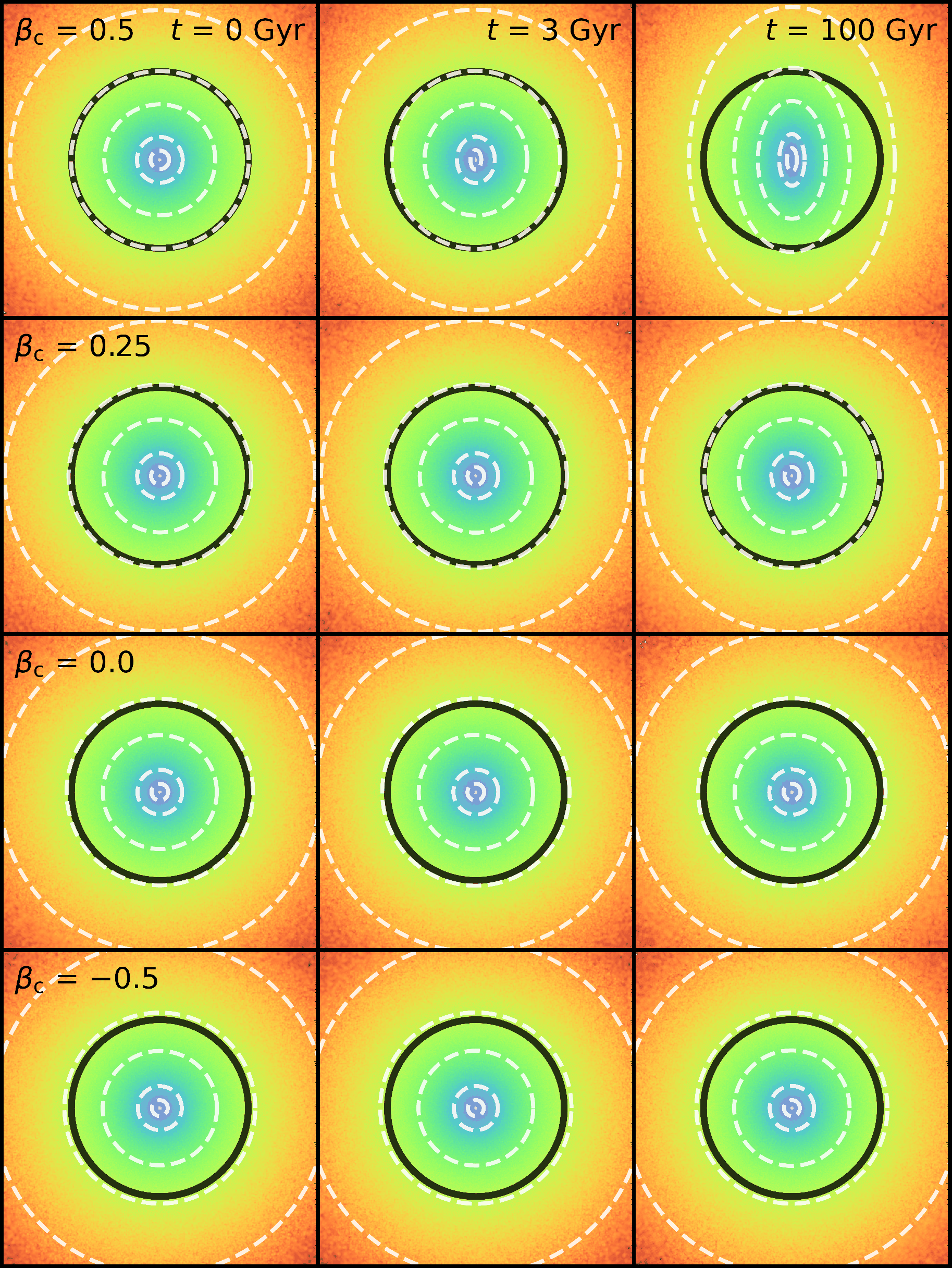}
	\caption{Time evolution of \PIANISTpyNS-generated NFW haloes in isolation with four different initial constant anisotropies $\betac = 0.5, 0.25, 0.0, -0.5$ (top to bottom rows) at $t = 0$, $3$, and $100$~Gyr (left to right columns), colour-coded by projected mass density on a shared colour scale. The projection is along the instantaneous minor axis at each epoch, with the $y$-axis aligned with the instantaneous major axis. White dashed elliptical contours trace the major and intermediate axis lengths of the 3D ellipsoids computed from the 5\%, 30\%, 50\%, 70\%, and 90\% most energetically bound particles; the black solid circle marks the initial virial radius.}
	\label{fig:Isolation_Comp_beta}
\end{figure}

To investigate how velocity anisotropy affects the dynamical stability of the innermost regions, we construct four Method-II-truncated NFW systems with constant anisotropy $\betac = 0.5$, $0.25$, $0.0$, and $-0.5$, all sharing the same potential--density pair shown in \fref{fig:DF_Trun_sigma}, and evolve them in isolation over 100~Gyr following the numerical setup of \S\ref{ssec:OP_Comparison}. At each snapshot, we quantify the instantaneous 3D shape by computing the distance-weighted inertia tensor
\begin{align}\label{eqn:I_ij}
    \mathcal{I}_{ij} \eee \Mt^{-1}
    \sum_{n=1}^{\Npar} m \Bigg[
    \frac{x_{i,n} x_{j,n}}{x_{1,n}^2+ x_{2,n}^2/(b/a)^2+ 
    x_{3,n}^2/(c/a)^2} \Bigg]\;,
\end{align}
where $i, j \in \{1, 2, 3\}$ index the major ($a$), intermediate ($b$), and minor ($c$) semi-axis directions with $a \geq b \geq c$, following the iterative scheme of \citet{Dubinski1991ApJ378496D}\footnote{ Starting from $b/a = c/a = 1$, we iteratively update the axis ratios and re-evaluate \eref{eqn:I_ij}, excluding particles outside the rescaled ellipsoid at each step. The iteration terminates when $|\zeta_k - \zeta_{k-1}| < 10^{-3}$ for all $\zeta \in \{b/a,\, c/a,\, c/b\}$.}. The three converged eigenvalues $\lambda_1 \geq \lambda_2 \geq \lambda_3$ then give the semi-axis ratios $b/a = \sqrt{\lambda_2/\lambda_1}$ and $c/a = \sqrt{\lambda_3/\lambda_1}$, from which the triaxiality
\begin{align}\label{eqn:Triaxiality}
	T \eee \frac{1-(b/a)^2}{1-(c/a)^2},
\end{align}
is computed \citep{Franx1991ApJ383112F}, with $0 < T < 1/3$ corresponding to oblate (pancake-like), $1/3 < T < 2/3$ to triaxial, and $2/3 < T < 1$ to prolate (cigar-like) morphologies. Shape parameters are measured from the 5\%, 30\%, 50\%, 70\%, and 90\% most energetically bound particles to probe the collective evolution at different radii.

\begin{figure}
        \centering
	\includegraphics[width=0.9\linewidth]{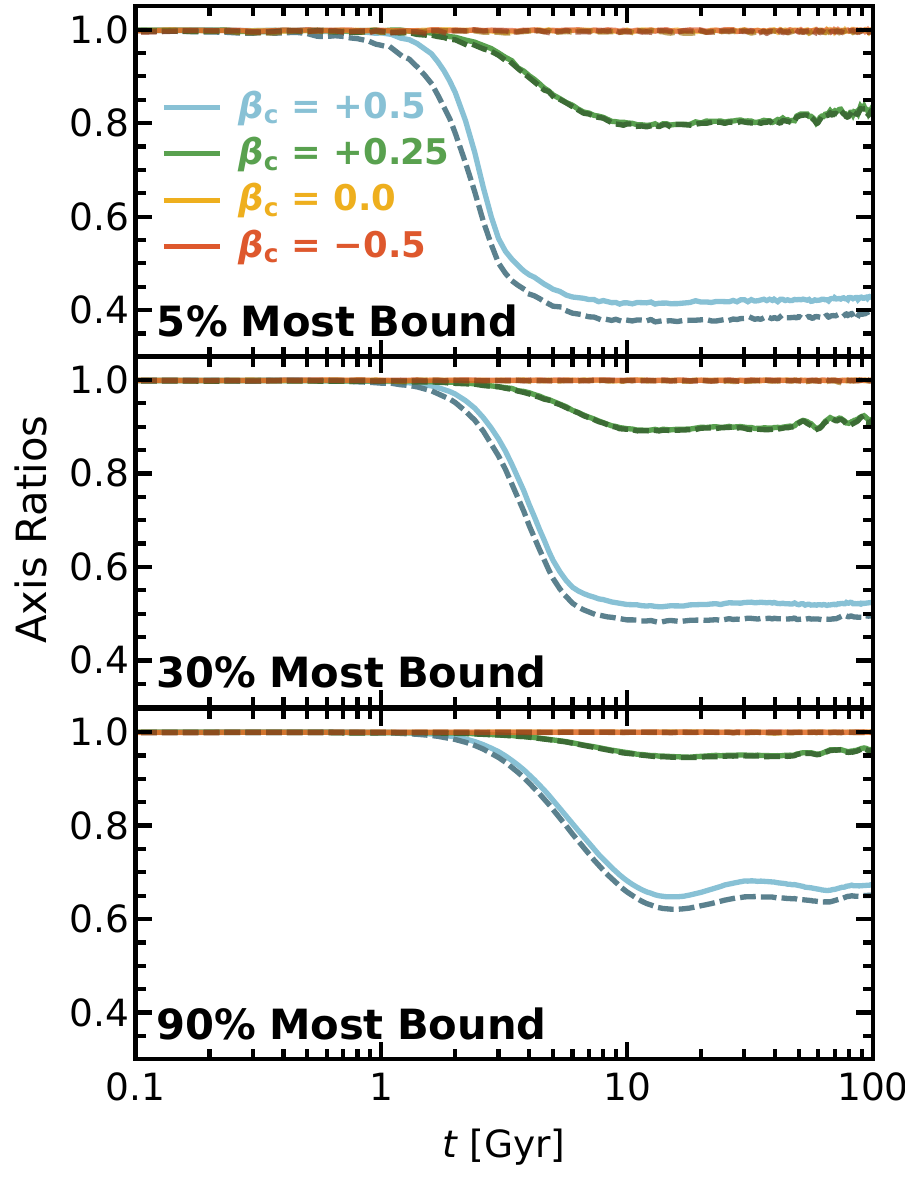}
	\caption{Time evolution of axis ratios $b/a$ (solid) and $c/a$ (dashed) for the systems shown in \fref{fig:Isolation_Comp_beta}, computed from the 5\%, 30\%, and 90\% most energetically bound particles (top, middle, and bottom panels, respectively). Different colours correspond to different initial constant anisotropies: $\betac = 0.5$ (cyan), $0.25$ (green), $0.0$ (yellow), and $-0.5$ (red).}
	\label{fig:Isolation_Comp_beta_Tri}
\end{figure}

\fref{fig:Isolation_Comp_beta} shows the time evolution of the initial conditions with constant anisotropy $\betac = 0.5$, $0.25$, $0.0$, and $-0.5$ (top to bottom panels), with mass density projected along the instantaneous minor axis at each epoch. For the isotropic ($\betac = 0$) and tangentially anisotropic ($\betac = -0.5$) cases, the systems remain visually spherical throughout the full 100~Gyr of evolution, with the elliptical contours showing no departure from their initial circular shapes at any bound particle fraction. In stark contrast, the radially anisotropic cases ($\betac = 0.5$ and $0.25$) develop a clear elongated bar-like structure in the innermost regions within the first few Gyr, which grows outward with time before saturating; this inside-out bar formation appears as a constant-density core in the shell-averaged density profile (see Fig.~12 of \citealt{Chiang2025}).

\fref{fig:Isolation_Comp_beta_Tri} quantifies this shape evolution by tracking the time evolution of $b/a$ (solid) and $c/a$ (dashed) for the 5\%, 30\%, and 90\% most energetically bound particles (top to bottom panels). For $\betac = 0$ and $-0.5$, both axis ratios remain essentially unity at all bound fractions throughout, confirming that the entire system remains spherically symmetric in isolation. For $\betac = 0.5$ and $0.25$, the innermost 5\% most bound particles (top panel) show a rapid drop in both $b/a$ and $c/a$ within the first $\sim$5~Gyr, settling to a prolate configuration ($b/a \simeq c/a < 1$) with $T \simeq 0.95$ at saturation. This evolution propagates outward, with the overall strength of such a prolate configuration increasing with the degree of central radial anisotropy. In contrast, the 90\% most bound particles show only a modest departure from sphericity, consistent with the bar-like structure being confined to the innermost regions visible in \fref{fig:Isolation_Comp_beta}. These numerical experiments demonstrate that this radial-orbit-instability (ROI)-driven inside-out bar formation is an intrinsic feature of sufficiently radially anisotropic systems, occurring spontaneously in the innermost regions even in the complete absence of external physical and truncation-induced perturbations, the latter being eliminated by \PIANISTpyNS's self-consistent phase-space truncation.

It is instructive to compare these results with the recent findings of \citet{Binney2026}, who investigated the stability of radially biased spherical systems using the \agamaNS-based action-space framework of \citet{BinneyWrightVasiliev2025}, in which self-consistent potential--density pairs are obtained via an axisymmetric iterative scheme. Working exclusively with DFs of the form $F(aJ_{\rm r} + \bigL)$, where $J_{\rm r}$ is the radial action, they find that radially biased models always converge to oblate rather than spherical configurations, and conclude that such systems are generically unstable to quadrupolar perturbations. However, the framework of \citet{BinneyWrightVasiliev2025} is explicitly axisymmetric\footnote{The action decomposition of \citet{BinneyWrightVasiliev2025} in cylindrical coordinates enforces axisymmetry ($b = a$ or $b = c$) statically, geometrically excluding genuinely triaxial configurations.}, restricting accessible deformation modes to either perfectly oblate ($T = 0$) or perfectly prolate ($T = 1$) configurations. This methodological limitation thus precludes the genuinely triaxial morphologies ($T \simeq 0.75$ to $0.95$, $a \neq b \neq c$) produced in our simulations that are also consistent with the well-established ROI signature \citep[e.g.,][]{Antonov1973, Henon1973, Trenti2006}. These contrasting findings further underscore the importance of initialising such systems without geometric constraints, as enabled by the $(Z, \bigL)$-based phase-space truncation framework introduced in this work.

\section{Velocity Structure of Anisotropic Systems}\label{sec:OP_Vel}

For observers, \PIANISTpy can serve as a forward-modelling tool for kinematic inference. For any truncated DF, \PIANISTpy can rapidly evaluate the local 3D velocity distribution from $V = 0$ to the local escape velocity $V_\text{esc}(R)$ at any 3D radius $R \leq \Rt$, enabling quantitative comparison of velocity structure across different anisotropy models at fixed potential--density pair. It further supports the computation of projected velocity dispersion profiles, line-of-sight velocity distributions (LOSVDs), and associated shape parameters such as Gauss--Hermite (GH) moments and kurtosis, all of which can be directly compared with observed kinematic data. In what follows, we illustrate these capabilities with concrete examples.

\subsection{Local 3D particle velocity distributions}\label{ssec:OP_Gaussian_VD}

\begin{figure}
	\includegraphics[width=0.99\linewidth]{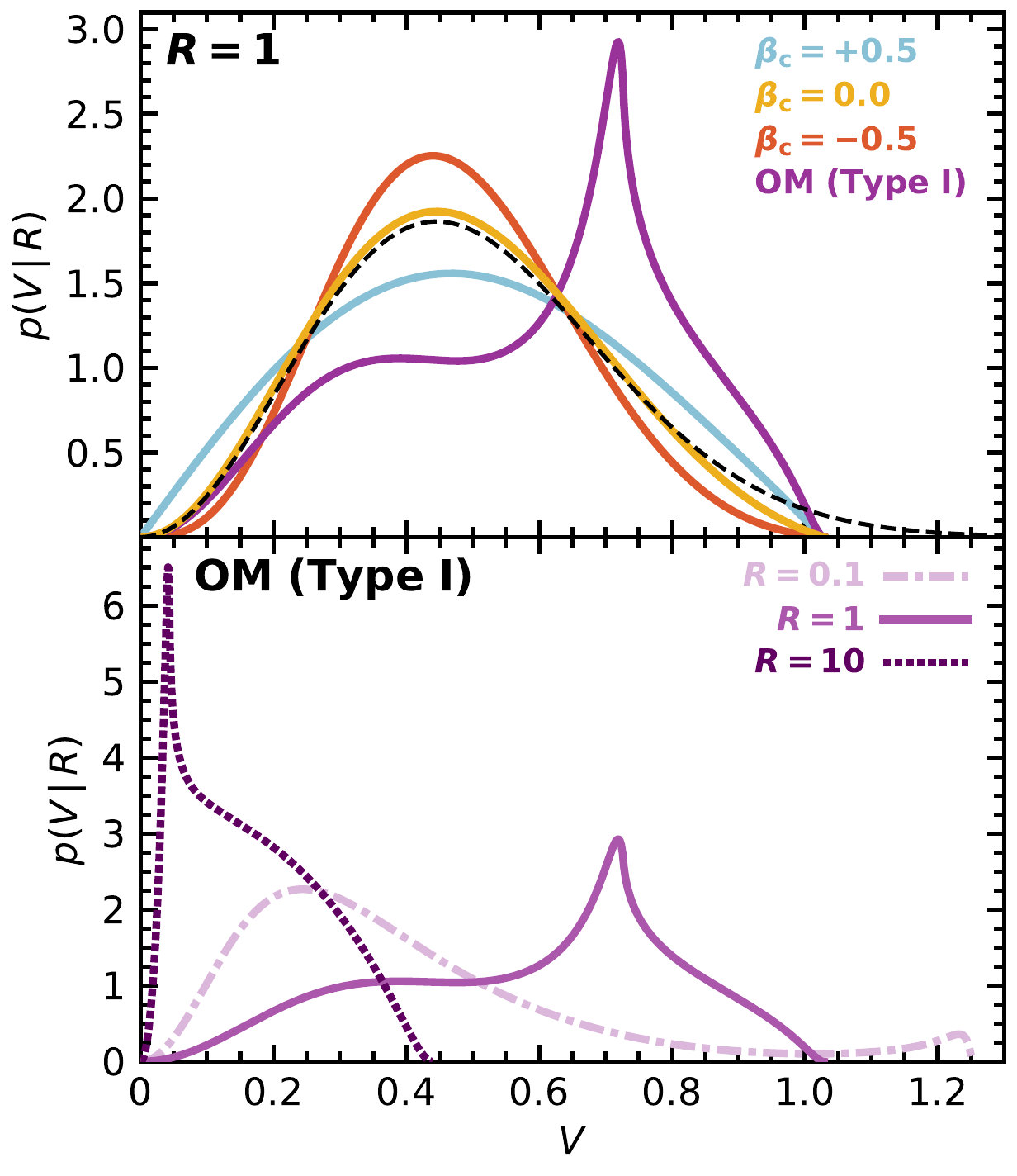}
    \caption{Probability density function of local 3D velocities (Eq.~[\ref{eqn:V_prob}]) for Method-II-truncated DFs constructed from the same isotropic potential--density pair as in \fref{fig:DF_Trun_sigma}. \textit{Top:} OM Type~I ($\Ra = 1$, purple) and constant-anisotropy models $\betac = +0.5$ (cyan), $0$ (yellow), and $-0.5$ (red), evaluated at a fixed radius $R = 1$. The black dashed curve shows the Maxwell\textendash Boltzmann distribution at the same peak location as the isotropic model. \textit{Bottom:} The OM Type~I model evaluated at $R = 0.1$, $1$, and $10$ (colour-coded as indicated). In both panels, each distribution terminates at the local escape velocity $V_\text{esc}(R)$.}
	\label{fig:V_shell}
\end{figure}

\fref{fig:V_shell} shows the local 3D velocity distributions for constant-anisotropy $\betac = +0.5$ (cyan), $0$ (yellow), $-0.5$ (red), and OM Type~I with $\Ra =1$ (purple), all constructed from the same Method-II-truncated isotropic potential--density pair as in \fref{fig:DF_Trun_sigma}. The top panel compares these models at a fixed radius $R = 1$, with the Maxwell--Boltzmann distribution (black dashed curve) at the same peak location as the isotropic case for reference. The bottom panel shows the OM Type~I model evaluated at three different radii $R = 0.1$, $1$, and $10$ (lightest to darkest purple). In both panels, the upper termination point of each distribution corresponds to the respective local escape velocity $V_\text{esc}(R)$.

Overall, radial anisotropy preferentially populates broader distributions with enhanced populations at both high and low velocities, reflecting the non-local nature of radial orbits. In contrast, tangential anisotropy produces more `peaky' distributions relative to the isotropic benchmark. The OM Type~I model exhibits a more complex shape, as further illustrated in the bottom panel of \fref{fig:V_shell}. At $R = \Ra = 1$, the distribution is bimodal: the low-velocity plateau is the isotropic orbits populated at the centre ($\betasigma \simeq 0$ for $R \ll \Ra$), Maxwell--Boltzmann-like in shape; the peak at $V_* = \sqrt{2P(R)/(1 + R^2/\Ra^2)} \simeq 0.72$ corresponds to the pile-up of orbits at the maximum angular momentum the OM model allows at their energy, above which only outer radial orbits remain and taper off exactly at $V_\text{esc}$. At $R = 10$, a similar sharp peak, now at $V_* \simeq 0.04$, bounds the isotropic population within a highly confined velocity range, above which the outer radial orbits dominate and populate a broad distribution that terminates at $V_\text{esc}$. At $R = 0.1$, the Maxwell--Boltzmann-shaped isotropic distribution dominates, with a much lower-amplitude peak at $V \simeq 1.2$ from the diminishing contribution of nearly radial outer orbits. 

Despite these variations, none of these anisotropic distributions is well described by the Maxwell--Boltzmann distribution that many theoretical studies routinely assume \citep[e.g.,][]{Chandrasekhar1943, Lacey1985ApJ, Hernquist:1993jf} and that terrestrial dark matter experiments adopt as the `standard halo model,' an isotropic Maxwell--Boltzmann distribution truncated at the escape speed \citep{Drukier:1986tm, Baxter2021} in data interpretation and setting dark matter constraints \citep[e.g.,][]{Adhikari:2018ljm, COSINE-100:2024nfa, LZ:2026axp, Freese:2026sga, Bose:2026ndd}. Indeed, \citet{Kazantzidis2004ApJ60137K} showed that such local-Maxwellian-approximation-based $N$-body realisations are not equilibrium solutions, where an initially isotropic cusp relaxes into a constant-density core with strong radial anisotropy within a crossing time.

\subsection{Projected velocity dispersions and distributions}\label{ssec:OP_LOSVD}

\begin{figure}
        \centering
	\includegraphics[width=0.85\linewidth]{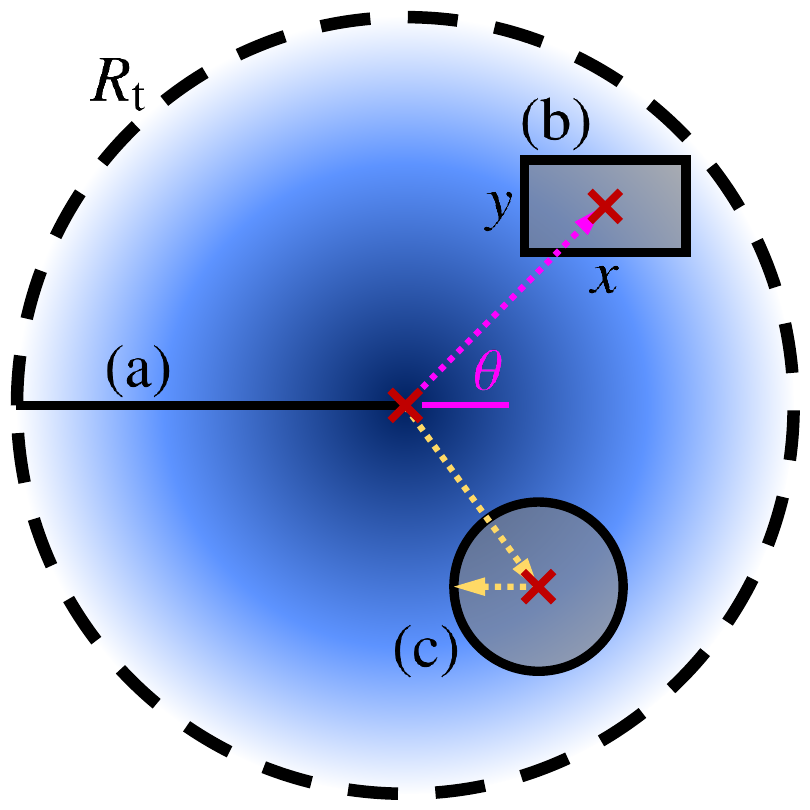}
        \caption{Schematic illustration of different projection configurations supported by \PIANISTpy for a spherical system truncated at $\Rt$ (dashed circle). \textbf{(a)} Line-of-sight projected radial profile, ideal for thin-slit observations with negligible thickness. \textbf{(b)} Rectangular-aperture-averaged projection, ideal for thin-slit observations with finite thickness. The orientations of $x$, $y$, and $\theta \in [0,\pi/2]$ are as indicated in the diagram. \textbf{(c)} Circular-aperture-averaged projection, ideal for fibre-fed observations. The aperture centre can either be co-centric with or off-centred from the system's CM. For aperture geometries beyond those shown, projection-averaged quantities can be computed via projected-surface-mass-density ($\Sigma$)-weighted  MCMC sampling.}
	\label{fig:Projection_Ops}
\end{figure}

\fref{fig:Projection_Ops} illustrates the projection configurations natively supported by \PIANISTpyNS: (a) a line-of-sight projected radial profile through the system's CM, (b) a rectangular aperture-averaged projection at position angle $\theta$ and offset $(x, y)$ from the CM, and (c) a circular aperture-averaged projection centred on or offset from the CM; technical details are provided in \aref{app:Velocity_prof_proj}. In what follows, we use configurations (a) and (c) to illustrate the projected velocity dispersion profiles and LOSVDs of four fiducial anisotropy models---constant-anisotropy $\betac = +0.5$ (cyan), $0$ (yellow), $-0.5$ (red), and OM Type~I with $\Ra = 1$ (purple)---all constructed from the same Method-II-truncated isotropic potential--density pair as in \fref{fig:DF_Trun_sigma}.

\begin{figure}
	\includegraphics[width=0.99\linewidth]{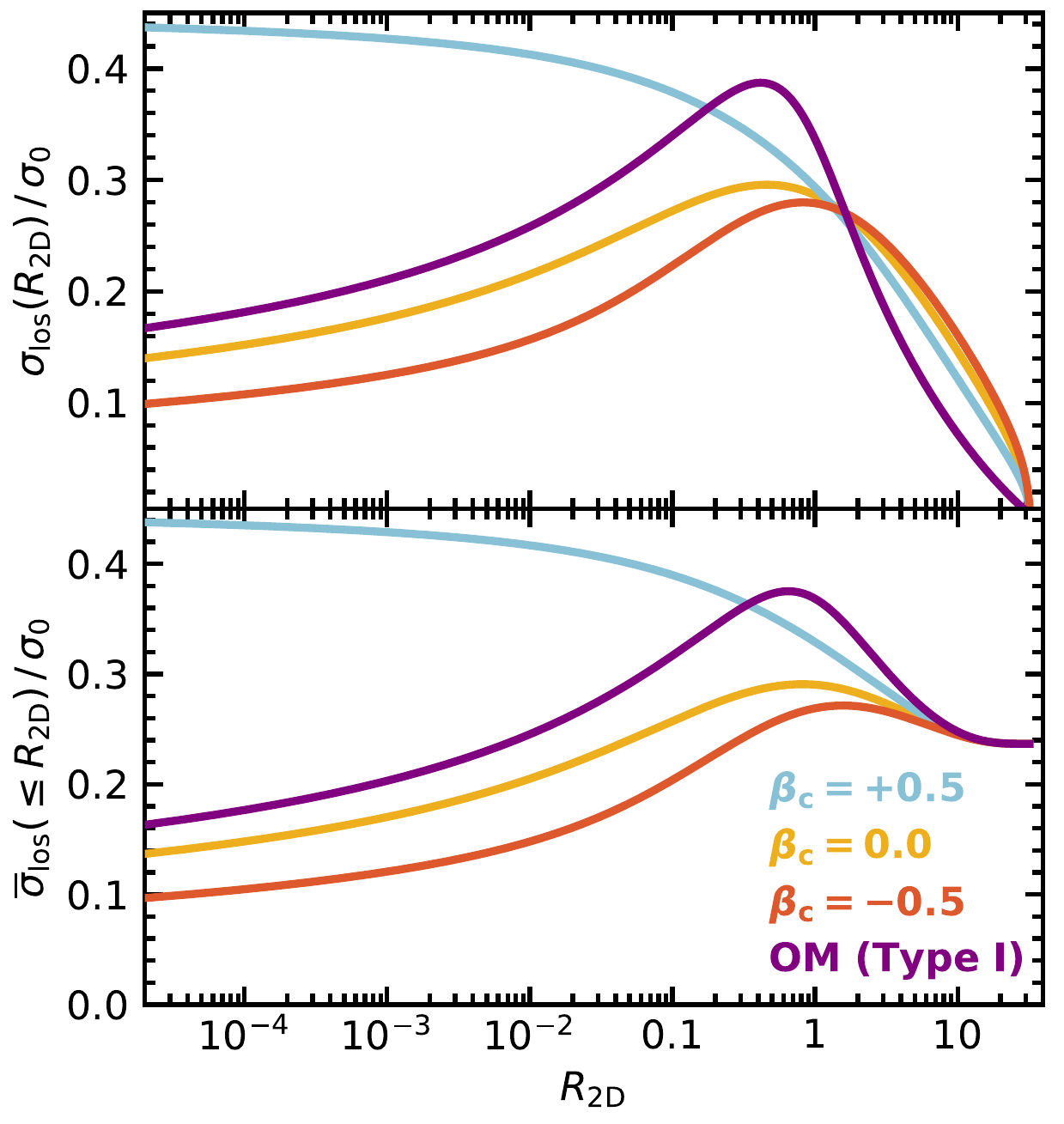}
    \caption{Line-of-sight projected velocity dispersion profiles (Eq.~[\ref{eqn:los_sigma_1}]) for constant-anisotropy $\betac = +0.5$ (cyan), $0$ (yellow), $-0.5$ (red), and OM Type~I with $\Ra = 1$ (purple), all constructed from the same potential--density pair as in \fref{fig:DF_Trun_sigma}. Here $\sigma_0 \eee \sqrt{G\Mz/\rz}$. \textit{Top:} Line-of-sight velocity dispersion profile as a function of projected radius $R_\text{2D}$ from the system's CM. \textit{Bottom:} Circular-aperture-averaged profile as a function of aperture radius $R_\text{2D}$.}
\label{fig:sigma_los_integrated}
\end{figure}

\fref{fig:sigma_los_integrated} shows the line-of-sight projected velocity dispersion profiles for our four fiducial anisotropy models. The top panel compares the line-of-sight velocity dispersion profile $\sigma_\text{los}$ as a function of projected radius $R_\text{2D}$, corresponding to a narrow long-slit observation as in configuration (a) of \fref{fig:Projection_Ops}. Overall, radially anisotropic models exhibit larger $\sigma_\text{los}$ at $R_\text{2D} \lesssim 1$ and smaller $\sigma_\text{los}$ for $1 \lesssim R_\text{2D} \leq \Rt$ relative to the isotropic benchmark, and vice versa for tangential anisotropy. All curves identically vanish at $R_\text{2D} = \Rt$ as required by the self-consistent truncation boundary condition $\hatP(\Rt) \eee 0$ imposed in \eref{eqn:Poissonr_Truncation_MT}. The OM Type~I model exhibits a prominent peak around $R_\text{2D} \simeq 0.5$, due to the competition between the increasing radial anisotropy at $R \gtrsim \Ra$ and the declining velocity dispersion as $R \to \Rt$. The bottom panel shows the circular-aperture-averaged dispersion $\overline{\sigma}_\text{los}(\leq R_\text{2D})$ as a function of aperture radius $R_\text{2D}$, corresponding to configuration~(c) centred on the CM. Here, radially (tangentially) biased models yield higher (lower) aperture-averaged dispersions at all radii relative to the isotropic benchmark. At $R_\text{2D} = \Rt$, the aperture encloses the entire system and $\overline{\sigma}_\text{los}(\leq\Rt)$ reduces to the global velocity dispersion, which is independent of anisotropy by the virial theorem; all four profiles expectedly converge to the same value.

\begin{figure}
	\includegraphics[width=0.99\linewidth]{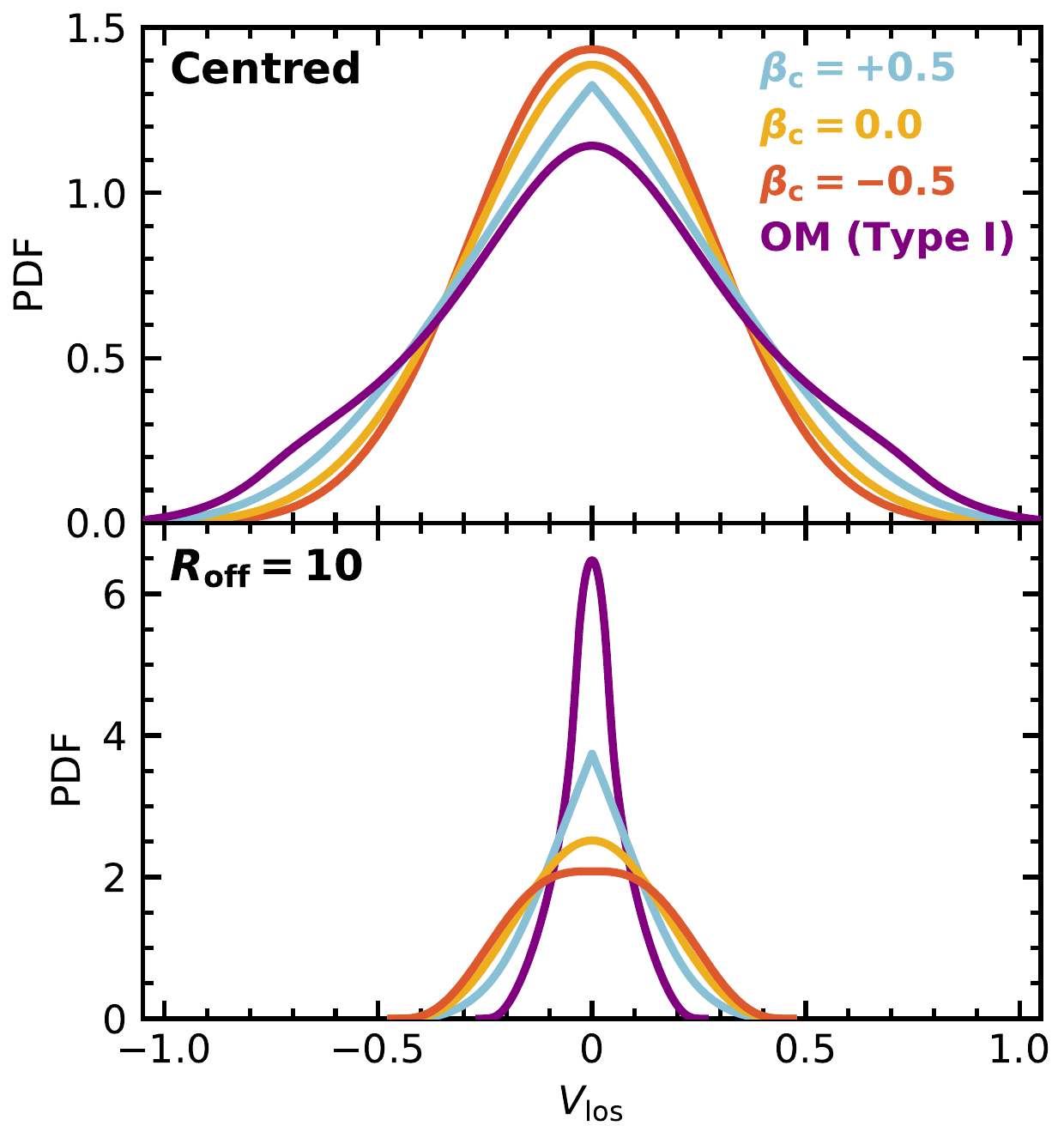}
    \caption{$\Sigma$-weighted unit-normalised line-of-sight velocity distributions (LOSVDs; Eq.~[\ref{eqn:losvd_4}]) within a circular aperture of radius $R_\text{ap} = 1$, centred on the system's CM (top panel) or offset by $R_\text{off} = 10$ from the CM (bottom panel), for the same four anisotropy models and potential--density pair as in \fref{fig:sigma_los_integrated}.}
	\label{fig:LOSVD}
\end{figure}

\fref{fig:LOSVD} shows the projected-surface-mass-density ($\Sigma$)-weighted LOSVDs within a circular aperture of radius $R_\text{ap} = 1$ centred on the system's CM (top panel) and offset by $R_\text{off} = 10$ from the CM (bottom panel), for the same four anisotropy models as in \fref{fig:sigma_los_integrated}. The sensitivity of the LOSVD shape to both the anisotropy model and the aperture position can be understood from the projection geometry. For a tracer at 3D radius $R$ and projected radius $R_\text{2D}$, the line-of-sight velocity receives contributions from the radial velocity weighted by $\sqrt{R^2 - R_\text{2D}^2}/R$ and the tangential velocity weighted by $R_\text{2D}/R$. Near the projected centre ($R_\text{2D} \rightarrow 0$), radial orbits project their full speed onto the line of sight, so radial anisotropy broadens the LOSVD. At large projected radii ($R_\text{2D} \rightarrow \Rt$), radial orbits move nearly perpendicular to the line of sight and contribute $V_\text{los} \rightarrow 0$, so radial anisotropy instead narrows the LOSVD. A centred aperture therefore probes the former regime, while an offset aperture at large $R_\text{off}$ probes the latter, leading to a reversal in the anisotropy-dependent ordering of LOSVD widths between the two configurations.

Beyond the overall width captured by $\sigma_\text{los}$, the shape of the LOSVD encodes additional kinematic information commonly quantified through GH moments $h_n$ \citep{vanderMarel1993, Gerhard1993}, where $h_4$ in particular measures symmetric deviations from a Gaussian shape (\aref{app:kurtosis}); \PIANISTpy computes these moments analytically for any LOSVDs. The centred aperture (top panel of \fref{fig:LOSVD}) shows LOSVDs centred at $V_\text{los} = 0$, with radially (tangentially) anisotropic models producing broader (narrower) distributions relative to the isotropic benchmark. Two nuances beyond this simple ordering are visible. First, the peak shape at $V_\text{los} = 0$ differs qualitatively between the constant-anisotropy and OM models: for $\betac = +0.5$, the DF factor $\bigL^{-2\betac}$ diverges as $\bigL \rightarrow 0$, producing a sharply peaked, leptokurtic profile ($h_4 > 0$; \aref{app:kurtosis}) that is imprinted at every projected radius since $\betasigma = \betac$ is radius-independent; the OM model is also narrowly peaked but with a smooth, bell-shaped apex, as its isotropic core ($\betasigma \rightarrow 0$ for $R \ll \Ra$) dominates the central line of sight and rounds the $V_\text{los} = 0$ peak. Second, the kurtosis ordering across the constant-anisotropy models follows the textbook expectation of \citet{Merritt1987} and \citet{vanderMarel1993}: $\betac = +0.5$ leptokurtic ($h_4 > 0$), isotropic near-Gaussian ($h_4 \simeq 0$), and $\betac = -0.5$ platykurtic ($h_4 < 0$).

In the offset aperture (bottom panel), the ordering reverses dramatically. The OM Type~I model, which showed the broadest distribution with radially-varying-anisotropy-induced secondary bumps in the centred aperture, now produces by far the sharpest and narrowest distribution, forming a striking spike at $V_\text{los} = 0$, while $\betac = -0.5$ now exhibits the broadest distribution. This reversal is a direct consequence of the projection geometry described above. At $R_\text{off} = 10$, the line of sight passes through the radially anisotropic outskirts where $R_\text{2D}/R \rightarrow 1$, suppressing the projected radial velocity contribution towards $V_\text{los} \rightarrow 0$. The picture established by \citet{vanderMarel1993}, who considered tracer populations in Keplerian or logarithmic host potentials along centred lines of sight, thus breaks down once off-centred apertures or systems with finite physical extent are considered. More generally, for any system with a radially varying $\betasigma(R)$, the LOSVD shape can develop complex features, such as secondary bumps and sharp central peaks illustrated in \fref{fig:LOSVD}, that are not fully captured by any finite set of GH moments $h_n$, however high the truncation order. While standard observational practice characterises the LOSVD through $\{\overline{V}_{\rm los}, \sigma, h_3, h_4$\}, or even higher order moments in some studies, the full information content of a generic LOSVD exceeds what any finite moment expansion can faithfully represent. These examples underscore the value of full LOSVD profile fitting, a capability that \PIANISTpyNS's forward-modelling framework is designed to provide.

\section{Possible Application Cases}\label{sec:User_Case}

The phase-space truncation framework and \PIANISTpy introduced in this work open up a range of applications beyond what other existing codes support. We highlight a few direct use cases below:
\begin{itemize}

\item \textbf{Particle initial conditions for idealised $N$-body simulations:}
The primary motivation for developing \PIANISTpy is the construction of dynamically stable anisotropic particle initial conditions with finite spatial extent. Prior tools have been restricted to isotropy and/or limited to truncation schemes that induce instabilities due to the inability to generate fully self-consistent boundary conditions (see \S\ref{ssec:OP_Comparison}). Unlike approximate iterative schemes \citep[e.g.,][]{Schwarzschild1979ApJ232236S, Yurin:2014pga}, our truncation framework (\S\ref{ssec:DF_Truncation}) is analytically exact, so \PIANISTpyNS-generated initial conditions require no further relaxation while preserving the intended velocity anisotropy profile. This is particularly valuable in idealised simulations that aim to cleanly isolate the effect of velocity anisotropy on galactic dynamics \citep[e.g.,][]{Chiang2025, Chiang2026}.

\item \textbf{Rapid local particle sampling:}
Beyond full initial condition construction, \PIANISTpy also supports rapid sampling of the local velocity distribution $p(V|R)$ at any fixed 3D radius $R \leq \Rt$. For example, on-the-fly halo particle sampling is utilised in the idealised self-interacting dark matter simulations of \citet{Klemmer2026} to more faithfully study host--subhalo scattering interactions. \PIANISTpy serves as a numerically efficient tool with high accuracy across the entire radial domain (see e.g., \fref{fig:DF_Comparison}), improving upon local Maxwell--Boltzmann approximations \citep[e.g.,][]{Zeng2022} or the assumption of velocity isotropy. Separately, Monte Carlo-based star-by-star globular cluster stream models by \citet{Panithanpaisal2026}, where escaped stars are sampled on-the-fly from isotropic distributions, represent another natural application. For even more efficient integration with C/C\texttt{++}-based simulation codes \citep[e.g.,][]{Teyssier2002, Springel2005, Springel2010, Hopkins2015, Schive:2017sdo}, one can feed \PIANISTpyNS-generated cumulative distribution function tables (\fref{fig:CDF_PDF_Comparison}) directly into low-level interpolation and sampling subroutines of the simulation code.

\item \textbf{Systematic exploration of anisotropic DFs:}
\PIANISTpy enables rapid DF computation over varying density profiles, anisotropy models, and truncation choices. Recently, \citet{Weinberg2023MNRAS5254962W} and \citet{Dattathri2025} demonstrated that any inflection $\rmd F/\rmd Z < 0$ in the isotropic DFs, whether due to overly sharp outer truncation or steep transition from inner to outer power-law slope in the density profile, always drives dipole instability; \citet{2026MNRAS.546ag050B} showed that truncation with density vanishing exactly at a finite radius can still admit an everywhere non-negative isotropic or OM DF, as long as the density tapering near the boundary is sufficiently gradual. \PIANISTpy enables such analyses to be extended to anisotropic DFs \citep[e.g.,][]{Antonov1973, Hjorth1994ApJ424106H, Meza1997ApJ490136M}, where the additional degree of freedom in velocity anisotropy presents a richer set of DF asymptotic behaviours in the limit $R \to 0$, as characterised by Eqs.~(\ref{eqn:CSCA},~\ref{eqn:kappa_prime}), which are directly relevant to the physicality and stability of the resulting systems but not yet explored in these works. Having finitely truncated initial conditions with exact equilibrium boundary conditions allows one to cleanly isolate these instabilities from truncation-induced artefacts, facilitating controlled comparisons across models with different velocity anisotropies.

\item \textbf{Flexible fitting to target anisotropy profiles:}
For a target velocity anisotropy profile, such as those directly measured from simulated dark matter haloes \citep[e.g.,][]{Ludlow2011MNRAS4153895L, Klypin2016MNRAS4574340K, He2024ApJ976187H} or inferred for cluster member galaxies \citep[e.g.,][]{Annunziatella2016, Stark2019, Mamon2019}, \PIANISTpy permits construction of highly diverse and previously uncharted anisotropy profiles (\fref{fig:Linear_Super_betasigma}) via linear superposition of Method-II-truncated DFs, using known analytical anisotropy models as basis functions. Such numerical evaluation requires no additional explicit DF computations beyond the initial truncated base potential--density pair and costs less than a millisecond per profile on a single workstation-grade CPU core (\S\ref{ssec:DF_Superposition}), making it ideal for efficient MCMC-based fitting. These exact DF-based solutions are fully physical by construction---unlike methods that directly solve the anisotropic Jeans equations \citep[e.g.,][]{Walker2009, Annunziatella2016} or assume parametric Gaussian local 3D velocity distributions \citep{Mamon2013, Mamon2019}, both of which can yield unphysical DF solutions with locally negative phase-space densities \citep[e.g.,][]{Ciotti1992, DeDeo2024}---and can be directly used to generate particle initial conditions and associated velocity observables (\aref{app:DF_superposition}).

\item \textbf{Kinematic inferences of observational data:}
\PIANISTpy produces full line-of-sight projected velocity dispersion profiles (\fref{fig:sigma_los_integrated}), LOSVDs (\fref{fig:LOSVD}), and associated velocity moments (\aref{app:kurtosis}) that can be further $\Sigma$-weighted by flexible aperture geometries and configurations (\fref{fig:Projection_Ops}). This is well-suited for direct MCMC-based fitting against observational kinematic data, providing a numerically efficient alternative to the computationally expensive direct Jeans equation solvers \citep[e.g.,][]{GeringerSameth2015, Goldstein2022}. Discrete kinematic analyses moreover commonly assume a Gaussian LOSVD at each projected radius, in unbinned Jeans fits to dwarf-galaxy data \citep[e.g.,][]{GeringerSameth2015} as in discrete Jeans models \citep[e.g.,][]{Watkins2013}, which discards the anisotropy information carried by the LOSVD shape required to break the mass--anisotropy degeneracy from line-of-sight data alone (e.g., \citealp{Read2021}; \S\ref{ssec:OP_LOSVD}). The flexible anisotropy profile construction (Figs.~\ref{fig:DF_Trun_DFs},~\ref{fig:Linear_Super_betasigma}) also eliminates the need to assume constant anisotropy profiles when marginalising over model parameters \citep[e.g.,][]{Newman2013, Mitra2024}.

\item \textbf{Forward modelling discrete tracer noise in observations:}
\PIANISTpy can construct mock kinematic observables that exactly reflect the phase-space structure of the model by direct particle sampling at fixed 3D or projected radius, directly quantifying the sampling variance arising from a limited number of kinematic tracers drawn from an underlying smooth DF. For example, in observed ultra-faint dwarf satellites \citep[e.g.,][]{Simon2007ApJ670, Simon2019, Cerny2026} or ultra-diffuse dwarf galaxies \citep[e.g.,][]{vanDokkum2019, Haacke2025, Keim2026}, the number of tracer stars or globular clusters per system is often (far) fewer than 100, making them susceptible to discrete tracer noise in their dynamical inferences \citep[e.g.,][]{Chang2021MNRAS5074715C}. \PIANISTpy can be used for bootstrapping analyses of the uncertainties associated with various velocity dispersion and LOSVD observables (e.g., Figs.~\ref{fig:sigma_los_integrated},~\ref{fig:LOSVD}).

\item \textbf{Self-consistent local velocity distributions of dark matter for terrestrial experiments:}
Dark matter detection experiments conventionally adopt the `standard halo model,' an isotropic Maxwell--Boltzmann distribution truncated at the escape speed \citep{Drukier:1986tm, Baxter2021}, both in setting dark matter constraints \citep[e.g.,][]{Adhikari:2018ljm, COSINE-100:2024nfa, LZ:2026axp} and in interpreting candidate events \citep{Freese:2026sga, Bose:2026ndd}. The detection rates can be highly sensitive to the high-velocity tail, for example, for light or inelastic dark matter searches \citep[e.g.,][]{Lewin:1995rx, Tucker-Smith:2001myb} or separately the low-velocity end for signals from solar-captured dark matter particles \citep[e.g.,][]{Press:1985ug, Jungman:1995df}. However, in generic velocity-anisotropic host haloes, the local velocity distributions of dark matter particles not only deviate strongly from this routinely assumed simplification but can even exhibit localised peaks in hosts with a radially varying velocity-anisotropy profile (\fref{fig:V_shell}). \PIANISTpy provides a highly flexible pipeline to evaluate the full distribution $p(V|R)$ (Eq.~[\ref{eqn:V_prob}]) from fully self-consistent solutions of the collisionless Boltzmann equation for a host with a generic target anisotropy profile (\fref{fig:Linear_Super_betasigma}).

\end{itemize}

\section{Summary}\label{sec:Summary}

Dark matter (sub)haloes and galaxies exhibit rich anisotropic velocity structures, as consistently demonstrated by cosmological simulations and observations. However, velocity isotropy is still generally assumed in idealised numerical and observational inference studies, primarily due to the lack of methodologies to construct anisotropic systems with finitely truncated, fully self-consistent boundary conditions. In this paper, we address this long-standing gap by presenting the first phase-space truncation framework for anisotropic spherical systems and the associated open-source \JAXNS-accelerated \PYTHON package \PIANISTpyNS.

The phase-space truncation framework directly re-parametrises anisotropic DFs of the partially separable form $F(\bigK, \bigL) = F_\bigK(\bigK) \times \bigL^{-2\betac}$, encompassing constant-anisotropy, Osipkov--Merritt, and Cuddeford--Osipkov--Merritt models, and solves the corresponding Poisson equation to obtain the truncated density profile (Method~I; \fref{fig:DF_Trun_varrho}). Provided that the input DF is everywhere physical and monotonic, the truncated DF inherits these well-behaved properties by construction. The tapering of the outer density profile is consequently self-consistently determined by the full phase-space distribution near the truncation boundary, which guarantees dynamical stability at the truncation radius. The complementary truncation Method~II can be used to construct DFs for systems with identical truncated density profiles but with different anisotropy profiles. This is achieved by directly inverting a Method-I-truncated potential--density pair into DFs of different anisotropy models. Both methods preserve the input anisotropy profile and inner density power-law slope by construction; physical constraints on the central DF (\S\ref{ssec:DF_Physicalness}) therefore apply equally to untruncated and truncated systems. Since all Method-II-constructed systems share the same potential--density pair, truncated DFs of all implemented anisotropy models can be self-consistently superposed, yielding composite DFs with highly flexible radially varying anisotropy profiles beyond the known analytical families.

\PIANISTpy implements the above framework in a \JAX just-in-time compiled and auto-vectorised architecture, supporting analytical double-power-law radial density profiles, a library of analytical anisotropy models, and outputting potential--density pairs, DFs, discrete particle initial conditions, velocity dispersion profiles, LOSVDs and associated velocity moments, as schematically illustrated in \fref{fig:flowchart}. \PIANISTpy also natively supports projected velocity outputs as flexible $\Sigma$-weighted averages over a range of aperture geometries that can be directly integrated into observational inference analyses. Numerical benchmarks show that \PIANISTpy achieves a roundtrip accuracy several orders of magnitude higher than \agama \citep{Vasiliev2019MNRAS4821525V} (\aref{app:num_DF}). Isolated relaxation and static-potential tests demonstrate that \PIANISTpyNS's truncation scheme produces initial conditions that are quantitatively more stable than either \galpy \citep{Bovy2015ApJS21629B} or \agamaNS, a direct consequence of the analytically exact truncation boundary conditions. These tests further identify radial orbit instability as an intrinsic physical mode of systems with central radial anisotropy that spontaneously drives prolate central bar formation within a few dynamical times, even in the absence of external or truncation-induced perturbations.

In terms of complementarity with existing widely used codes, \PIANISTpyNS, \galpyNS, and \agama each reflect distinct architectural philosophies optimised for different modelling tasks (see \S\ref{ssec:OP_Comparison}): \galpy operates in classical $(Z, \bigL)$ space without a native Poisson solver; \agama provides a fully self-consistent action-based framework at the cost of losing direct access to the separable $F(Z,\bigL)$ structure required for phase-space truncation; \PIANISTpy combines $(Z,\bigL)$-based phase-space accessibility with a self-consistent Poisson solver and the novel truncation framework introduced here. More fundamentally, \galpy and \agama are `density-first,' obtaining the DF from a given density and potential with finite spatial extent achieved by direct modification in density (or potential). \PIANISTpy is `DF-first,' where the truncation acts directly on the DF at a dimensionless $\bigKt$, from which the resulting density and potential are self-consistently derived to guarantee the equilibrium boundary condition at $\Rt$. The anisotropy families available differ accordingly: \agama provides isotropic, constant-$\betac$, OM, and COM DFs restricted to $-1/2 \leq \betac < 1$ and \galpy offers no COM form, whereas \PIANISTpy provides all these analytical families and the flexible linear superposition thereof (\S\ref{ssec:DF_Superposition}) together with their existence conditions (\S\ref{ssec:DF_Physicalness}); the \JAXNS-native implementation also permits rapid end-to-end differentiable profile or observable fitting with analytic gradients. These tools are best viewed as complementary. For example, \PIANISTpyNS-generated initial conditions can be directly combined with \galpyNS's or \agamaNS's orbit integration, stream modelling, and action-angle analysis. Separately, \agama also supports $F(\mathcal{E}, L)$-type spherical DFs via its \texttt{QuasiSpherical} framework, but natively evaluates all DF inversions and downstream velocity observables purely numerically; \PIANISTpyNS's analytically exact solutions therefore provide a natural validation baseline for \agamaNS's numerical implementation for the class of systems considered in this work.

The scope of this initial release is worth stating: \PIANISTpy currently supports spherical, non-rotating, single-component systems with analytical double-power-law density profiles as the untruncated input. Looking ahead, planned future extensions include the construction of self-consistently truncated multi-component DFs and particle initial conditions (e.g., halo and central stellar population), as well as support for generic numerical input density profiles such as those extracted from cosmological simulations.

\section*{Acknowledgements}
We thank the referee, Gary A. Mamon, for an exceptionally detailed and constructive report that helped improve the clarity and presentation of this paper. We also thank Eugene Vasiliev and Moritz Fischer for useful discussions. BC and FvdB are supported by the National Science Foundation (NSF) through grants AST-2307280 and AST-2407063. This research is partially supported by the National Science and Technology Council (NSTC) of Taiwan under Grant No. NSTC 111-2628-M-002-005-MY4 and the National Taiwan University (NTU) Academic Research-Career Development Project under Grant No. NTU-CDP-113L7729. We use \texttt{NumPy} \citep{numpy} and \texttt{SciPy} \citep{scipy} for data analysis, and \texttt{Matplotlib} \citep{matplotlib} and \texttt{yt} \citep{yt} for data visualisation. We also construct particle initial conditions with \galpy v1.11.0 \citep{Bovy2015ApJS21629B} and \agama v1.0 \citep{Vasiliev2019MNRAS4821525V} for code comparison. The final stage of package assembly, the online documentation, and the proofreading of the manuscript were carried out with the assistance of Anthropic Claude (Fable 5 and Opus 5), under the direction of BC, who reviewed and verified all output.

\section*{Data Availability}

The public \PIANISTpy online documentation and Jupyter notebooks to reproduce all figures in this paper are available at \url{https://pianistpy.com}. Full source code is hosted at \url{https://github.com/barrytchiang/PIANISTpy} (Zenodo: \doi{10.5281/zenodo.23548188}).

\bibliographystyle{mnras}
\bibliography{MyBibTeX1} 

\appendix\onecolumn
\section{Unit Normalisation of Double-power-law Density Profiles}\label{app:Double_Power_Law_Rho}

For any spherical density profile, the corresponding dimensionless relative potential can be written as (following \citealt{Zhao1996MNRAS278488Z})
\begin{align}\label{eqn:P_def_raw}
P(R;C)
&\equiv \frac{\Psi(r)}{4\pi G \rhow \rz^2}
= \frac{1}{R}\int_{0}^{R}\varrho(R')\,R'^2\,\rmd R'
\;+\;\int_{R}^{\infty}\varrho(R')\,R'\,\rmd R',
\end{align}
where $\varrho(R)=\rho(r)/\rhow$ is the dimensionless density and $R=r/\rz$; the quantity $C$ is the density prefactor defined in \eref{eqn:double_power_law_MT}. The unit-normalisation scheme of \citet{Widrow2000ApJS13139W} fixes the dimensionless scaled relative potential energy $P(R)$ defined in \eref{eqn:relative_potential_energy} to be unit-normalised at the centre,
\begin{equation}\label{eqn:P_center_unit}
P(R = 0; C) \eee 1 = \int_{0}^{\infty}\varrho(R')\,R'\,\rmd R',
\end{equation}
with the outer boundary condition $P(R\to\infty)=0$ for non-truncated systems, or $P(\Rt)=0$ for truncated systems with $\varrho(\Rt)=0$. For a given shape of the dimensionless density profile $\varrho(R)$, this requirement translates to a unique value for the density prefactor $C$ and for the potential zero-point $\Phi_0$ so that the central depth of the adopted dimensionless relative potential is unity. Below, we present closed-form expressions for $C(\alpha, \beta, \gamma)$ for the family of double-power-law density profiles given by \eref{eqn:double_power_law_MT}\footnote{See also \citet{Widrow2000ApJS13139W}, who presented $C(\alpha, \beta, \gamma)$ for a small subset of parameter combinations for which $P(R)$ is known analytically.}.

For the family of double-power-law density profiles (Eq.~[\ref{eqn:double_power_law_MT}]), the dimensionless relative potential is given by
\begin{align}\label{app:P_beta}
P(R;C)
&= \frac{C}{\alpha}\Bigg\{\frac{1}{R}\, \bigB\!\left(z;\frac{3-\gamma}{\alpha},\frac{\beta-3}{\alpha}\right)
{\;-\;
\bigB\!\left(z;\frac{2-\gamma}{\alpha},\frac{\beta-2}{\alpha}\right) +
\bigB\!\left(\frac{\beta-2}{\alpha},\frac{2-\gamma}{\alpha}\right)} \Bigg\},
\end{align}
\citep[e.g., \citealt{Zhao1996MNRAS278488Z}; \S5.4.7 of][]{Mo2010gfebookM}. Here,
\begin{equation}\label{eqn:x_def}
z \equiv R^\alpha/(1+R^\alpha),
\end{equation}
$\bigB(z;a,b)$ is the incomplete beta function, $\bigB(a,b)$ is the beta function, and $\Gamma(a)$ is the gamma function \citep[see][for definitions]{Gradshteyn1980tispbookG}. Recall that the physicality of the double-power-law density profile requires $\alpha \geq 1/2$, $\beta > 2$, and $0 \leq \gamma < 2$ (see \S\ref{ssec:Units}).
Evaluating the analytical expression of the relative potential in the limit $P(R\rightarrow 0;C)$, where the two incomplete-beta terms in \eref{app:P_beta} vanish for $\gamma<2$ and the complete-beta term remains finite for $\beta>2$, we have \citep[cf. Appendix A of][]{Zhao1997}
\begin{equation}\label{eqn:P0_beta}
P(0;C)=\frac{C}{\alpha}\,
\bigB\!\left(\frac{\beta-2}{\alpha},\frac{2-\gamma}{\alpha}\right).
\end{equation}
Imposing the unit-normalisation condition $P(R=0; C)=1$ therefore implies that
\begin{equation}\label{eqn:C_fix}
C \equiv \frac{\alpha}{\bigB\!\left(\frac{\beta-2}{\alpha},\frac{2-\gamma}{\alpha}\right)}
= \frac{\alpha\,\Gamma\!\left(\frac{\beta-\gamma}{\alpha}\right)}
{\Gamma\!\left(\frac{\beta-2}{\alpha}\right)\Gamma\!\left(\frac{2-\gamma}{\alpha}\right)}.
\end{equation}

The corresponding dimensionless enclosed-mass profile for the analytical double-power-law family is
\begin{align}\label{eqn:double_power_law_mass}
    \bigM(R)
    \;\eee\;
    \frac{M(r)}{4\pi\rhow\rz^3}
    \;=\;
    \int_0^{R}\varrho(R')\,R'^2\,\rmd R'
    \;=\;
    \frac{C}{\alpha}\,
    \bigB\bigg(\frac{R^\alpha}{1 + R^\alpha}; \frac{3-\gamma}{\alpha}, \frac{\beta - 3}{\alpha}\bigg),
\end{align}
where $R\eee r/\rz$. The conversion constant $q$ introduced in \eref{eqn:M_nor_1} can be written directly in terms of $\bigM$ as
\begin{equation}\label{eqn:q_beta_closed}
    q(c_\rho;\alpha,\beta,\gamma)
    \;\eee\;
    \frac{\bigM(c_\rho)}{C}
    \;=\;
    \frac{1}{\alpha}\,
    \bigB\bigg(\frac{c_\rho^\alpha}{1 + c_\rho^\alpha}; \frac{3-\gamma}{\alpha}, \frac{\beta - 3}{\alpha}\bigg).
\end{equation}

\subsection{Inner ($R\ll 1$) and outer ($R \gg 1$) asymptotics for double-power-law profiles}
\label{app:inner_asymptotics}

All double-power-law profiles reduce to a single power law at sufficiently small radii, $\varrho(R\ll 1)\propto R^{-\gamma}$, and separately at sufficiently large radii, $\varrho(R\gg 1)\propto R^{-\beta}$. First, we derive the corresponding inner asymptotic behaviour of the potential and enclosed-mass profiles, retaining the leading-order term and the next-to-leading-order correction in the $P\to 1$ tail (untruncated DFs) or the $\hat{P}\to 1-\bigKt$ tail (truncated DFs). We then derive the outer asymptotic behaviour of the potential and the corresponding inversion $R(P\to 0)$. These analytical expressions are used for numerical regularisation and solution patching near the boundaries of the inversion domain (see \aref{app:num_DF}).

Expanding the analytical double-power-law profile at small $R\rightarrow0$ yields
\begin{align}\label{eqn:inner_density_expansion}
\varrho(R\ll 1)
= C\,R^{-\gamma}\left[1+\left(\frac{\gamma-\beta}{\alpha}\right)R^\alpha\right]\,.
\end{align}
Next, by using that $z (R\rightarrow 0) = R^\alpha - R^{2\alpha} + \mathcal{O}(R^{3\alpha})$ and that
\begin{equation}\label{eqn:beta_smallz}
\bigB(z\rightarrow0;a,b)=\frac{z^a}{a}\left[1+\frac{a(1-b)}{a+1}\,z+\mathcal{O}(z^2)\right]\,,
\end{equation}
\citep[e.g.,][]{Gradshteyn1980tispbookG}, we can similarly expand \eref{app:P_beta} in the limit $R\to 0$ to obtain
\begin{align}\label{eqn:inner_potential_expansion}
P(R\ll1) &= P_0 -\frac{C}{(3-\gamma)(2-\gamma)}\,R^{2-\gamma}
\left[1+{\frac{(\gamma-\beta)(2-\gamma)(3-\gamma)}{\alpha(2-\gamma+\alpha)(3-\gamma+\alpha)}}\,R^\alpha \right]\,,
\end{align}
where $P_0 \equiv P(R=0)$ is unity for untruncated DFs and equals $1-\bigKt$ for truncated systems. Finally, from \eref{eqn:double_power_law_mass} we obtain
\begin{align}\label{eqn:inner_mass_expansion}
\bigM(R\ll 1)
&= \frac{C}{3-\gamma}\,R^{3-\gamma}
\left[ 1+\frac{(\gamma-\beta)(3-\gamma)}{\alpha(3-\gamma+\alpha)}\,R^\alpha \right]\,.
\end{align}

Next, we consider the limit $R \gg 1$, or equivalently $P\to0$. Here, $z \equiv R^\alpha/(1+R^\alpha) \rightarrow 1$, and the first term in \eref{app:P_beta} dominates at large $R$. Using the standard large-argument expansion of the incomplete beta function near $z\to 1$ yields, to leading order,
\begin{align}\label{eqn:outer_pot_expansion}
    P(R \gg 1) =
    \begin{cases}
    {\dfrac{\bigM(R\rightarrow\infty)}{R} - \dfrac{C}{(\beta-3)(\beta-2)}\,R^{-(\beta - 2)}} & \text{($\beta >3$)},\\
    C R^{-1} \ln R & \text{($\beta =3$)}.
    \end{cases}
\end{align}
The corresponding inversion, to leading order in the limit $P\to 0$, gives
\begin{align}\label{eqn:outer_R_of_Psi}
    R(P \rightarrow 0)
    =
    \begin{cases}
    {\dfrac{\bigM(R\rightarrow\infty)}{P}}
 & \;\;\;\;\:\,\text{($\beta > 3$)},\\
    {\dfrac{C}{P}\,\ln\!\left(\dfrac{C}{P}\right)}
 & \;\;\;\;\:\,\text{($\beta = 3$)}.
    \end{cases}
\end{align}

\section{Distribution Functions and Phase-space Truncation}\label{app:DF}

\subsection{DF preliminaries and model definitions}\label{app:DF_Prelim}

The DF, $F(\bX,\bV)$, represents the probability density within a six-dimensional phase-space volume centred on $(\bX,\bV)$ for an arbitrary dynamical system. For systems exhibiting spherical symmetry\footnote{The analytical complexity in solving for DFs non-trivially increases from spherical to triaxial systems (e.g., see \citealt{Dejonghe1991MNRAS252606D}), and we leave such an implementation in \PIANISTpy for future work.} in both their density distribution and velocity structure (hence with no net rotation), the associated potential--density pairs reduce to functions of radius $R$ alone, and the system is governed by two isolating integrals of motion in a time-invariant potential: the relative specific energy $Z$ and the magnitude of the specific angular momentum $\bigL$ such that $F(\bX,\bV) = F(Z, \bigL)$, thereby reducing the dimensionality of the associated DF from six down to two \citep[e.g.,][]{Dejonghe1987MNRAS22413D}. This symmetry also allows for the advantageous decomposition of the instantaneous 3D velocity of individual member particles, $V = |\bV|$, into the radial $\Vr$ and tangential (transverse) components $\Vt$ with respect to the system's CM through the velocity angle $\varphi \in [0,\pi]$
\begin{align}\label{eqn:velocity_angle}
	\Vr \eee V \cos\varphi; \;\;\;\; \Vt \eee V \sin\varphi.
\end{align}
Consequently, the density profile, as the integral of the DF over velocity space defined in \eref{eqn:rho_DF_Poisson}, can be expressed more explicitly as
\begin{align}\label{eqn:rho_f_anisotropic}
	\varrho(R)={2\pi}\int \rmd\Vr \, \rmd \Vt \, \Vt \, F(R,\Vr,\Vt) = 2\pi\int_0^\pi \rmd\varphi\sin\varphi \int_0^{\sqrt{2P}} \rmd V \, V^2 \, F\left(P - \frac{V^2}{2}, RV\sin\varphi\right).
\end{align}

In the special restrictive case of velocity isotropy, the DF is uniformly distributed in and independent of $\bigL$, and the corresponding physical mass density profile is bijectively mapped to the one-dimensional ergodic DF $F(Z)$ such that  \eref{eqn:rho_f_anisotropic} reduces to
\begin{align}\label{app:Den_Def}
	\varrho(P) = 4\pi \int_0^{P} \rmd Z \, F(Z) \, \sqrt{2(P-Z)},
\end{align}
which via the Poisson equation (Eq.~[\ref{eqn:rho_DF_Poisson}]) gives $\varrho(R) \eee \varrho(P(R))$. Consequently, given a potential--density pair, the isotropic DF is uniquely determined via the Eddington inversion
\begin{align}\label{eqn:isotropic_DF}
	F(Z) = \frac{1}{\sqrt{8}\pi^2}\bigg[\int_0^Z \frac{\rmd P}{\sqrt{Z-P}}\frac{\rmd^2\varrho}{\rmd P^2}+\frac{1}{\sqrt{Z}}\Big(\frac{\rmd\varrho}{\rmd P}\Big)\Big|_{P=0}\bigg]\,,
\end{align}
\citep{Eddington1916MNRAS76572E}.

More generally, for anisotropic systems the DFs are bivariate (with two independent input variables), and \eref{eqn:rho_f_anisotropic} can be written as
\begin{align}\label{app:Aug_Den_Def}
	\varrho_{\text{aug}}(P,R) = 4\pi \int^{P}_0 \rmd Z \int^{\sqrt{2(P-Z)}}_{0} \rmd\Vt \, \Vt \, \frac{F(Z, R\,\Vt)}{\sqrt{2(P-Z)-\Vt^2}}\,,
\end{align}
which explicitly highlights the bivariate nature of the \textit{augmented} density $\varrho_{\text{aug}}(P,R)$, a concept first introduced by \citet{Dejonghe1986PhR133217D} and later coined by \citet{Dejonghe1989ApJ343113D}\footnote{Throughout, we reserve the term `augmented density' to refer exclusively to the \textit{bivariate} function $\varrho_{\text{aug}}(P,R)$ obtained from integrated bivariate (anisotropic) DFs over all velocity space. This is conceptually distinct from the physical one-parameter density profile $\varrho(R(P)) = \varrho(P)$, which is the convention adopted in some recent literature \citep[e.g.,][]{Buyle2007MNRAS375773B, Ciotti2010MNRAS4081070C, BovyDynamics2026}. Alternatively, some authors refer to the zeroth-moment functions of DFs (Eq.~[\ref{eqn:rho_f_anisotropic}]) parameterised by potential $P$ as augmented densities \citep[e.g.,][]{Dejonghe1998CeMDA72219D, Baes2002AA393485B, Baes2022MNRAS5122266B}. Note that these two conventions are not mutually exclusive, as in the special case of velocity isotropy, $\varrho_{\text{aug}}(P,R) = \varrho_{\text{aug}}(P) = \varrho(R(P))$.}. Importantly, the augmented density $\varrho_{\text{aug}}(P,R)$ maps bijectively to the corresponding bivariate DF, which together maps uniquely to the physical $\varrho(R)$ with the Poisson equation as closure \cite[e.g.,][]{Baes2004MNRAS35118B}. However, the inverse mapping is not unique. The physical density profile $\varrho(R)$ does not uniquely determine $\varrho_{\text{aug}}(P,R)$, as infinitely many augmented densities (and thus bivariate DFs) can satisfy the constraint
\begin{align}\label{app:Aug_to_Phy_Den}
\varrho(R) = \varrho_{\text{aug}}(P(R),R).
\end{align}
Intuitively, this non-uniqueness reflects the additional degrees of freedom in the velocity structure (i.e., anisotropy) of a physical system that remain even after its spatial structure $\varrho(R)$ is fully specified \citep{Baes2007AA471419B}.

A major technical difficulty going beyond the assumption of velocity isotropy lies in the construction of anisotropic DFs $F(Z, \bigL)$, given a desired physical density profile $\varrho(R)$. Unlike the isotropic case where the DF can always be uniquely computed from $\varrho(R)$ via \eref{eqn:isotropic_DF}, such closed-form analytical inversion is generally not known for generic anisotropy profiles $\beta_\sigma(R)$. Pioneering work by \citet{Dejonghe1986PhR133217D} demonstrated that evaluating bivariate anisotropic DFs by directly inverting \eref{app:Aug_Den_Def} via the Laplace-Mellin integral transformation is complex and almost always numerically unstable for generic $\varrho_{\text{aug}}$ and $\beta_\sigma$. 

An alternative and fruitful approach involves a three-step modelling strategy: (1) seeking a parameterisation of the DF through specific combinations of the isolating integrals, (2) analytically computing the corresponding augmented density, and (3) identifying a suitable functional form for an \textit{auxiliary} density $\varrho_\text{aux}(P)$. This auxiliary density is constructed such that the resulting DF can be derived from the derivatives of $\varrho_\text{aux}(P)$, effectively mirroring the simplicity of the isotropic Eddington inversion. While the augmented density is tied to the physical profile via the potential $P(R)$ through the Poisson equation, the auxiliary density remains a conceptually distinct mathematical construct. By choosing an expression for $\varrho_\text{aux}(P)$ that allows the anisotropic DF to retain a functional form nearly identical to the isotropic case in \eref{app:Den_Def}, this methodology enables an analytical inversion that is structurally analogous to the isotropic Eddington formula.

Here we briefly review the classes of analytical anisotropic models implemented in \PIANISTpy whose $\varrho_{\text{aug}}$-to-DF inversion formulae are analytically known:
\begin{itemize}
	\item \textbf{Constant Anisotropy:} Separable DFs of the form $F(Z,\bigL) = F_{Z}(Z) \bigL^{-2\betac}$ \citep{Henon1973} represent a class of spherically symmetric solutions with radius-independent anisotropy $\beta_\sigma(R) = \betac \in (-\infty,1)$. For such systems, the density profile can be straightforwardly computed from \eref{eqn:rho_f_anisotropic} by using that $\int_0^\pi \rmd\varphi \sin^{1-2\betac} \varphi = {\sqrt{\pi}\,\Gamma(1-\betac)/\Gamma(3/2-\betac)}$ so that
    \begin{align}\label{eqn:const_beta_DF}
			\varrho_\text{aug}(P,R) = \frac{2\pi^{3/2}\,\Gamma(1-\betac)}{R^{2\betac}\,\Gamma({3/2}-\betac)} \int_0^P \rmd Z \, \frac{F_Z(Z)}{[2(P - Z)]^{\betac - 1/2}}\,,
    \end{align}
    \citep[e.g., \S4.3.2 of][]{Binney&Tremaine2008}. Note that this expression reduces to \eref{app:Den_Def} for $\betac = 0$. To analytically solve for the `energy portion' of the DF $F_{Z}(Z)$, we first define the suitable auxiliary density
    \begin{align}\label{eqn:const_beta_aux}
		\varrho_\text{aux}(P) \eee \varrho_\text{aug}(P,R) \times R^{2\betac}\big|_{R = R(P)} = \frac{2\pi^{3/2}\Gamma(1-\betac)}{\Gamma({3/2}-\betac)} \int_0^P \rmd Z \frac{F_Z(Z)}{[2(P - Z)]^{\betac - 1/2}}\,,
    \end{align}
	\citep{Cuddeford1991MNRAS253414C}, which can be Abel-inverted to give
    \begin{align}\label{eqn:F_bigE_constant_beta_gen}
		F_Z(Z) = \frac{2^\betac}{(2\pi)^{3/2}\,\Gamma(1-n)\,\Gamma(1-\betac)}\bigg[\int_0^Z \rmd P \, \frac{\big(\rmd^{m+1}\varrho_\text{aux}/\rmd P^{m+1}\big)}{(Z-P)^n} + \frac{1}{Z^n}\frac{\rmd^m\varrho_\text{aux}}{\rmd P^m}\Big|_{P=0}\bigg],
    \end{align}
    \cite[e.g.,][]{An2006AJ131782A}, where $m \eee \lfloor 3/2-\betac \rfloor$ and $n \eee 3/2 - \betac - m$. Here $\lfloor y \rfloor$ denotes the floor function that returns the largest integer less than or equal to $y$. In what follows, we refer to \eref{eqn:F_bigE_constant_beta_gen} as the \textit{Cuddeford inversion operator}. The surface terms evaluated at $P=0$ in these inversion formulae vanish when $\varrho_\text{aux}$ falls off more rapidly than $P^m$ as $P \to 0$. For untruncated profiles, $\beta > 3$ suffices for all constant-anisotropy models and for (C)OM Type~I models with $\betac > -1/2$; (C)OM Type~I models with $\betac \leq -1/2$ instead require $\beta > m+2$.

    For $-0.5 < \betac {\leq 0.5}$, \eref{eqn:F_bigE_constant_beta_gen} simplifies to
    \begin{align}\label{eqn:F_bigE_constant_beta}
		F_Z(Z) = \frac{2^\betac}{(2\pi)^{3/2}\,\Gamma(\frac{1}{2}+\betac)\,\Gamma(1-\betac)}\bigg[\int_0^Z \rmd P \, \frac{\big(\rmd^{2}\varrho_\text{aux}/\rmd P^{2}\big)}{(Z-P)^{\frac{1}{2}-\betac}} + \frac{1}{Z^{\frac{1}{2}-\betac}}\frac{\rmd\varrho_\text{aux}}{\rmd P}\Big|_{P=0}\bigg].
	\end{align} 
    Further algebraic simplification is possible for half-integer values of $\betac$. For example, for $\betac = \pm0.5$, we have
        \begin{align}
		&
    		\begin{cases}\label{eqn:const_beta0p5_DF}
			F_Z(Z) = \frac{1}{2\pi^2} \frac{\rmd\varrho_\text{aux}}{\rmd P}\scalebox{0.9}{\Big|}_{P=Z}\\
			\varrho(R) = \varrho_\text{aux}(P)\times R^{-1}{\big|}_{P=P(R)} = \frac{2\pi^2}{R} \int_0^P \rmd {Z} \, F_Z(Z)
		\end{cases} && \;\;\;\;\;\;\text{($\betac = 0.5$),}\\
		&
		\begin{cases}\label{eqn:const_betam0p5_DF}
			F_Z(Z) = \frac{1}{2\pi^2} \frac{\rmd^2\varrho_\text{aux}}{\rmd P^2}\scalebox{0.9}{\Big|}_{P=Z}\\
			\varrho({R}) = \varrho_\text{aux}(P)\times R{\big|}_{P=P(R)} = 2\pi^2 R \int_0^P \rmd {Z} \, F_Z(Z)\,(P - Z)
		\end{cases} && \;\;\;\;\;\;\text{($\betac = -0.5$)}.
	\end{align}
	
\medskip

\item \textbf{OM Type I:} These DFs are functions of the single parameter $\bigQ_{-}$, defined in \eref{eqn:unit_nor_4}, and have a radius-dependent anisotropy profile
    \begin{align}
		\betasigma(R) = \frac{1}{1+\Ra^2/R^2} {= \frac{R^2}{R^2+\Ra^2}},
	\end{align}
    that transitions from being isotropic $\betasigma(R\ll\Ra) \rightarrow 0$ near the centre to purely radial-orbit-dominated $\betasigma(R\gg\Ra) \rightarrow 1$ at large radii \citep{Osipkov1979PAZh577O, Merritt1985AJ901027M}. Note that the anisotropy radius $\Ra$ has a physical lower limit that depends on $P(R)$ \citep{Carollo1995}. Rewriting \eref{eqn:rho_f_anisotropic} by the following change of variable $\rmd\bigQ_{-} = -[1+(R/\Ra)^2\sin^2\varphi]V\rmd V$, the augmented density reads
    \begin{align}\label{eqn:OM_DF_1}
		\varrho_\text{aug}(P, R) = 2\pi \int_0^\pi \rmd\varphi \sin\varphi \int_0^P \rmd\bigQ_{-} \, F(\bigQ_{-}) \frac{\sqrt{2(P-\bigQ_{-})}}{[1+(R/\Ra)^2\sin^2\varphi]^{3/2}}= \frac{4\pi}{1+(R/\Ra)^2} \int_0^P \rmd\bigQ_{-} \, F(\bigQ_{-}) \, \sqrt{2(P - \bigQ_{-})},
	\end{align}
    where we have used that $\int_0^\pi \rmd\varphi \sin\varphi\,[1+(R/\Ra)^2\sin^2\varphi]^{-3/2} = 2/[1+(R/\Ra)^2]$. By defining the following suitable auxiliary density  
    \begin{align}\label{eqn:OM_DF_2}
		\varrho_\text{aux}(P) \eee \bigg(1+\frac{R^2}{\Ra^2}\bigg) \, \varrho_\text{aug}(P,R)\bigg|_{R=R(P)}= 4\pi \int_0^{P} \rmd\bigQ_{-} \, F(\bigQ_{-}) \, \sqrt{2(P - \bigQ_{-})},
	\end{align}
    we recover the identical functional form as in the isotropic case (Eq.~[\ref{app:Den_Def}]), such that the associated DF can be analytically inverted via the following Eddington-like formula 
    \begin{align}\label{eqn:OM_DF_3}
		F(\bigQ_{-}) = \frac{1}{\sqrt{8}\pi^2}\bigg[\int_0^{\bigQ_{-}}\frac{\rmd P}{\sqrt{\bigQ_{-}-P}}\frac{\rmd^2\varrho_\text{aux}}{\rmd P^2} + \frac{1}{\sqrt{\bigQ_{-}}}\bigg(\frac{\rmd\varrho_\text{aux}}{\rmd P}\bigg)\bigg|_{P=0}\bigg],
	\end{align}
    \citep{Merritt1985AJ901027M}.

\medskip
    
    \item \textbf{OM Type II:}
    These DFs are functions of the alternative parameter $\bigQ_{+}$ defined in \eref{eqn:relative_potential_energy}, which yields a radius-dependent anisotropy profile 
    \begin{align}\label{eqn:OM_TypeII_beta}
		\betasigma(R) = \frac{1}{1-\Ra^2/R^2} {= \frac{R^2}{R^2-\Ra^2}},
	\end{align}
    that becomes completely tangentially biased $\betasigma(R\rightarrow\Ra) \rightarrow -\infty$ at the anisotropy radius from below \citep{Merritt1985AJ901027M}. From \eref{eqn:rho_f_anisotropic}, we find
    \begin{align}\label{eqn:OM_TypeII_DF_1}
		\varrho_\text{aug}(P, R) = 2\pi \int_0^\pi \rmd\varphi \sin\varphi \int_0^P \rmd\bigQ_{+} \, F(\bigQ_{+}) \, \frac{\sqrt{2(P-\bigQ_{+})}}{[1-(R/\Ra)^2\sin^2\varphi]^{3/2}}\,.
	\end{align}
    Using the suitable auxiliary density
    \begin{align}\label{eqn:OM_TypeII_DF_2}
		\varrho_\text{aux}(P) \eee \bigg(1-\frac{R^2}{\Ra^2}\bigg)\,\varrho_\text{aug}(P,R)\bigg|_{R=R(P)}= 4\pi \int_0^P \rmd\bigQ_{+} \, F(\bigQ_{+}) \, \sqrt{2(P - \bigQ_{+})},
	\end{align}
    the resulting DF inversion has the same functional form as its Type~I counterpart:
    \begin{align}\label{eqn:OM_TypeII_DF_3}
		F(\bigQ_{+}) = \frac{1}{\sqrt{8}\pi^2}\bigg[\int_0^{\bigQ_{+}}\frac{\rmd P}{\sqrt{\bigQ_{+}-P}}\frac{\rmd^2\varrho_\text{aux}}{\rmd P^2} + \frac{1}{\sqrt{\bigQ_{+}}}\bigg(\frac{\rmd\varrho_\text{aux}}{\rmd P}\bigg)\bigg|_{P=0}\bigg]\,.
	\end{align}
    As pointed out by \citet{Merritt1985AJ901027M}, this class of tangentially biased solutions is discontinuous at $\Ra$ and unphysical for $R \geq \Ra$, as is evident from the factor $[1-(R/\Ra)^2\sin^2\varphi]^{-3/2}$ on the RHS of \eref{eqn:OM_TypeII_DF_1}. The OM Type~II DFs are thus formally defined for only $\bigQ_{+} > P(\Ra)$, the relative potential at the anisotropy radius. Numerically, \PIANISTpy writes valid DF values only on the grid over $\bigQ_{+} > P(\Ra)$ and gates the truncation value $\bigKt > P(\Ra)$. Nonetheless, physical and dynamically stable solutions can still be obtained by truncating the density profile at some radius $\Rt < \Ra$ \citep[e.g.,][see also \fref{fig:DF_Trun_varrho}]{Baes2022MNRAS5122266B, Baes2023MNRAS5196065B}.

\medskip

\item \textbf{COM:} \citet{Cuddeford1991MNRAS253414C} generalised the constant-anisotropy and OM models by introducing the functional form $F(\bigQ_\pm,\bigL) = F_{\bigQ_\pm}(\bigQ_\pm) \bigL^{-2\betac}$, which results in the following radius-dependent anisotropy profiles
    \begin{align}\label{eqn:COM_beta}
		\betasigma(R; \bigQ_\pm) = \frac{1\mp\betac(\Ra/R)^2}{1\mp(\Ra/R)^2} = \frac{(R/\Ra)^2\mp\betac}{(R/\Ra)^2\mp1}.
	\end{align}
    Compared to the original OM model, this extension modifies only the central anisotropy to a constant, but non-zero $\betac$. At larger radii, the model converges to the OM model of Type~I ($\betasigma \rightarrow 1$ as $R \gg \Ra$) or Type~II ($\betasigma \rightarrow -\infty$ as $R \rightarrow \Ra$). Since the DF functional form matches that of the constant-anisotropy parameterisation, the inverted DF follows from the same Abel inversion as \eref{eqn:F_bigE_constant_beta_gen}, written here in its general form\footnote{The prefactor in Eq.~(43) of \citet{Vasiliev2018arXiv180208255V} is incorrect; the second Gamma functions in the denominators should be $\Gamma(-1/2+\beta)$ and $\Gamma(1/2+\beta)$ respectively.}:
    \begin{align}\label{eqn:COM_DF_gen}
	{F_{\bigQ_\pm}(\bigQ_\pm) = \frac{2^\betac}{(2\pi)^{3/2}\, \Gamma(1-n)\,\Gamma(1-\betac)}\,\frac{\rmd}{\rmd\bigQ_\pm}\int_0^{\bigQ_\pm} \rmd P \, \frac{\rmd^{m}\varrho_\text{aux}/\rmd P^{m}}{(\bigQ_\pm-P)^n}},
	\end{align}
    where $m$ and $n$ are as defined in \eref{eqn:F_bigE_constant_beta_gen}\footnote{\label{fn:COM_IBP}Integrating \eref{eqn:COM_DF_gen} by parts recovers the analytically simpler form of \eref{eqn:F_bigE_constant_beta_gen} only if $\rmd^m\varrho_\text{aux}/\rmd P^m$ is continuous on $[0,\bigQ_\pm]$ and $\int_0^{\bigQ_\pm}|\rmd^{m+1}\varrho_\text{aux}/\rmd P^{m+1}|\,(\bigQ_\pm-P)^{-n}\,\rmd P<\infty$, neither of which is guaranteed for COM models. For half-integer $\betac$, $n=0$ and \eref{eqn:COM_DF_gen} reduces to $F_{\bigQ_\pm}=\frac{2^\betac}{(2\pi)^{3/2}\,\Gamma(1-\betac)}\,\rmd^m\varrho_\text{aux}/\rmd P^m|_{P=\bigQ_\pm}$. For its numerical implementation, \PIANISTpy evaluates these simpler forms wherever they apply for the given user inputs, and otherwise defaults to \eref{eqn:COM_DF_gen}.}. The suitable auxiliary density now takes the form 
    \begin{align}\label{eqn:COM_aux_gen}
		\varrho_\text{aux}(P; \bigQ_\pm) \eee \bigg(1\mp\frac{R^2}{\Ra^2}\bigg)^{1-\betac}\,\varrho_\text{aug}(P,R; \bigQ_\pm) \times R^{2\betac}\bigg|_{R=R(P)} = \frac{2\pi^{3/2}\,\Gamma(1-\betac)}{\Gamma({3/2}-\betac)}\int_0^P \rmd\bigQ_\pm \, \frac{F_{\bigQ_\pm}(\bigQ_\pm)}{[2(P - \bigQ_\pm)]^{\betac - 1/2}},
	\end{align}	
    \citep{Cuddeford1991MNRAS253414C}. In the formal limit $\Ra \rightarrow \infty$, the $R/\Ra$ factors in the (C)OM DFs vanish such that $\betasigma(R) = \betac$. Hence, the COM (OM) model reduces to the constant-anisotropy (isotropic) model in the limit $\Ra \rightarrow \infty$. For COM Type~II models, the same type of discontinuity as in OM Type~II is now set by the factor $(1 - R^2/\Ra^2)^{1-\betac}$. Unlike OM Type~II ($\betac = 0$), whose integer-power factor $1 - R^2/\Ra^2$ is a polynomial and regular at $P(\Ra)$, COM Type~II models generally have a non-integer power, so their DFs are singular and diverge at $\bigQ_{+} = P(\Ra)$. Analytically, the derivative of order $m$ of the auxiliary density in \eref{eqn:COM_DF_gen} carries the non-analytic term $|P - P(\Ra)|^{\,1-\betac-m}$, and the inversion lowers its exponent by $n$ to exactly $-1/2$, so the DF diverges as $[\bigQ_{+} - P(\Ra)]^{-1/2}$ for every non-integer $\betac$. Indeed, the bottom panels of \fref{fig:DF_Trun_DFs} show that the untruncated OM Type~II DFs decrease smoothly and monotonically toward $\bigQ_{+} = P(\Ra)$ and simply end there, whereas all COM examples with non-integer powers ($\betac = \pm0.25$) show a highly localized positive divergence just above the respective essential discontinuity. Furthermore, although (C)OM Type~II models are undefined for $R \geq \Ra$, the inversion of \eref{eqn:COM_DF_gen} for the physical regime $R < \Ra$ still integrates over $R > \Ra$, where $(1 - R^2/\Ra^2)^{1-\betac}$ is complex-valued for non-integer $1 - \betac$. One practical consequence of this structural discontinuity is that (C)OM Type~II DF solutions are not unique, even within the physical regime $\bigQ_{+} > P(\Ra)$, but depend on the convention adopted for the analytical continuation into the unphysical regime. \PIANISTpy adopts the standard continuation of a real-valued power with the odd branch $|u|^{1-\betac}\,\mathrm{sgn}\,(u)$ with $u \eee 1 - R^2/\Ra^2$ (e.g., Chapter~I, \S3.3 of \citealp{GelfandShilov1964}); see the online technical documentation\footnote{\label{fn:technotes}\url{https://pianistpy.com/docs/reference/notes}} for the specific detail.

\end{itemize}

\subsubsection{$P\to0$ asymptotics of (C)OM Type~I DFs}
\label{app:DF_lowK_asymp}

As shown in \fref{fig:DF_Trun_DFs}, the low-$P$ behaviour of (C)OM DFs is richer than that of the constant-anisotropy model, whose DFs always decay to zero as $P \rightarrow 0$. Here, we focus on (C)OM Type~I models and derive the analytical power-law scaling of the DF in the limit $\bigQ_{-} \rightarrow 0$. This scaling provides the low-$\bigK$ analytical anchor used in the numerical tail treatment described in \aref{app:num_DF_low_tail}.

First, in the limit $R \gg \Ra$, the auxiliary density expressions of the OM and COM
Type~I models are simplified to $(1+R^2/\Ra^2) \simeq R^2/\Ra^2$ at leading order, giving $\varrho_{\rm aux} \simeq \varrho\,R^{2}/\Ra^{2(1-\betac)}$ for all $\betac$. Substituting the physical density $\varrho \propto R^{-\beta}$ and the outer potential inversion of \eref{eqn:outer_R_of_Psi} then gives
\begin{align}\label{eqn:rho_aux_OM_COM}
    \varrho_{\rm aux}(P \rightarrow 0)
    \propto
    \begin{cases}
        {P^{\,\beta-2}} & \qquad\qquad\text{((C)OM Type~I; $\beta>3$)},\\
        {P\,[\ln(C/P)]^{-1}} & \qquad\qquad\text{((C)OM Type~I; $\beta=3$)},\\
    \end{cases}
\end{align}
where the OM Type~I exponent is a special case of COM Type~I with $\betac=0$. The expression $\rmd^{m+1}\varrho_{\rm aux}/\rmd P^{m+1}$ in the Cuddeford inversion of \eref{eqn:F_bigE_constant_beta_gen} then has the same power-law exponent as \eref{eqn:rho_aux_OM_COM} subtracted by $(m+1)$. The Abel kernel further carries a factor $(\bigQ_{-} - P)^{-n}$ with $n = 3/2 - \betac - m$. Evaluating the dominant contribution to the integral in \eref{eqn:F_bigE_constant_beta_gen} as $\bigQ_{-} \rightarrow 0$ via a standard power-counting argument gives\footnote{For $\beta = 3$, the factor $[\ln(C/\bigQ_{-})]^{-1}$ in $F$ is slowly varying relative to $\bigQ_{-}^{\kappa_{\rm (C)OM}}$ and leaves the sign of $\kappa_{\rm (C)OM}$ and the qualitative divergence classification unchanged.}
\begin{align}\label{eqn:F_lowK_asymp}
    F_{\bigQ_{-}}(\bigQ_{-} \rightarrow 0)
    \;\propto\;
    \bigQ_{-}^{\;\kappa_{\rm (C)OM}},
    \qquad
    \kappa_{\rm (C)OM} =
    \begin{cases}
        {\beta - \frac{7}{2}}
 & \qquad\text{(OM Type~I)},\\
        {\beta + \betac - \frac{7}{2}}
 & \qquad\text{(COM Type~I; $\beta>3$)},\\
        {\betac - \frac{1}{2}}
 & \qquad\text{(COM Type~I; $\beta=3$)}.
    \end{cases}
\end{align}
Thus for OM Type~I, the low-$\bigQ_-$ tail is positively divergent for $\beta < 7/2$ and vanishes for $\beta > 7/2$. For instance, the NFW profile with $\beta = 3$ has $\kappa_{\rm (C)OM} = -1/2$, and Hernquist with $\beta = 4$ has $\kappa_{\rm (C)OM} = +1/2$. For COM Type~I, asymptotic divergence ($\kappa_{\rm (C)OM} < 0$) requires $\betac < 7/2 - \beta$ for $\beta > 3$, while for $\beta = 3$ it requires $\betac < 1/2$. In particular, the NFW DF examples with $\betac = \pm 0.25$ shown in \fref{fig:DF_Trun_DFs} have $\kappa_{\rm (C)OM} = -3/4$ and $-1/4$ and indeed exhibit positive divergence in the $\bigQ_{-} \rightarrow 0$ limit. We note, for generic COM DFs, this asymptotic analysis at $R\to\infty$ is decoupled from whether the DF remains non-negative and monotonic at $\bigQ_{-}>0$ (at finite radius), which can only be inspected and determined numerically.

\subsection{Truncation of partially separable (an)isotropic DFs}\label{app:DF_Truncation}

Here we present the self-consistent DF truncation framework implemented in \PIANISTpy for non-rotating spherical systems with finite central potential, $|\Phi(r=0)|<\infty$ (i.e., not restricted to the double-power-law density profiles discussed in \aref{app:Double_Power_Law_Rho}). We describe the technical details and summarise the advantages and limitations of the two methods illustrated in \fref{fig:Trun_Illustration}. Throughout, truncated quantities are denoted with a caret (e.g., $\hat{F}$ denotes the truncated DF).

\subsubsection{Method I:}\label{app:DF_trun_I}

Consider the family of partially separable (an)isotropic DFs of the form
\begin{align}\label{eqn:DF_Functional_form}
	F(\bigK, \bigL; \betac) = F_\bigK(\bigK)\,\bigL^{-2\betac}.
\end{align}
The key observation is that the auxiliary density $\varrho_\text{aux}(P)$ uniquely determines, and is uniquely determined by, the one-parameter function $F_\bigK(\bigK)$ for a fixed $\betac$; i.e., the mapping $\varrho_\text{aux}(P)\leftrightarrow F_\bigK(\bigK)$ is bijective. The dimensionless coordinate $\bigK\in[0,1]$\footnote{This unit-normalised range is guaranteed by construction under the normalisation scheme of \citet{Widrow2000ApJS13139W}; see \S\ref{ssec:Units} and \aref{app:Double_Power_Law_Rho}.} is an energy-like variable, and equals $Z$ or $\bigQ_\pm$ depending on the adopted anisotropy model. The corresponding expressions for each model are given in Eqs.~(\ref{eqn:const_beta_DF}, \ref{eqn:F_bigE_constant_beta_gen}) (constant anisotropy), Eqs.~(\ref{eqn:OM_DF_2}, \ref{eqn:OM_DF_3}) (OM~I), Eqs.~(\ref{eqn:OM_TypeII_DF_2}, \ref{eqn:OM_TypeII_DF_3}) (OM~II), and Eqs.~(\ref{eqn:COM_DF_gen}, \ref{eqn:COM_aux_gen}) (COM). This algebraic structure, identical to the isotropic case, permits self-consistent phase-space truncation by re-parameterising only $F_\bigK(\bigK)$ along $\bigK$, thereby providing a direct generalisation of the isotropic energy-truncation framework of \citet{Drakos2017MNRAS4682345D}.

Specifically, for a chosen truncation value $\bigKt\in(0,1)$, we introduce the shifted variable
$\hat{\bigK}\equiv \bigK-\bigKt\in[0,1-\bigKt]$
and define the truncated DF $\hatF$ as
\begin{align}\label{eqn:DF_Truncation}
	\hatF(\hat{\bigK}, \bigL) \equiv
	\begin{cases}
		F(\hat{\bigK}+\bigKt,\bigL)-F(\bigKt,\bigL), & \hat{\bigK}\ge 0,\\
		0, & \hat{\bigK}<0,
	\end{cases}
\end{align}
where $F(\bigK,\bigL)$ is the original (untruncated) DF. This redefinition is always possible via an appropriate shift of the reference constant $\Phi_0$ defined in \eref{eqn:unit_nor_4}. If $F(\bigK,\bigL)$ is everywhere physical (non-negative) and monotonic in $\bigK$, then truncation via this re-parameterisation guarantees preservation of these properties in $\hatF$ while enforcing self-consistent boundary conditions. In particular, provided the input DF is free of instability-driving inflection points in $\bigK$ \citep[see][]{Weinberg2023MNRAS5254962W, Dattathri2025, 2026MNRAS.546ag050B}, the transformation $F_\bigK(\bigK)\rightarrow \hatF_\bigK(\hat{\bigK})$ constitutes a shift along the $\bigK$-axis and therefore does not change the topology of the bivariate DF $F(\bigK,\bigL)$; consequently, the truncated DF retains both non-negativity and monotonicity.

The truncated relative potential $\hatP(R)$ and density profile $\hatvarrho(R)$ are then obtained by solving the Poisson equation
\begin{align}\label{eqn:Poissonr_Truncation}
	\frac{\rmd^2\hatP}{\rmd R^2} + \frac{2}{R} \frac{\rmd\hatP}{\rmd R} = -\hatvarrho(R),
\end{align}
coupled to the velocity integral of $\hatF$,
\begin{align}\label{eqn:Poissonr_Truncation_rho}
	\hatvarrho(R) \equiv \hatvarrho[\hatP(R)] = \int \rmd^3 {\bV}\,\hatF(\hat{\bigK},\bigL),
\end{align}
with boundary conditions $\hatP(R=0)=1-\bigKt$ and the corresponding value of the central slope $\rmd\hatP/\rmd R|_{R\rightarrow0}$, integrated outward to the truncation radius $\Rt$ where $\hatP(\Rt)=0$. Specifically, this central slope is set by the inner density slope $\gamma$ and the density prefactor $C$ of \eref{app:P_beta},
{
\begin{equation}\label{eqn:central_slope}
\frac{\rmd\hatP}{\rmd R}\bigg|_{R\to 0} = -\frac{C\,R^{1-\gamma}}{3-\gamma} \to
\begin{cases}
0 & \qquad\,(0 \leq \gamma < 1),
\\
-C/2 & \qquad\,(\gamma = 1),
\\
-\infty & \qquad\,(1 < \gamma < 2),
\end{cases}
\end{equation}
}
that vanishes, $\rmd\hatP/\rmd R|_{R=0} = 0$, exactly for a cored or shallow ($\gamma<1$) central profile as considered in the seminal work of \citet{King1966AJ7164K}, but was incorrectly adopted in \citet{Drakos2017MNRAS4682345D} for cuspy NFW profiles. In practice, for all $\gamma$ (and necessarily for $\gamma \geq 1$), \PIANISTpy instead computes the equivalent expression $\hatP(R)-\hatP(0) \to -C\,R^{2-\gamma}/[(3-\gamma)(2-\gamma)]$, the leading term of the inner series expansion \eref{eqn:inner_potential_expansion}, at the first non-zero central grid radius for closure. For any anisotropy model of the form in \eref{eqn:DF_Functional_form}, the velocity integral $\int \rmd^3V\,\hatF(\hat{\bigK},\bigL)$ can be algebraically reduced to a one-dimensional integral over $\hatP\in[0,1-\bigKt]$ (see \aref{app:DF_trun_II}). Consequently, these boundary conditions enforce a self-consistent phase-space cutoff at $\Rt$. 

As a trade-off, while the resulting $\hatvarrho(\hatP)$ generally corresponds to a smoothly truncated spherical system with finite spatial extent and convergent total mass\footnote{\label{fn:method_I_issue}A notable exception occurs for Type~I (C)OM models at sufficiently small $\bigKt$. Because these models comprise \textit{purely} radial orbits at large radii, $\bigKt$ must be chosen large enough (in a manner that depends on $\Ra$) to remove a sufficient fraction of outer radial orbits such that the truncation reaches the transitional region comprising a wider range of orbits and $\hatvarrho$ can self-consistently taper to zero at a finite radius. Otherwise, the self-consistent $\hatvarrho$ still extends to spatial infinity ($\Rt\to\infty$), albeit with a suppressed outer density relative to the untruncated profile. This limitation can be circumvented by the Method-II-based truncation.},
the detailed tapering of $\hatvarrho(R)$ near $\Rt$ is set by $\hatF(\hat{\bigK},\bigL)$ and therefore depends on the adopted anisotropy model, even when starting from the same untruncated density profile $\varrho(R)$. Additional minor caveats are addressed below on a model-by-model basis.

\subsubsection{Method II:}\label{app:DF_trun_II}

This method is introduced to overcome the limited control that Method~I affords over the density taper near the truncation radius. It allows for the construction of models with identical truncated density profiles but different velocity anisotropy profiles. Starting from a self-consistently truncated reference model (e.g., an isotropic NFW truncated at $\Zt$), Method~II uses its $\hatvarrho(\hatP)$--$\hatP(R)$ pair as input to construct an auxiliary profile $\hatvarrho_{\rm aux}(\hatP)$ appropriate for a chosen anisotropy model, and then invert for the corresponding DF. For bivariate DFs of the form given by \eref{eqn:DF_Functional_form}, the inverted $\hatF(\hat{\bigK},\bigL)$ spans exactly $\hat{\bigK}\in[0,1-\bigKt]$, matching the truncated potential range $\hatP\in[0,1-\bigKt]$. This procedure guarantees that all output models share the same physical $\hatvarrho(R)$, at the cost that the inverted $\hatF$ may develop sharp inflection features or unphysical kinks near the truncation boundary, particularly when an aggressively truncated input density is paired with strong outer radial anisotropy. Therefore, in general, independent inspection of $\hatF$ and/or numerical stability tests may be required. Below, we compare Methods~I and~II across the DF models available in \PIANISTpyNS.
\begin{itemize}
    \item \textbf{Isotropy:} The DF of the isotropic model $\betasigma = 0$ is uniform in angular-momentum space and depends only on energy, such that $F(Z,\bigL) = F(Z) \equiv F_Z(Z)$. For a chosen truncation energy $\Zt \in (0,1)$, the resulting truncated physical density profile is
    \begin{align}\label{eqn:Iso_DF_Density}
		\hatvarrho(\hatP) = \hatvarrho_\text{aux}(\hatP) = 4\pi \int_0^{\hatP} \rmd\hatZ \, \hatF(\hatZ) \, \sqrt{2(\hatP-\hatZ)}\,,
    \end{align}
    (cf. Eq.~[\ref{app:Den_Def}]), where $\hatZ \eee Z-\Zt \in [0,1-\Zt]$. The dimensional density profile is simply $\rho(r) = \rhow \times \hatvarrho(\hatP(\rz R))$ for $R \in [0,\Rt]$.

    \medskip
    
    \item \textbf{Constant-anisotropy:} The DF of constant (radius-independent) $\betasigma$ models takes the form $F(Z,\bigL) = F_Z(Z) \bigL^{-2\betac}$, and the Method~I phase-space-truncated density profile is given by
    \begin{align}\label{eqn:Constant_beta_DF_Density}
    	\hatvarrho(\hatP) = \hatvarrho_\text{aux}(\hatP)\times R^{-2\betac} \Big|_{R=R(\hatP)} =\bigg\{ \frac{2\pi^{3/2}\,\Gamma(1-\betac)}{\Gamma({3/2}-\betac)} \int_0^{\hatP} \rmd\hatZ \, \frac{\hatF_Z(\hatZ)}{[2(\hatP - \hatZ)]^{\betac - 1/2}}\bigg\}\times R^{-2\betac} \bigg|_{R=R(\hatP)}.
    \end{align}
    Again, the above expression can be considerably simplified for half-integer $\betac$. For example, $\hatvarrho(\hatP) = (2\pi^2/R) \int_0^{\hatP} \rmd\hatZ \, \hatF_Z(\hatZ)$ at $\betac=0.5$, while $\hatvarrho(\hatP) = (2\pi^2R) \int_0^{\hatP} \rmd\hatZ \, \hatF_Z(\hatZ) \times (\hatP-\hatZ)$ at $\betac=-0.5$.
    
    The alternative Method~II truncation approach starts from an already Method-I-truncated potential--density pair, say at a truncation energy $\Zt$ generated from the isotropic model. The corresponding truncated DF of constant $\betac \in (-\infty, 1)$ can then readily be computed from
    \begin{align}\label{eqn:hatF_bigE_constant_beta}
    	\hatF_Z(\hatZ) = \frac{2^\betac}{(2\pi)^{3/2}\Gamma(1-n)\Gamma(1-\betac)}\bigg[\int_0^{\hatZ} \rmd\hatP \frac{\big(\rmd^{m+1}\hatvarrho_\text{aux}/\rmd\hatP^{m+1}\big)}{(\hatZ-\hatP)^n} + \frac{1}{\hatZ^n}\frac{\rmd^m\hatvarrho_\text{aux}}{\rmd\hatP^m}\Big|_{\hatP=0}\bigg],
    \end{align}
    over precisely the same range $\hatZ \in [0,1-\Zt]$; here, $m$ and $n$ are as defined in \eref{eqn:F_bigE_constant_beta_gen}.

    \medskip
    
    \item \textbf{(C)OM:} Similar to the constant-anisotropy model, the Method-I-truncated density profile of the OM or COM anisotropy model can be straightforwardly computed as
    \begin{align}\label{eqn:COM_DF_Density}
    	\hatvarrho(\hatP; \hat{\bigQ}_\pm) =
    	\begin{cases}
    		\hatvarrho_\text{aux}(\hatP; \hat{\bigQ}_\pm)\,{\big[1\mp(R/\Ra)^2\big]^{-1}}
    		= {\dfrac{4\pi}{1\mp(R/\Ra)^2}} \int_0^{\hatP} \rmd \hat{\bigQ}_\pm \, \hatF_{\bigQ_\pm}(\hat{\bigQ}_\pm)  \, \sqrt{2(\hatP-\hat{\bigQ}_\pm)}
 & \text{(OM)},\\
    		\hatvarrho_\text{aux}(\hatP; \hat{\bigQ}_\pm)\,{R^{-2\betac}\big[1\mp(R/\Ra)^2\big]^{-(1-\betac)}}
    		=\dfrac{2\pi^{3/2}\,\Gamma(1-\betac)}{{R^{2\betac}[1\mp(R/\Ra)^2]^{1-\betac}\,}\Gamma({3/2}-\betac)} \int_0^{\hatP} \rmd\hat{\bigQ}_\pm \frac{\hatF_{\bigQ_\pm}(\hat{\bigQ}_\pm)}{[2(\hatP - \hat{\bigQ}_\pm)]^{\betac - 1/2}}
 & \text{(COM)},
    	\end{cases}
    \end{align}
    where $\hat{\bigQ}_\pm \in [0,1-\bigKt]$, and the radius is evaluated at $R=R(\hatP)$, in the same functional form as \eref{eqn:OM_DF_1} for OM and \eref{eqn:COM_aux_gen} for COM.

    Although this re-parameterisation approach is fully self-consistent in phase space, it does suffer from a few additional caveats when applied to (C)OM models. First, Type~I models may fail to yield a $\hatvarrho$ that truncates at a finite radius with a converged total mass for sufficiently small $\bigKt$, due to the non-local nature of highly radial orbits. Furthermore, prior to any truncation, these DFs can exhibit an instability-associated inflection point followed by a divergence in the low-$\bigQ_{-}$ tail, $F(\bigQ_{-}\rightarrow0)\rightarrow+\infty$. To ensure that the post-truncation DFs are everywhere non-negative and monotonic in $\bigQ_{-}$ for dynamical stability, Method~I truncation is restricted to values $\bigKt \ge \bigQ_{-}^\text{min}$, where $\bigQ_{-}^\text{min}$ is the location of the global minimum of $F(\bigQ_{-})$. An example of such a minimum is evident in the lower panel of the second column of \fref{fig:DF_Trun_DFs}. Note that the value of $\bigQ_{-}^\text{min}$ can be quite substantial; for example, $\bigQ_{-}^\text{min} = 0.278$ for an untruncated NFW density profile with an OM DF with $\Ra=1$.

    Separately, all DFs of Type~II (C)OM models exhibit an \textit{essential discontinuity} in velocity space at $R=\Ra$. Namely, as $R$ approaches $\Ra$ from below, $\betasigma(R)\rightarrow -\infty$, and the solutions are unphysical for $R\geq\Ra$ \citep{Merritt1985AJ901027M}. Examples of this behaviour are shown in the third and fifth columns of \fref{fig:DF_Trun_varrho}. These DFs, therefore, require sufficiently large $\bigKt$ to excise the unphysical portion of phase space and have the truncation fall inside the physical regime, which in practice yields a truncation radius $\Rt<\Ra$. The discontinuity-gated truncation lower bound is $\bigKt > P(\Ra)$, the relative potential at the anisotropy radius, which keeps the truncated DF within the range where the untruncated model is defined, $\bigQ_{+} > P(\Ra)$. This lower bound, which increases for smaller $\Ra$ and depends on the input density profile $\varrho(R)$, can be quite substantial (e.g., $P(\Ra) = 0.693$ for the NFW profile with $\Ra = 1$, and $P(\Ra) = 0.240$ for $\Ra = 10$ as shown in \fref{fig:DF_Trun_DFs}). For both Type~I and Type~II models, the highly constrained range of $\bigKt$ values viable within the Method~I framework can leave the maximum truncated total mass well below $\Mz$. \PIANISTpy enforces the physical lower limit on $\bigKt$ from the untruncated DF itself: it locates numerically any zero crossing $\bigQ^{0}$ and any global minimum $\bigQ^\text{min}$ and requires $\bigKt \ge \max(\bigQ^\text{min}, \bigQ^{0})$; for Type~II models, whose DF is defined on $\bigQ_{+} > P(\Ra)$ only, the two features are located there and the limit is $\bigKt \ge \max(\bigQ^\text{min}, \bigQ^{0}, P(\Ra))$.

    To circumvent this inherent restriction, the alternative Method~II allows the construction of initial conditions for these anisotropy models from a given input $\hatvarrho(R)$. In this approach, \PIANISTpy constructs the truncated (C)OM DFs directly from an already Method-I-truncated $\hatvarrho$--$\hatP$ pair at truncation value $\bigKt$, i.e.,
    \begin{align}\label{eqn:hatF_bigE_COM}
    	\hatF_{\bigQ_{\pm}}(\hat{\bigQ}_{\pm}) =
    	\begin{cases}
    		\dfrac{1}{\sqrt{8}\pi^2}\bigg[\int_0^{\hat{\bigQ}_\pm}\dfrac{\rmd\hatP}{\sqrt{\hat{\bigQ}_\pm-\hatP}}\dfrac{\rmd^2\hatvarrho_{\text{aux}}}{\rmd\hatP^2}
    		+ \dfrac{1}{\sqrt{\hat{\bigQ}_\pm}}\Big(\dfrac{\rmd\hatvarrho_{\text{aux}}}{\rmd\hatP}\Big)\bigg|_{\hatP=0}\bigg] & \qquad\qquad\text{(OM)},\\
    		{\dfrac{2^\betac}{(2\pi)^{3/2}\,\Gamma(1-n)\,\Gamma(1-\betac)}\,\dfrac{\rmd}{\rmd\hat{\bigQ}_\pm}\int_0^{\hat{\bigQ}_\pm} \rmd\hatP \, \dfrac{\rmd^{m}\hatvarrho_{\text{aux}}/\rmd\hatP^{m}}{(\hat{\bigQ}_\pm-\hatP)^n}} & \qquad\qquad\text{(COM)},
    	\end{cases}
    \end{align}
    over the same numerical range $\hat{\bigQ}_\pm \in [0,1-\bigKt]$; here, $m$ and $n$ are as defined in \eref{eqn:F_bigE_constant_beta_gen}. Note, however, that DFs (and thus particle initial conditions) constructed in this fashion are not guaranteed to be everywhere physical and monotonic (and hence dynamically stable), unlike the case with Method~I.
\end{itemize}

For every DF of the form of \eref{eqn:DF_Functional_form_MT}, which covers all anisotropy models implemented in \PIANISTpyNS, Method~II is additionally subject to one existence condition that does not apply to Method~I. Consider an already truncated input density that vanishes at the truncation radius as $\hatvarrho \propto \hatP^{\,\lambda_\rmt}$. A Method-II-inverted DF of this form exists and remains positive at the truncation boundary $\hat{\bigK}\to 0^+$, if and only if $\betac > 1/2 - \lambda_\rmt$. This follows from the integral expressions of $\hatvarrho_{\rm aux}$ in terms of $\hatF$, which share the same kernel $(\hatP-\hat{\bigK})^{1/2-\betac}$, as in \eref{eqn:COM_DF_Density}. As such, a DF that scales as $\hatF\propto\hat{\bigK}^{\,p}$ near the truncation boundary then gives $\hatvarrho_{\rm aux} \propto \int_0^{\hatP} \rmd\hat{\bigK}\, \hat{\bigK}^{\,p}\, (\hatP-\hat{\bigK})^{1/2-\betac} \propto\hatP^{\,p+3/2-\betac}$ with a positive constant of proportionality, where the integral of $\hat{\bigK}^{\,p}$ converges if and only if $p > -1$. Because $\hatvarrho_{\rm aux}$ differs from $\hatvarrho$ only by a factor that is positive and finite at $\Rt$, the very requirement that the Method-II-inverted DF reproduce the input density fixes the asymptotic power-law slope $p$ of $\hatF$ by the condition $\lambda_\rmt = p + 3/2 - \betac$. Combined with $p > -1$, this gives the condition $\betac > 1/2 - \lambda_\rmt$ stated above; it is one-sided because integrability bounds $p$ only from below, so that an input density may vanish arbitrarily fast but not more slowly than $\hatP^{\,1/2-\betac}$. The uniqueness of the Abel inversion mapping from $\hatvarrho_{\rm aux}$ to $\hatF$, \eref{eqn:hatF_bigE_COM}, also implies that there exists no integrable DF of this form that can reproduce an input with $\lambda_\rmt \leq 1/2 - \betac$. This analytical constraint is an algebraic property for DFs of the form \eref{eqn:DF_Functional_form_MT}, where $\betac$ enters the `global' kernel $(\hatP-\hat{\bigK})^{1/2-\betac}$ at all radii via the exponent of the angular momentum factor $\bigL^{-2\betac}$; the condition therefore does not carry universal implication for DFs outside this functional form. In practice, for any Method-I-truncated input with a central anisotropy $\beta_\text{c}^\text{in}$, its truncated DF vanishes linearly in $\hat{\bigK}$ ($p = 1$) whenever the untruncated DF has a positive slope at the truncation value, $(\rmd F_\bigK/\rmd\bigK)|_{\bigKt} > 0$ (cf. Eq.~[\ref{eqn:DF_Truncation}]), so that $\lambda_\rmt = 5/2 - \beta_\text{c}^\text{in} \;\Rightarrow\;\betac - \beta_\text{c}^\text{in} > -2$, independent of the input density profile. When $\bigKt$ is instead placed at a local minimum of the untruncated DF, the truncated DF has zero slope there and vanishes as $\hat{\bigK}^{2}$ ($p = 2$), so that $\lambda_\rmt = 7/2 - \beta_\text{c}^\text{in}$, and the condition relaxes to $\betac - \beta_\text{c}^\text{in} > -3$. For the isotropic input adopted throughout this work, Method~II therefore requires $\betac > -2$ in general, and the resulting $\hatF \propto \hat{\bigK}^{\,1+\betac}$ diverges to $+\infty$ as $\hat{\bigK}\to 0^+$ for $-2 < \betac < -1$. This condition concerns the truncation boundary alone; non-negativity of the DF over the full range of $\hat{\bigK}$ remains model dependent and is verified numerically in \PIANISTpyNS.

\subsubsection{$\hatP \to P_0$ asymptotics of (un)truncated DFs}
\label{app:DF_inner_asymp}

A key mathematical property of the phase-space re-parameterisation of \eref{eqn:DF_Truncation} is that the inner density slope of all such self-consistently truncated systems, $\hatvarrho(R\to0)\propto R^{-\gamma}$, remains identical to that of the untruncated input $\varrho(R\to0)\propto R^{-\gamma}$. We can exploit this property to derive analytical expressions for the asymptotic DF slope as $\hat\bigK \to P_0$ ($P_0 = 1$ for the untruncated DF and $P_0 = 1 - \bigKt$ for the truncated DF) in terms of the inner density power-law exponent $\gamma$; these results apply universally to both truncated and untruncated cases without requiring separate treatment.

In the formal limit $R \to 0$, all analytical anisotropy models considered in \aref{app:DF_Prelim} reduce to constant anisotropy $\betac$. Given this and the inner potential expansion of \eref{eqn:inner_potential_expansion}, we substitute $R \propto (P_0 - \hatP)^{1/(2-\gamma)}$ into the auxiliary density, yielding $\hatvarrho_\mathrm{aux} \propto (P_0 - \hatP)^{(2\betac-\gamma)/(2-\gamma)}$. The Cuddeford inversion operator of \eref{eqn:F_bigE_constant_beta_gen}, combining an $m$-th ordinary derivative of $\varrho_\mathrm{aux}$ with an Abel convolution kernel $(\hatP - \hat\bigK)^{-n}$, reads as a Riemann--Liouville fractional derivative of total order $m+n = 3/2 - \betac$ \citep{Cuddeford1991MNRAS253414C}. Applying the Riemann--Liouville power rule $D^{m+n}(P_0-\hatP)^{\mu} \propto (P_0-\hatP)^{\mu-(m+n)}$ then yields the central scaling
\begin{equation}\label{eqn:df_asymp_final}
    \hatF_\bigK(\hat\bigK) \propto 
    (P_0 - \hat\bigK)^{\kappa}, \qquad 
    \kappa \eee \frac{2\betac - \gamma}{2-\gamma} - (m+n) 
    = \frac{\betac(4-\gamma) + \frac{1}{2}\gamma - 3}{2-\gamma},
\end{equation}
which satisfies $\kappa < 0$ whenever $\gamma > 2\betac$, i.e., precisely the regime in which the DF diverges as $\hat\bigK \to P_0$.

The boundary case $\gamma = 2\betac$, which corresponds to the upper limit permitted by the Cusp Slope--Central Anisotropy theorem (\citealp{An2006ApJ642752A}; see also \S\ref{ssec:DF_Physicalness}), requires a separate treatment since the leading-order exponent in $\varrho_\mathrm{aux}$ vanishes as $\hatP \to P_0$. Expanding via \eref{eqn:inner_density_expansion} and \eref{eqn:inner_potential_expansion} up to next-to-leading order gives $\hatvarrho_\mathrm{aux}(\hatP) \simeq C_1 + C_2(P_0 - \hatP)^{\alpha/(2-\gamma)}$. The Cuddeford inversion operator annihilates $C_1$ identically (for $\betac \leq 1/2$, or equivalently $m \geq 1$), so the DF behaviour as $\hat\bigK \to P_0$ is governed entirely by the next-to-leading term, yielding
\begin{equation}\label{eqn:df_asymp_critical}
    \hatF_\bigK(\hat\bigK) \propto 
    (P_0 - \hat\bigK)^{\kappa'}, \qquad 
    \kappa' \eee \frac{\alpha}{2-\gamma} - (m+n) 
    = \frac{\alpha}{2-\gamma} - 
    \left(\frac{3}{2} - \betac\right).
\end{equation}
In the general case of a non-integer exponent $\lambda_\rmc \eee \alpha/(2-\gamma)$, for which $\kappa' = \lambda_\rmc - (m+n)$, the central DF diverges as $(P_0-\hat\bigK)^{\kappa'}$ for $\kappa'<0$, diverges logarithmically for $\kappa'=0$, and tends to a finite value for $\kappa'>0$. The divergence occurring whenever $\lambda_\rmc \leq m+n$ can be positive (when $\lambda_\rmc<1$) or negative (when $\lambda_\rmc>1$, giving an unphysical central DF); the integer $\lambda_\rmc = 1$ is the special case discussed below. For (C)OM models, the expansion of $\hatvarrho_\mathrm{aux}$ above misses one additional term at next-to-leading order, $C_3\,(P_0-\hatP)^{2/(2-\gamma)}$, as incurred from the anisotropy factor $(1\pm R^2/\Ra^2)^{1-\betac}$ of $\hatvarrho_\mathrm{aux}$. The role of $\lambda_\rmc$ above is then replaced by the smaller non-integer exponent between $\alpha/(2-\gamma)$ and $2/(2-\gamma)$; if both are integers, the central DF is finite. In the special case of a positive integer $\lambda_\rmc$, the next-to-leading term $C_2\,(P_0-\hatP)^{\lambda_\rmc}$ reduces to an ordinary polynomial in $\hatP$ and is hence smooth at $\hatP = P_0$; the central DF is then always finite, irrespective of $\kappa'$. Consistently, the amplitude of the divergent term in the general case vanishes as $\lambda_\rmc$ approaches an integer. For example, an NFW profile ($\alpha = \gamma = 1$) at $\betac = 1/2$ has $\lambda_\rmc = 1$ and a finite central DF, as shown in the leftmost columns of Figs.~\ref{fig:DF_Trun_DFs}, \ref{fig:DF_Trun_varrho}, and \ref{fig:DF_Trun_sigma}.

\subsection{Velocity distributions and dispersion profiles}\label{app:Velocity_prof}

Provided a truncated $\hatvarrho$--$\hatP$ pair and the corresponding truncated DF $\hat F$, we derive the corresponding velocity distributions, both in 3D space and in projection.

\subsubsection{3D velocity quantities}\label{app:Velocity_prof_3D}

At fixed radius $R$, we rewrite the density integral of \eref{eqn:rho_f_anisotropic} to separate the dependence on the velocity angle $\varphi$ defined in \eref{eqn:velocity_angle} and the 3D velocity magnitude $V$
\begin{align}\label{eqn:rho_f_anisotropic_scale}
	\hatvarrho(R) = 2\pi\int_0^\pi \rmd\varphi \; p_\varphi(\varphi|R)
\int_0^{V_\text{esc}(R)} \rmd V \; p_V(V\,|\,R,\varphi),
\end{align}
where $V_\text{esc}(R)\eee\sqrt{2\hatP(R)}$ denotes the local escape velocity. At fixed $R$, $p_\varphi(\varphi|R)$ gives an unnormalised probability density in $\varphi$, and at fixed $(R,\varphi)$, $p_V(V|R,\varphi)$ denotes the corresponding unnormalised conditional density in $V$. The corresponding analytical expressions for each anisotropy model can be worked out directly from \pCAeref{eqn:const_beta_DF}\pCBeref{eqn:OM_DF_1}\pCBeref{eqn:OM_TypeII_DF_1}\pCCeref{eqn:COM_aux_gen} to yield
\begin{align}\label{eqn:varphi_prob}
        p_\varphi(\varphi|R) \propto 
	\begin{cases}
		\sin^{1-2\betac}\varphi & \:\,\text{(constant-anisotropy)},\\
		\sin\varphi\big[1+(R/\Ra)^2\sin^2\varphi\big]^{-3/2} & \:\,\text{(OM Type I)},\\
		\sin\varphi\big[1-(R/\Ra)^2\sin^2\varphi\big]^{-3/2} & \:\,\text{(OM Type II)},\\
		\sin^{1-2\betac}\varphi \big[1 + (R/\Ra)^2\sin^2\varphi\big]^{-(3/2-\betac)} & \:\,\text{(COM Type I)},\\
		\sin^{1-2\betac}\varphi \big[1 - (R/\Ra)^2\sin^2\varphi\big]^{-(3/2-\betac)} & \:\,\text{(COM Type II)},
	\end{cases}
\end{align}
and
\begin{align}\label{eqn:V_prob}
        p_V(V|R, \varphi) \propto 
		\begin{cases}
			V^{2-2\betac}_\text{aux} \, \hatF_Z(\hatP(R)-V_\text{aux}^2/2) & \;\;\;\;\;\;\;\,\,\text{(constant-anisotropy)},\\
			V_\text{aux}^2 \, \hatF(\hatP(R)-V_\text{aux}^2/2) & \;\;\;\;\;\;\;\,\,\text{(OM)},\\
			V^{2-2\betac}_\text{aux} \, \hatF_{\bigQ_{\pm}}(\hatP(R)-V_\text{aux}^2/2) & \;\;\;\;\;\;\;\,\,\text{(COM)},
		\end{cases}
\end{align}
where the auxiliary velocity $V_\text{aux}\equiv V_\text{aux}(V;R,\varphi)$ is defined as
\begin{align}\label{eqn:Vaux_def}
        V_\text{aux}(V; R, \varphi) \eee 
		\begin{cases}
		V & \;\;\;\;\;\;\;\;\;\;\;\:\:\:\:\text{(constant-anisotropy)},\\
		V\sqrt{1+(R/\Ra)^2\sin^2\varphi} & \;\;\;\;\;\;\;\;\;\;\;\:\:\:\:\text{((C)OM Type~I)},\\ 
		V\sqrt{1-(R/\Ra)^2\sin^2\varphi} & \;\;\;\;\;\;\;\;\;\;\;\:\:\:\:\text{((C)OM Type~II)}, 
		\end{cases}
\end{align}
which is used as notational shorthand. Thus, the local velocity distribution at radius $R$ is specified by the angular factor $p_\varphi(\varphi|R)$ together with the conditional speed density $p_V(V\,|\,R,\varphi)$. The latter is integrated over $V\in[0,V_\text{esc}(R)]$ at fixed $(R,\varphi)$ to yield the conditional velocity CDF used for inverse-CDF sampling of $V$ when generating particle initial conditions, after $\varphi$ has already been sampled from $p_\varphi(\varphi|R)$.

Separately, we can also compute the velocity-dispersion profiles directly from the definitions of the velocity moments \citep{Dejonghe1986PhR133217D}
\begin{subequations}\label{eqn:linearity_1}
\begin{align}
		\hatvarrho_{\text{aug}}(\hatP,R)
		&= 4\pi \int^{\hatP}_0 \rmd\hat{\bigK} \int^{\sqrt{2(\hatP-\hat{\bigK})}}_{0} \rmd \Vt \,\Vt \,
		\frac{\hatF(\hat{\bigK}, R\,\Vt)}{\sqrt{2(\hatP-\hat{\bigK})-\Vt^2}},\label{eqn:linearity_1a}\\
		\sigmar^2(\hatP, R)\,\hatvarrho_{\text{aug}}(\hatP,R)
		&= 4\pi \int^{\hatP}_0 \rmd\hat{\bigK} \int^{\sqrt{2(\hatP-\hat{\bigK})}}_{0} \rmd \Vt \,\Vt \,
		\hatF(\hat{\bigK}, R\,\Vt)\sqrt{2(\hatP-\hat{\bigK})-\Vt^2},\label{eqn:linearity_1b}\\
		\sigmat^2(\hatP, R)\,\hatvarrho_{\text{aug}}(\hatP,R)
		&= 4\pi \int^{\hatP}_0 \rmd\hat{\bigK} \int^{\sqrt{2(\hatP-\hat{\bigK})}}_{0} \rmd \Vt \,\Vt^3 \,
		\frac{\hatF(\hat{\bigK}, R\,\Vt)}{\sqrt{2(\hatP-\hat{\bigK})-\Vt^2}}.\label{eqn:linearity_1c}
\end{align}
\end{subequations}
With the change of variable $\bigL = R\,\Vt$ such that $\rmd\Vt=\rmd\bigL/R$ at fixed $(\hatP,R)$, the velocity moments can be rewritten in a more computationally convenient form as follows
\begin{align}\label{eqn:linearity_2}
	\begin{aligned}
        \sigmar^2(\hatP,R)
		&= \frac{1}{\hatvarrho_{\text{aug}}(\hatP,R)}
		\int^{\hatP}_0 \rmd\hat{\bigK}\,\hatvarrho_{\text{aug}}(\hat{\bigK}, R),\\
		\sigmat^2(\hatP,R)
		&= \frac{2}{\hatvarrho_{\text{aug}}(\hatP,R)}
		\int^{\hatP}_0 \rmd\hat{\bigK}\,
		\frac{\partial}{\partial R^2}\!\Big[R^2\hatvarrho_{\text{aug}}(\hat{\bigK}, R)\Big].
	\end{aligned}
\end{align}
Here $\partial/\partial R^2$ denotes single differentiation with respect to the variable $R^2$, i.e., $\partial/\partial R^2 \equiv (2R)^{-1}\partial/\partial R$.

For the isotropic case, where $\partial\hatvarrho_{\text{aug}}(\hatP,R)/\partial R^2 = \partial\hatvarrho_{\text{aug}}(\hatP)/\partial R^2 = 0$, we recover $\sigmar^2 = \sigmat^2/2$ and $\beta_\sigma(R)\equiv 1 - \sigmat^2/(2\sigmar^2) = 0$, as required. Conversely, it is precisely the non-vanishing explicit $R$-dependence, $\partial\hatvarrho_{\text{aug}}(\hatP,R)/\partial R^2 \neq 0$, that gives rise to non-zero velocity anisotropy. For the analytical anisotropic models considered in \aref{app:DF_Prelim}, the adopted (partially separable) DF of the form \eref{eqn:DF_Functional_form} implies a separable augmented density, $\hatvarrho_{\text{aug}}(\hatP,R)=\hatvarrho_{\text{aux}}(\hatP)\,g(R)$. Accordingly, we define the following \textit{separability factors}
\begin{align}\label{eqn:linearity_beta_1}
    g(R) \equiv \frac{\hat\varrho_{\rm aug}(\hat P,R)}{\hat\varrho_{\rm aux}(\hat P)}\bigg|_{\hat P=\hat P(R)}=
    \begin{cases}
        1 & \text{(isotropic)},\\
        R^{-2\beta_c} & \text{(constant-anisotropy)},\\
        \left(1+\dfrac{R^2}{\Ra^2}\right)^{-1} & \text{(OM Type I; }Q_- \text{)},\\
        \left(1-\dfrac{R^2}{\Ra^2}\right)^{-1} & \text{(OM Type II; }Q_+ \text{)},\\
        R^{-2\beta_c}\left(1+\dfrac{R^2}{\Ra^2}\right)^{-(1-\beta_c)} & \text{(COM Type I; }Q_- \text{)},\\
        R^{-2\beta_c}\left(1-\dfrac{R^2}{\Ra^2}\right)^{-(1-\beta_c)} & \text{(COM Type II; }Q_+ \text{)},
    \end{cases}
\end{align}
i.e., the explicit $R$-dependent factor that satisfies $\hat\varrho_{\rm aug}(\hat P,R)=g(R)\times \hat\varrho_{\rm aux}(\hat P)$ when evaluated under the condition $\hatP=\hatP(R)$ at the physical radius $R$. With this separation, \eref{eqn:linearity_2} can be simplified to write
\begin{align}\label{eqn:linearity_beta_2}
	\beta_\sigma(R) \equiv 1 -\frac{\sigmat^2}{2\sigmar^2}
	&= 1-\bigg[1+\frac{R^2}{g(R)}\frac{\partial g(R)}{\partial R ^2} \bigg]
	= -\frac{R^2}{g(R)}\frac{\partial g(R)}{\partial R^2}
	= -R^2\,\frac{\partial}{\partial R^2}\ln g(R),
\end{align}
where we have used that $\hatvarrho_{\text{aux}}(\hatP)$ is independent of $R$. A straightforward substitution of \eref{eqn:linearity_beta_1} into \eref{eqn:linearity_beta_2} then recovers the corresponding velocity anisotropy profile of each model. This formalism provides an advantageous route to analytically express, and straightforwardly evaluate numerically, the anisotropy profile via $g(R)$ for linear superpositions of truncated DFs, as discussed further in \aref{app:DF_superposition}.

\subsubsection{Projected velocity quantities}\label{app:Velocity_prof_proj}

To connect the intrinsic velocity moments derived above to observable quantities, we now consider the line-of-sight projected velocity dispersion. For the calculation of projected observables here and in \aref{app:kurtosis}, all quantities entering the projection are evaluated at the physical radius $R$ via $\hatP=\hatP(R)$. For a system with finite truncation radius $\Rt$ and a constant mass-to-light ratio, the line-of-sight (LOS) velocity dispersion at a projected distance $R_\text{2D} \eee r_\text{2D}/\rz$ from the centre is given by
\begin{equation}\label{eqn:los_sigma_1}
	\sigma_\text{los}^2(R_\text{2D}) = \frac{2}{\Sigma(R_\text{2D})}\Bigg[\int_{R_\text{2D}}^{\Rt}\frac{\hatvarrho(R)\sigmar^2(R)R}{(R^2-R_\text{2D}^2)^{1/2}}\rmd R 
        - R_\text{2D}^2\int_{R_\text{2D}}^{\Rt} \frac{\beta_\sigma(R)\hatvarrho(R)\sigmar^2(R)}{R(R^2-R_\text{2D}^2)^{1/2}}\rmd R\Bigg]\,,
\end{equation}
\citep{Binney1982MNRAS200361B}, where
\begin{equation}
\Sigma(R_\text{2D}) = 2 \int_{R_\text{2D}}^{\Rt} \frac{\hatvarrho(R) R}{(R^2-R_\text{2D}^2)^{1/2}}\rmd R\,,
\end{equation}
is the projected surface mass density under our phase-space unit-normalisation scheme. This expression for the LOS velocity dispersion can be directly compared with measurements from thin-slit observations. Although written here for a constant mass-to-light ratio, \eref{eqn:los_sigma_1} holds for any constant weighting (mass, luminosity, or tracer number density), provided the velocity dispersions are weighted consistently with the corresponding density.

Beyond local (radially resolved) kinematics, many surveys report aperture-integrated quantities for different aperture geometries. In general, observational apertures need not be centred on the system's projected CM. Here we present analytical expressions for the $\Sigma$-weighted, aperture-averaged LOS velocity dispersion squared, $\overline{\sigma}_\text{los}^2$, for rectangular and circular apertures. Throughout, we use coordinates centred on the aperture and evaluate the projected $\Sigma$ and $\sigma_\text{los}^2$ at the projected distance to the system's projected CM. We define $R_\text{off}$ as the projected separation between the aperture centre and the system's projected CM, and $R_\Delta$ as the projected distance from a point within the aperture to the system's projected CM.

For thin-slit observations with a finite slit width, we consider the rectangular aperture in Fig.~\ref{fig:Projection_Ops}~(b). In that schematic, $x$ denotes the full length of the slit (along the slit direction) and $y$ denotes its full thickness (across the slit). Introducing slit-aligned Cartesian coordinates $(\mathcal{X},\mathcal{Y})$ centred on the aperture, with $\mathcal{X}$ along the slit and $\mathcal{Y}$ across the dispersion direction, the $\Sigma$-weighted rectangular-aperture average is
\begin{align}\label{eqn:los_sigma_rect}
\overline{\sigma}_\text{los}^2(x,y;R_\text{off},\theta)
=\frac{\int_{-x/2}^{x/2}\int_{-y/2}^{y/2}
\Sigma(R_\Delta)\,\sigma_\text{los}^2(R_\Delta)\,\rmd \mathcal{Y}\,\rmd \mathcal{X}}
{\int_{-x/2}^{x/2}\int_{-y/2}^{y/2}
\Sigma(R_\Delta)\,\rmd \mathcal{Y}\,\rmd \mathcal{X}},
\end{align}
where $\theta\in[0,\pi/2]$ is the angle between the slit direction and the projected offset direction, and
\begin{align}\label{eqn:RDelta_rect}
R_\Delta(\mathcal{X},\mathcal{Y};R_\text{off},\theta) \eee
\sqrt{\mathcal{X}^2+\mathcal{Y}^2+R_\text{off}^2
-2R_\text{off}\left(\mathcal{X}\cos\theta+\mathcal{Y}\sin\theta\right)}.
\end{align}
The centred case corresponds to $R_\text{off}=0$, for which $R_\Delta=\sqrt{\mathcal{X}^2+\mathcal{Y}^2}$.

Fibre-fed spectroscopic observations, such as the Sloan Digital Sky Survey \citep[SDSS,][]{Blanton2003AJ1252276B}, typically use circular apertures such as the one depicted in \fref{fig:Projection_Ops}~(c). The corresponding aperture-averaged LOS velocity dispersion, for an aperture with radius $R_\text{ap} \eee r_\text{ap}/\rz$ centred on the projected CM, is
\begin{align}\label{eqn:los_sigma_2}
\overline{\sigma}_\text{los}^2(R_\text{ap})
=\frac{\int_0^{R_\text{ap}} \Sigma(R_\text{2D})\,\sigma_\text{los}^2(R_\text{2D})\, R_\text{2D}\,\rmd R_\text{2D}}
{\int_0^{R_\text{ap}} \Sigma(R_\text{2D})\, R_\text{2D}\,\rmd R_\text{2D}}.
\end{align}
For an aperture of the same size centred instead at a projected offset $R_\text{off}$ from the system's projected CM, we have
\begin{align}\label{eqn:los_sigma_3}
\overline{\sigma}_\text{los}^2(R_\text{ap}; R_\text{off})
=\frac{\int_0^{2\pi}\!\!\int_0^{R_\text{ap}}
\Sigma(R_\Delta)\,\sigma_\text{los}^2(R_\Delta)\,R_\text{2D}\,\rmd R_\text{2D}\,\rmd\psi}
{\int_0^{2\pi}\!\!\int_0^{R_\text{ap}}
\Sigma(R_\Delta)\,R_\text{2D}\,\rmd R_\text{2D}\,\rmd\psi},
\end{align}
where $(R_\text{2D},\psi)$ are polar coordinates centred on the aperture, with $\psi$ measured from the projected offset direction. In this geometry, $R_\Delta$ reads
\begin{align}\label{eqn:los_Rdelta}
R_\Delta \eee \sqrt{R_\text{off}^2 + R_\text{2D}^2 - 2R_\text{off}R_\text{2D}\cos\psi}.
\end{align}
We note that, for the isotropic, constant-anisotropy and Osipkov--Merritt families, the line-of-sight and aperture velocity dispersions can alternatively be written as single radial integrals with closed-form kernels as shown in \citet{2005MNRAS.362...95M, 2005MNRAS.363..705M, 2006MNRAS.370.1582M}, against which we have validated \PIANISTpyNS.

The line-of-sight velocity distribution (LOSVD) is given by
\begin{align}\label{eqn:losvd_1}
        D(R_\text{2D},V_\text{los}) = \frac{2}{\Sigma(R_\text{2D})}\int_{R_\text{2D}}^{\Rt}  \frac{R \, \rmd R}{(R^2-R_\text{2D}^2)^{1/2}} \iint \rmd V_{\perp1} \, \rmd V_{\perp2} \, \hatF(\hatZ, \bigL),
\end{align}
\citep[e.g.,][]{Merritt1993ApJ40975M}, where $V_{\perp1}$ and $V_{\perp2}$ denote the two orthogonal (plane-of-sky) velocity components perpendicular to the line-of-sight direction; the double integral is over the region of $(V_{\perp1},V_{\perp2})$ for which the truncated DF is positive. This expression can be further simplified by considering the following change of variables $V_{\perp1} \eee w \sin\eta$ and $V_{\perp2} \eee w \cos\eta$ \citep{Mahdavi2004ApJ607202M}, where $w$ is the magnitude of the plane-of-sky velocity, so that
\begin{align}\label{eqn:losvd_2}
    D(R_\text{2D},V_\text{los})=\frac{2}{\Sigma(R_\text{2D})}\int_{R_\text{2D}}^{\min\left(\Rt,\,R_\text{max}\right)}  \frac{R\,\rmd R}{(R^2-R_\text{2D}^2)^{1/2}} \int_0^{\sqrt{2\hatP(R)-V_\text{los}^2}} \rmd w\,w\, \int_0^{2\pi} \hatF(\hatZ,\bigL) \, \rmd\eta\,.
\end{align}
Here,
\begin{align}
\hatZ = \hatP(R)-\frac{w^2+V_\text{los}^2}{2}\,,
\end{align}
\begin{align}
   \bigL = R\, w\, \bigg\{{\cos^2\eta + \Big[\sin\eta\frac{(R^2-R_\text{2D}^2)^{1/2}}{R} - \frac{V_\text{los}}{w}\frac{R_\text{2D}}{R}\Big]^2}\bigg\}^{1/2}\,,
\end{align}
and $R_\text{max}(V_\text{los})$ is the kinematic turning-point radius, i.e., the maximum 3D radius at which a bound particle can still carry a line-of-sight velocity $V_\text{los}$ (i.e., with the local escape velocity satisfying $V_\text{esc}(R) \geq |V_\text{los}|$), defined implicitly by
\begin{align}
{\hatP(R_\text{max}) = V_\text{los}^2/2}.
\end{align}

These parametric expressions for the relative energy $\hatZ$ and the specific angular momentum $\bigL$ can be substituted into $\hatF(\hatZ,\bigL)$ for each class of analytical anisotropy models discussed in \aref{app:DF_Prelim}. The LOSVD averaged over a fixed aperture size $R_\text{ap}$ centred on the system's CM is given by
\begin{align}\label{eqn:losvd_4}
        \overline{D}(R_\text{ap},V_\text{los})=
        \frac{\int_0^{R_\text{ap}} \Sigma(R_\text{2D}) \, D(R_\text{2D},V_\text{los}) \, R_\text{2D} \, \rmd R_\text{2D}}
        {\int_0^{R_\text{ap}} \Sigma(R_\text{2D}) \, R_\text{2D} \, \rmd R_\text{2D}}.
\end{align}
Following \eref{eqn:los_sigma_3}, for an aperture of the same size centred instead at a distance $R_\text{off}$ away from the true projected CM of the system, we instead have
\begin{align}\label{eqn:losvd_5}
        \overline{D}(R_\text{ap},V_\text{los}; R_\text{off})
        =\frac{\int_0^{2\pi} \int_0^{R_\text{ap}} \Sigma(R_\Delta) \, D(R_\Delta,V_\text{los}) \, R_\text{2D} \, \rmd R_\text{2D} \, \rmd\psi}
        {\int_0^{2\pi} \int_0^{R_\text{ap}} \Sigma(R_\Delta) \, R_\text{2D} \, \rmd R_\text{2D} \,\rmd\psi},
\end{align}
where $R_\Delta$ is as defined in \eref{eqn:los_Rdelta}. For any LOSVD outputted by \PIANISTpy (e.g., $D$ or $\overline{D}$ at fixed geometric parameters), we enforce the unit normalisation
\begin{align}\label{eqn:losvd_nor}
\int_{-\infty}^{\infty} \mathrm{LOSVD}(V_\text{los})\,\rmd V_\text{los} \eee 1.
\end{align}

\subsection{Gauss--Hermite $h_n$ and standardised central $s_n$ moments}\label{app:kurtosis}

It is standard observational practice to quantify a given LOSVD through its velocity moments. Often, this is limited to the first and second order moments, typically characterised through the mean and dispersion of the best-fit Gaussian. However, it is also common to include higher-order moments that characterise the non-Gaussianity of the LOSVD, which can be done through either the Gauss--Hermite (GH) moments $h_n$ or the  standardised central moments $s_n$. \PIANISTpy can compute all these various moments for a given LOSVD, as described below. In \aref{app:kurtosis_basis} we first discuss the adopted normalisation conventions and the practical merits of both $h_n$ and $s_n$. \aref{app:kurtosis_imp} presents the general expressions for the higher-order moments that are applicable for arbitrary anisotropy models of spherical non-rotating systems.

\subsubsection{Definitions and normalisation convention}\label{app:kurtosis_basis}

The GH expansion provides a complete orthogonal basis for any square-integrable function on $\mathbb{R}$, independent of any assumption of Gaussianity. Adopting the Hermite polynomial normalisation convention of \citet{vanderMarel1993}\footnote{This convention has become standard in kinematic studies and is adopted, for example, in the widely used full-spectrum fitting code \textsc{pPXF} \citep{Cappellari2004,Cappellari2017}. Alternative choices are possible \citep[e.g.,][]{Gerhard1993}, which result in conversion coefficients with respect to the standardised central moments that differ from \eref{eqn:GH_exact_conversion}.}, where the GH polynomial of order $n\in\{0, 1, 2, ...\}$ and the associated orthogonality relation read
\begin{align}\label{eqn:Hermit_polys}
H_n(y) \eee \frac{\rme^{y^2}}{\sqrt{n!}} \bigg[ \bigg( -\frac{1}{\sqrt{2}} \frac{\text{d}}{\text{d}y} \bigg)^n \rme^{-y^2} \bigg] \qquad\Leftrightarrow\qquad     \int_{-\infty}^{\infty}
    H_m(y)\,H_n(y)\,\bigg(\frac{\rme^{-y^2/2}}{\sqrt{2\pi}}\bigg)^2\,\rmd y
    = \frac{1}{2\sqrt{\pi}}\delta_{mn},
\end{align}
where $\delta_{mn}$ denotes the Kronecker delta. Explicitly, $H_0 = 1$, $H_1 = \sqrt{2}\,y$, $H_2 = (2y^2-1)/\sqrt{2}$, $H_3 = (2y^3-3y)/\sqrt{3}$, and $H_4 = (4y^4-12y^2+3)/(2\sqrt{6})$, identical to eqs.~(A4)--(A6) of \citet{vanderMarel1993}; this normalisation is related to the physicists' convention by $H_n(y) = 2^{-n/2}(n!)^{-1/2}\,H_n^{\rm phys}(y)$. Any LOSVD $D(R_\mathrm{2D}, V_\mathrm{los})$ can  be represented as an infinite series of Hermite polynomials
\begin{equation}\label{eqn:B_GH_expansion}
    D(R_\mathrm{2D}, V_\mathrm{los})
    = \frac{\rme^{-y^2/2}}{\sqrt{2\pi}\,\sigma_*}
      \sum_{n=0}^{\infty} h_n(R_\mathrm{2D})\,H_n(y), \qquad y \equiv \frac{V_\mathrm{los} - V_*}{\sigma_*},
\end{equation}
where $V_*$ and $\sigma_*$ are reference parameters (not necessarily the true mean or dispersion of $D$, as discussed below). The corresponding GH moments $h_n$ are computed by projecting the LOSVD onto the orthogonal basis
\begin{equation}\label{eqn:B_hn_def}
    h_n(R_\mathrm{2D})
    =2\sqrt{\pi}
      \int_{-\infty}^{\infty}
      D(R_\mathrm{2D}, V_\mathrm{los})\,
      H_n(y)\,\frac{\rme^{-y^2/2}}{\sqrt{2\pi}}\,\rmd V_\mathrm{los}.
\end{equation}
For these GH moments, $h_0 = 1$ sets the overall normalisation; $h_1$ encodes the fractional shift of the mean velocity relative to the reference $V_*$; $h_2$ encodes the fractional difference between the true dispersion and the reference $\sigma_*$, with $h_2 = 0$ if and only if $\sigma_* = \sigma_\mathrm{los}$; $h_3$ measures odd (antisymmetric) deviations from a Gaussian shape, with $h_3 \neq 0$ signalling asymmetric LOSVDs due to, e.g., streaming motions in a disc or non-equilibrium infall. Finally, the fourth GH moment $h_4$ measures symmetric deviations from a Gaussian shape, with $h_4 > 0$ indicating a more centrally peaked, heavy-tailed LOSVD (leptokurtic) and $h_4 < 0$ indicating a flat-topped, light-tailed LOSVD (platykurtic) \citep[e.g.,][]{vanderMarel1993}. In practice, the GH expansion is truncated at $n = 4$ (retaining $h_0, h_1, h_2, h_3, h_4$ only), which is the standard in observational stellar kinematics \citep{vanderMarel1993, Gerhard1993, Bender1994}. For any LOSVD with finite variance, let $\mu(R_\mathrm{2D}) \equiv \langle V_\mathrm{los}\rangle$ denote the true mean and $\sigma_\mathrm{los}^2 \equiv \langle(V_\mathrm{los}-\mu)^2\rangle$ the true variance given by \eref{eqn:los_sigma_1}. Setting $V_* = \mu$ and $\sigma_* = \sigma_\mathrm{los}$ is equivalent to imposing $h_1 = h_2 = 0$, which fixes the reference Gaussian to be the best-fit Gaussian to $D$; this is the choice adopted in the forward-modelling pipeline of \PIANISTpyNS, and the normalisation $\int D\,\rmd V_\mathrm{los} = 1$ fixes $h_0 = 1$ exactly.

Separately, and independent of any basis expansion, for a given LOSVD one can always compute the associated \textit{standardised central moments} $s_n$ by direct integration over $D$, without any assumption about its shape. The skewness (third standardised central moment) is
\begin{equation}
    s_3(R_\mathrm{2D})
    \equiv
    \frac{\langle(V_\mathrm{los} - \mu)^3\rangle}
         {[\sigma_\mathrm{los}^2(R_\mathrm{2D})]^{3/2}} = \frac{\int_{-\infty}^\infty (V_\mathrm{los} - \mu)^3 \, D(R_\mathrm{2D}, V_\mathrm{los}) \, \rmd V_\mathrm{los}}
         {[\sigma_\mathrm{los}^2(R_\mathrm{2D})]^{3/2}},
    \label{eqn:B_skewness_def}
\end{equation}
and measures the asymmetry of the LOSVD about its mean. The \textit{kurtosis} (fourth standardised central moment) is
\begin{equation}
    s_4(R_\mathrm{2D})
    \equiv
    \frac{\langle(V_\mathrm{los} - \mu)^4\rangle}
         {[\sigma_\mathrm{los}^2(R_\mathrm{2D})]^2}
    {\;=\; \frac{\int_{-\infty}^\infty (V_\mathrm{los} - \mu)^4 \, D(R_\mathrm{2D}, V_\mathrm{los}) \, \rmd V_\mathrm{los}}
         {[\sigma_\mathrm{los}^2(R_\mathrm{2D})]^2}},
    \label{eqn:B_kurtosis_def}
\end{equation}
and measures the weight of the tails relative to a Gaussian. For a Gaussian distribution, the kurtosis is $s_4 = 3$; more generally, the Cauchy--Schwarz inequality $\langle V_\mathrm{los}^4\rangle \geq \langle V_\mathrm{los}^2\rangle^2$ implies a fundamental lower bound of $s_4 \geq 1$.

In observational inferences, the choice between characterising the LOSVD shape via GH moments (`local' to the central distribution as weighted by a Gaussian kernel) or standardised central moments (sensitive to the `global' distribution) has been driven by the respective pathological features:
\begin{itemize}
    \item \textbf{GH moments} are mutually orthogonal by construction (Eq.~[\ref{eqn:Hermit_polys}]), which reduces parameter covariance in $\chi^2$ minimisation fits to spectral line profiles. This allows $h_3$ and $h_4$ to be fitted nearly independently of $V_*$ and $\sigma_*$ to first order—an advantage that makes GH moments the standard in observational stellar kinematics \citep[e.g.,][]{vanderMarel1993,Gerhard1993} and the native output of \textsc{pPXF} \citep{Cappellari2004,Cappellari2017}. The practical choice of truncating the infinite series at $h_4$ can mathematically introduce unphysical non-positive values in the wings of the LOSVD \citep[see, e.g.,][Fig.~1]{vanderMarel1993}. Furthermore, the orthogonality is exact only when the true $V_*$ and $\sigma_*$ are known; in a fit where these parameters are simultaneously estimated with $h_3$ and $h_4$, a typical 10\%-level residual covariance is also present between $h_4$ and $\sigma_*$ \citep[e.g.,][]{Bender1994, Cappellari2004}.
    \item \textbf{Standardised central moments} are well-defined, real, and bounded below ($s_4 \geq 1$) for any $D$ with a finite fourth moment, which ensures $D \geq 0$ throughout the reconstruction. However, in observational data fitting, the $V^4$ weighting makes $s_4$ highly sensitive to noise in the wings of the LOSVD (e.g., due to low S/N per spectral pixel or finite stellar tracers), which introduces inherent biases and a skewed sampling distribution into the statistical estimator of $s_4$, necessitating careful bias correction \citep{Lokas2003}. Conversely, when the LOSVD remains well-populated into the far wings, the kurtosis and velocity anisotropy can be robustly measured, as demonstrated by \citet{Mitra2024} for satellite kinematics inferred from SDSS.
\end{itemize}

In the forward-modelling pipeline of \PIANISTpyNS, the full DF $\hat{F}\geq 0$ and LOSVD $D \propto \int\hat{F}\,\mathrm{d}^2V_\perp \geq 0$ are known exactly. As such, the choice of either representation becomes merely a matter of convention. \PIANISTpy evaluates $\langle V_\mathrm{los}^4\rangle$ and $\sigma_\mathrm{los}^2$ through direct 1D quadrature of smooth analytic integrands rather than iterative fitting, thereby eliminating parameter covariance and wing-noise sensitivity. Under the truncation $h_{n>4}=0$ and the identification $V_* = \mu$ and $\sigma_* = \sigma_{\text{los}}$, the conversions read, to first order in $h_3$ and $h_4$,
\begin{equation}\label{eqn:GH_exact_conversion}
{s_3 = 4\sqrt{3}\,h_3 \qquad \text{and} \qquad s_4 = 3 + 8\sqrt{6}\,h_4,}
\end{equation}
identical to eq.~(18) of \citet{vanderMarel1993} as commonly adopted in the literature \citep[e.g.,][]{Cappellari2017}.

\subsubsection{General projection formula for spherically symmetric non-rotating systems}
\label{app:kurtosis_imp}

For the equilibrium, spherically symmetric, non-rotating systems considered throughout this work, the DF $\hat{F}(\hat{Z},\bigL)$ is an even function of all velocity components at every spatial point; this follows because both $\hat{Z} = \hat{P}(R) - V^2/2$ and $\bigL^2 = R^2 \Vt^2$ are even in velocity. Consequently, $D(R_\mathrm{2D}, V_\mathrm{los})$ is even in $V_\mathrm{los}$, and all odd moments vanish identically: $\langle V_\mathrm{los}^{2k+1}\rangle(R_\mathrm{2D}) = 0$ for $k \in \{0, 1, 2, \dots\}$. Specifically, zero net rotation implies $\mu = \langle V_\mathrm{los}\rangle = 0$, and hence $V_* = \mu = 0$. It follows that $\langle V_\mathrm{los}^3\rangle = 0$, implying that $s_3 = h_3 = 0$ at all projected radii \citep[e.g., \S4.1.2 of][]{Binney&Tremaine2008}.

The only non-trivial shape parameter in the GH expansion truncated at $n=4$ is therefore $h_4(R_\mathrm{2D})$, which we evaluate numerically by substituting $H_4$ as defined in \eref{eqn:Hermit_polys} into \eref{eqn:B_hn_def}
\begin{equation}\label{eqn:h4_spherical}
    h_4(R_\mathrm{2D})
    =
      \int_{-\infty}^{\infty}
      D(R_\mathrm{2D}, V_\mathrm{los})\,
      \,\Big(\frac{4y^4-12y^2+3}{2\sqrt{3}}\Big) \, \rme^{-y^2/2}\,\rmd V_\mathrm{los}, \qquad y = V_\mathrm{los}/\sigma_\mathrm{los}.
\end{equation}
The kurtosis from \eref{eqn:B_kurtosis_def} now simplifies to
\begin{equation}\label{eqn:kappalos_spherical}
    s_4(R_\mathrm{2D})
    = \frac{\langle V_\mathrm{los}^4\rangle(R_\mathrm{2D})}{[\sigma_\mathrm{los}^2(R_\mathrm{2D})]^2} = \frac{\int_{-\infty}^{\infty}
    V_\mathrm{los}^4\,D(R_\mathrm{2D}, V_\mathrm{los})\,
    \rmd V_\mathrm{los}}{\Big[\int_{-\infty}^{\infty}
    V_\mathrm{los}^2\,D(R_\mathrm{2D},V_\mathrm{los})\,
    \rmd V_\mathrm{los}\Big]^2},
\end{equation}
where the second equality follows from $\sigma_\mathrm{los}^2 = \langle V_\mathrm{los}^2\rangle$. Physically, radially anisotropic systems ($\betasigma > 0$) generally produce $h_4 > 0$ (more centrally peaked, heavier tails than Gaussian), while tangentially anisotropic systems ($\betasigma < 0$) generally produce $h_4 < 0$ (flatter, lighter-tailed than Gaussian), as established for constant-anisotropy models in \citet{vanderMarel1993} \citep[see also][]{Merritt1987}. However, this simple correspondence between $h_4$ and $\betasigma$ does not hold in general for systems with radially varying $\betasigma(R)$ or off-centred apertures; see \S\ref{ssec:OP_LOSVD} for explicit examples and discussion. Also, a pure Gaussian LOSVD is generically inconsistent with an equilibrium solution of the collisionless Boltzmann equation for anisotropic systems \citep{Richardson2013}; hence, non-zero $h_4$ is physically expected and necessary. Indeed, even the LOSVDs of isotropic spherical systems, while typically very close to Gaussian, are not exactly Gaussian, not even locally (cf. \fref{fig:V_shell} and \fref{fig:LOSVD}).

\subsection{Anisotropy profiles of linearly superposed DFs}\label{app:DF_superposition}

Since the space of (integrable) DFs on phase space is a vector space and the DF is an additive density, a system composed of components with DFs $\hat F_i$ and weights $c_i$ is described by the linear superposition $\hat F=\sum_i c_i\hat F_i$ (with $\sum_i c_i=1$ and $\forall c_i>0$). Moreover, because both the collisionless Boltzmann equation and the velocity-moment operators are linear in $\hat F$, raw moments (e.g., $\hat\varrho$ and second moments such as $\langle V_{\rm los}^2\rangle$ or $\hat\varrho\sigma^2$) superpose linearly. As discussed in \S\ref{ssec:DF_Superposition}, we can exploit this property to construct self-consistent superpositions of Method-II-truncated DFs (all generated from the same input pair $\hat\varrho$-$\hat P$); here we extend the single-DF expressions in Appendix~\ref{app:Velocity_prof} and \ref{app:kurtosis} to this superposed case.

Consider a unit-normalised superposition of truncated DFs over $\hat{\bigK}\in[0,1-\hat{\bigK}_{\rm t}]$, all constructed from the same $\hatvarrho(R)$--$\hatP(R)$ pair
\begin{align}\label{eqn:sup}
		\hatF(\hat{\bigK}, R\,\Vt) \eee \sum_i c_i \hatF_i(\hat{\bigK}, R\,\Vt).
\end{align}
In the Method-II construction, each component DF $\hat F_i$ is obtained by inverting an auxiliary profile $\hat\varrho_{{\rm aux},i}(\hat P)$ that is derived from the same physical input pair $(\hat\varrho,\hat P)$. Consequently, when evaluating under the condition $\hatP=\hatP(R)$, we have 
\begin{align}\label{eqn:sup_rhoonshell}
\hat\varrho_i(R)
\equiv \hat\varrho_{{\rm aug},i}(\hat P(R),R)
= \hat\varrho(R)
\qquad \forall\, i.
\end{align} 
Therefore, the superposed DF also reproduces $\hat\varrho(R)$ in the same potential $\hat P(R)$, since $\hat\varrho(R)=\sum_i c_i \hat\varrho_i(R)$ and $\hat\varrho_i(R)\equiv\hat\varrho(R)$.

For individual particle velocity sampling, \eref{eqn:rho_f_anisotropic_scale} can be written as
\begin{align}\label{eqn:rho_f_anisotropic_scale_sup}
	\hatvarrho(R) = \sum_i c_i\hatvarrho_i(R)= 2 \pi \,\sum_i c_i \, \int_0^\pi \rmd\varphi \; p_{\varphi,i}(\varphi|R)
\int_0^{V_\text{esc}(R)} \rmd V \; p_{V,i}(V\,|\,R,\varphi),
\end{align}
where $p_{\varphi,i}$ and $p_{V,i}$ for each model are given by Eqs.~(\ref{eqn:varphi_prob}) and~(\ref{eqn:V_prob}), respectively. Here, $V_\text{esc}(R)\eee\sqrt{2\hatP(R)}$ is independent of the choices for $c_i$ given that the potential profile is identical for all $i$.

Next, for the `composite' velocity dispersion profiles $\sigmar^2$ and $\sigmat^2$ sourced by \eref{eqn:sup}, we start from \eref{eqn:linearity_1} evaluated under the condition $\hatP=\hatP(R)$ at fixed $R$ to write
\begin{align}
	\begin{aligned}
		\hatvarrho_{\text{aug}}(\hatP(R), R)
		= \sum_i c_i \hatvarrho_{\text{aug},i}(\hatP(R), R)  &= 4 \pi \sum_i c_i \, \int^{\hatP}_0 \rmd\hat{\bigK} \int^{\sqrt{2(\hatP-\hat{\bigK})}}_{0} \rmd \Vt \,\Vt \,
		\frac{\hatF_i(\hat{\bigK}, R\,\Vt)}{\sqrt{2(\hatP-\hat{\bigK})-\Vt^2}}\\
        &= 4\pi  \int^{\hatP}_0 \rmd\hat{\bigK} \int^{\sqrt{2(\hatP-\hat{\bigK})}}_{0} \rmd \Vt \,\Vt \, \frac{\hatF(\hat{\bigK}, R\,\Vt)}{\sqrt{2(\hatP-\hat{\bigK})-\Vt^2}},
    \end{aligned}
\end{align}
such that 
\begin{align}
	\begin{aligned}
		&\hatvarrho_{\text{aug}}\,\sigmar^2
		= 4\pi \int^{\hatP}_0 \rmd\hat{\bigK} \int^{\sqrt{2(\hatP-\hat{\bigK})}}_{0} \rmd \Vt \,\Vt \,
		\sum_i c_i \hatF_i\sqrt{2(\hatP-\hat{\bigK})-\Vt^2}
        = 
		\int^{\hatP}_0 \rmd\hat{\bigK}\,\sum_ic_i\hatvarrho_{\text{aug},i}
        = 
		\sum_ic_ig_i(R)\int^{\hatP(R)}_0 \rmd\hat{\bigK}\,\hatvarrho_{\text{aux},i}(\hatP),\\
        &\hatvarrho_{\text{aug}}\,\sigmat^2
		= 4\pi \int^{\hatP}_0 \rmd\hat{\bigK} \int^{\sqrt{2(\hatP-\hat{\bigK})}}_{0} \rmd \Vt \,\Vt^3 \,
		\frac{\sum_i c_i\hatF_i}{\sqrt{2(\hatP-\hat{\bigK})-\Vt^2}} = 2
		\int^{\hatP}_0 \rmd\hat{\bigK}\,
		\frac{\partial}{\partial R^2}\!\Big[R^2\sum_ic_i\hatvarrho_{\text{aug},i}\Big]\\
        &\qquad\qquad\qquad\qquad\qquad\qquad = 2
		\int^{\hatP}_0 \rmd\hat{\bigK}\,
		\frac{\partial}{\partial R^2}\!\bigg[R^2\sum_ic_ig_i(R)\hatvarrho_{\text{aux},i}({\hat{\bigK}})\bigg]
        = 2\sum_i c_i\bigg[g_i(R)+R^2\frac{\partial g_i(R)}{\partial R^2} \bigg]
		\int^{\hatP(R)}_0 \rmd\hat{\bigK} \hatvarrho_{\text{aux},i}({\hat{\bigK}}),
	\end{aligned}
\end{align}
where we have used the results of \eref{eqn:linearity_2}. The resulting anisotropy profile can then be expressed as
\begin{align}\label{eqn:beta_sigma_sup}
	\betasigma(R) = 1- \frac{\sigmat^2(R)}{2\sigmar^2(R)}  = 1 - \frac{\sum_i c_i\Big[g_i(R)+R^2\frac{\partial g_i(R)}{\partial R^2} \Big] \int^{\hatP(R)}_0 \rmd\hat{\bigK} \hatvarrho_{\text{aux},i}(\hat{\bigK})}{\sum_i c_ig_i(R)\int^{\hatP(R)}_0 \rmd\hat{\bigK}\hatvarrho_{\text{aux},i}(\hat{\bigK})} = 1 - \frac{\sum_i c_i\Big[g_i+R^2\frac{\partial g_i}{\partial R^2} \Big] \int^{\hatP(R)}_0 \rmd\hat{\bigK} g_i^{-1}\hatvarrho_{\text{aug},i}}{\sum_i c_ig_i\int^{\hatP(R)}_0 \rmd\hat{\bigK}g_i^{-1}\hatvarrho_{\text{aug},i}},
\end{align}
where $\hatvarrho_{\text{aux},i}(\hatP)$ and $\hat\varrho_{\text{aug},i}(\hatP(R),R)=\hat\varrho_{\text{aug}}(\hatP(R),R)=\hat\varrho(R)$ for all $i$ are fixed by the common input $\hatvarrho(R)$--$\hatP$ pair output by \PIANISTpyNS. Importantly, linearly superposing anisotropic DFs does not translate to a linear superposition of $\betasigma(R)$ profiles, owing to the non-linear division operator in the definition of $\betasigma(R)$. Note that the final expression in \eref{eqn:beta_sigma_sup} is particularly powerful numerically: with a fixed $\hatvarrho(R)$--$\hatP(R)$ pair, evaluating $\betasigma(R)$ for an arbitrary superposition of analytic basis models $\{g_i\}$ reduces to inexpensive 1D quadratures in $\hat{\bigK}$ and does not require additional explicit computation of any truncated DFs $\hatF_i$. This allows for rapid likelihood evaluations when fitting non-trivial anisotropy profiles.
 
If the linear superposition involves only constant-anisotropy models, then $g_i(R)=R^{-2\beta_{\text{c},i}}$ (see Eq.~[\ref{eqn:linearity_beta_1}]), which implies that $g_i + R^2\partial g_i/\partial R^2 = (1-\beta_{\text{c},i})g_i$. In this case, \eref{eqn:beta_sigma_sup} simplifies to
\begin{align}\label{eqn:beta_sigma_const_simp}
    \betasigma(R) = \frac{
    \sum_i c_i \, \beta_{\text{c},i} \, g_i(R) \,
        \int_0^{\hat P(R)} \rmd\hat{\mathcal{K}} \, \hat\varrho_{\mathrm{aux},i}(\hat{\mathcal{K}})}
   {\sum_i c_i \, g_i(R) \,
        \int_0^{\hat P(R)} \rmd\hat{\mathcal{K}} \, \hat\varrho_{\mathrm{aux},i}(\hat{\mathcal{K}})},
\end{align}
from which it is straightforward to work out the limiting behaviour for both small and large radii. In particular, we have that
\begin{align}
\betasigma(R) =
            \begin{cases}
                \max_i (\beta_{\text{c},i}) \quad &\text{(}R \rightarrow 0,\ \max_i \beta_{\text{c},i} \geq \gamma - 1\text{)},\\
                (\gamma - 1) - \big[\sum_i c_i/(\gamma - 1 - \beta_{\text{c},i})\big]^{-1} \quad &\text{(}R \rightarrow 0,\ \max_i \beta_{\text{c},i} < \gamma - 1\text{)},\\
                \sum_i c_i\,\beta_{\text{c},i} \quad &\text{(}R \rightarrow \Rt\text{)}.
            \end{cases}
\end{align}
Note that this behaviour is accurately recovered in the anisotropy profile of Example~1 shown in 
\fref{fig:Linear_Super_betasigma}.

Projected quantities can be treated analogously. Because $\hat\varrho(R)$ and $\hat P(R)$ are identical for all components, so are their projected surface densities, i.e.,
$\Sigma(R_{\rm 2D})=\Sigma_i(R_{\rm 2D})$ for all $i$. Moreover, the DF enters the LOSVD only through the linear velocity-space integral of \eref{eqn:losvd_1}. Hence, at fixed $(R_{\rm 2D},V_{\rm los})$, the LOSVD of the composite system is simply the weighted sum of the LOSVD components:
\begin{align}\label{eqn:losvd_sup} 
D(R_{\rm 2D},V_{\rm los}) = \sum_i c_i \, D_i(R_{\rm 2D},V_{\rm los}), 
\end{align}
where $D_i$ is obtained from Eq.~(\ref{eqn:losvd_2}) by replacing $\hat F\rightarrow \hat F_i$, while keeping $\Sigma(R_{\rm 2D})$, $\hat P(R)$, $V_{\rm esc}(R)=\sqrt{2\hat P(R)}$, and $R_{\rm max}(R_{\rm 2D},V_{\rm los})$ unchanged.

The same linearity carries over to any $\Sigma$-weighted aperture average of the LOSVD: for example, Eqs.~(\ref{eqn:losvd_4},~\ref{eqn:losvd_5}) remain valid with $\overline{D}=\sum_i c_i\,\overline{D}_i$, where each $\overline{D}_i$ is computed using $D_i$ in place of $D$ and the same (component-independent) $\Sigma$. In the case of the LOS velocity dispersion $\sigma_{\rm los}^2(R_{\rm 2D})$, it is first necessary to evaluate $\sigma_{{\rm los},i}^2(R_{\rm 2D})$ for each component using the corresponding $\beta_{\sigma,i}(R)$ and then combine the unnormalised LOS second moments using
\begin{align}\label{eqn:siglos_sup}
\Sigma(R_{\rm 2D})\,\sigma_{\rm los}^2(R_{\rm 2D})
=\sum_i c_i\,\Sigma(R_{\rm 2D})\,\sigma_{{\rm los},i}^2(R_{\rm 2D})\,.
\end{align}
Since $\Sigma$ is common to all components, this implies that $\sigma_{\rm los}^2(R_{\rm 2D})=\sum_i c_i\,\sigma_{{\rm los},i}^2(R_{\rm 2D})$. The same applies to aperture averages such as those in Eqs.~(\ref{eqn:los_sigma_rect}--\ref{eqn:los_sigma_3}). Finally, the unit-normalisation condition in Eq.~(\ref{eqn:losvd_nor}) is preserved under the superposition provided each $D_i$ is individually normalised and $\sum_i c_i=1$.

\section{Numerical Implementation and Optimisation in \PIANISTpy}\label{app:JAX_imp}

This Appendix briefly documents the \JAXNS-accelerated implementation of some of the major modules of \PIANISTpyNS; the full technical implementation details can be found in the companion online documentation (see footnote~\ref{fn:technotes}). Overall, the core numerical solvers are \texttt{jit}-compiled (just-in-time compiled) and \texttt{vmap}-vectorised (auto-vectorised across batches), replacing scalar loops with array operations. Since \JAX compiles the same kernels for CPUs and GPUs, the entire \PIANISTpy pipeline runs either CPU-only or GPU-accelerated, with results validated across our full benchmark set to agree to better than $10^{-10}$ in relative terms. Only the stages that batch a large volume of repetitive computations gain from a GPU and are GPU-enabled; the remaining implementation is CPU-only. Global \texttt{float64} precision is enforced throughout. Conceptually, the code follows a three-layer architecture separating \textit{Numerics} (model-agnostic \JAX primitives and helper functions), \textit{Physics} (all model-specific \texttt{jit}-compiled blocks and bulk numerical calculations), and \textit{Workflows} (module coordination and I/O management), where `model' denotes any explicit dependence on the input density profile, anisotropy parameters, or truncation method. This separation ensures that model-specific operations are localised within \textit{Physics}, while \textit{Numerics} operates on mathematically well-defined inputs independently of the physical context; \textit{Workflows} contains no physical model logic. Below, we summarise the numerical details pertaining to DF computations (\aref{app:num_DF}) and particle sampling (\aref{app:num_par}).

\subsection{DF pipeline}\label{app:num_DF}

This pipeline comprises two distinct components. First, the Abel-type inversion integrals Eqs.~(\ref{eqn:F_bigE_constant_beta_gen},~\ref{eqn:OM_DF_3},~\ref{eqn:OM_TypeII_DF_3},~\ref{eqn:COM_DF_gen}) are evaluated on a multi-zone hybrid $\hatP$-grid to construct the DF: in Method~I this produces the untruncated base DF $F_\bigK(\bigK)$, while in Method~II the input is an already-truncated $\hatvarrho$--$\hatP$ pair and the output is directly the truncated $\hatF_\bigK(\hat\bigK)$. Second, in Method~I only, the base DF is re-parameterised in phase space \eref{eqn:DF_Truncation} and the self-consistent truncated Poisson equation \eref{eqn:Poissonr_Truncation} is solved as a second-order ODE in the dimensionless radial coordinate $R$, reduced to a first-order system in $(\hatP,\,R\,\rmd\hatP/\rmd R)$ in $\ln R$ with boundary conditions $\hatP(0)=1-\bigKt$ and $\rmd\hatP/\rmd R|_{R\rightarrow0}$ set by the inner density slope (Eq.~[\ref{eqn:central_slope}]), integrated outward to the truncation radius $\Rt$ where $\hatP(\Rt)=0$. The ODE is adaptively integrated using \texttt{diffrax} \citep{Kidger2021} with the Dormand--Prince RK45 stepper (\texttt{Dopri5}) at tolerances $\texttt{rtol} = \texttt{atol} = 10^{-11}$. The pre-computed tabulated representation of $\hatvarrho[\hatP]$ (cf.~Eq.~[\ref{eqn:Poissonr_Truncation_rho}]) is passed to the compiled solver as an array argument. Tabulated arrays are interpolated with a monotone cubic Hermite (PCHIP) interpolant, which is shape-preserving on the steep near-boundary segments where an unconstrained cubic spline would overshoot and introduce spurious oscillations.

The Abel-type integrals share a common $(\hatP - \hat\bigK)^{-n}$ singularity at the upper integration limit $\hat\bigK \to \hatP$, where $n$ is the kernel exponent as defined in \eref{eqn:F_bigE_constant_beta_gen} and $\hat\bigK$ is the energy-like integration variable. This singularity is present at every evaluation point $\hatP$ and becomes numerically irresolvable as the integration window collapses to zero width near $\hatP \to P_0$, where $P_0$ is the central potential of the profile being inverted (specifically, $P_0 = 1$ for the untruncated base DF and $P_0 = 1 - \bigKt$ for the truncated DF). We address this with a purpose-built three-sub-interval Gauss--Legendre scheme. For each $\hatP$, the domain $[0, \hatP]$ is split at $\hat\bigK_\mathrm{cut} = 0.9\,\hatP$. The lower interval $[0, \hat\bigK_\mathrm{cut}]$ is divided at $\delta = 0.1\,\hat\bigK_\mathrm{cut}$: sub-interval~1a covers $[0^+,\,\delta]$ with log-spaced nodes to resolve the steep integrand near $\hat\bigK \to 0$, and sub-interval~1b covers $[\delta,\,\hat\bigK_\mathrm{cut}]$ with Gauss--Legendre nodes. The upper interval $[\hat\bigK_\mathrm{cut},\,\hatP]$ employs the feature-informed composite map $\hat\bigK = P_0 - (P_0 - \hatP)\,\exp[\tau^{1/(1-n)}]$, evaluated at Gauss--Legendre nodes in $\tau$. Specifically, the exponential factor places nodes in $\log(P_0 - \hat\bigK)$ to resolve integrands that steepen as $(P_0 - \hat\bigK)^{\kappa}$ with $\kappa < 0$ as $\hat\bigK \rightarrow P_0$ (\aref{app:DF_inner_asymp}), while the standard change-of-variable stretch $\tau^{1/(1-n)}$ for algebraic endpoint singularities \citep[e.g.,][]{2007nras.book.....P} removes the $(\hatP - \hat\bigK)^{-n}$ kernel singularity exactly for every kernel exponent $|n| \leq \tfrac12$ (see below). For $n > \tfrac12$, the integral of the Cuddeford inversion operator, \eref{eqn:F_bigE_constant_beta_gen}, is evaluated after one further analytical integration by parts, which moves one derivative onto $\varrho_\mathrm{aux}$ and lowers the kernel exponent to $n - 1 \in (-\tfrac12, 0)$; the composite map in $\hat\bigK$ introduced above is then applied with $|n| \leq \tfrac12$ in every case. When $n = 0$ (for the half-integer cases $\betac = \tfrac{3}{2} - m$ with integer $m \geq 1$, i.e., $\betac = +\tfrac12, -\tfrac12, -\tfrac32, \ldots$), the kernel carries no singularity and the DF is obtained directly from the $m$-th derivative of $\varrho_\mathrm{aux}$ evaluated at $\hatP$, Eqs.~(\ref{eqn:const_beta0p5_DF},~\ref{eqn:const_betam0p5_DF}), bypassing the quadrature entirely; this closed form is adopted for every half-integer $\betac$. Separately, the derivatives of $\varrho_\mathrm{aux}$ are computed directly from the input analytical profile in Method~I, or numerically from the already-truncated input profile in Method~II. The integral in \eref{eqn:F_bigE_constant_beta_gen} starts at the lowest energy grid value $\hat\bigK_{\rm min}$ $= 10^{-10}$ of the numerical $\hatvarrho_\mathrm{aux}(\hatP)$ profile, not exactly at $\hat\bigK = 0$, and its surface term is evaluated at $\hat\bigK_{\rm min}$. \PIANISTpy also explicitly accounts for the contribution over $0 \leq \hat\bigK < \hat\bigK_{\rm min}$, as computed from the analytical profile in Method~I or from the Taylor expansion of the input truncated DF about its formal truncation boundary $\hat\bigK = 0$ in Method~II.

Dedicated numerical treatments are required to maintain the relative-error goal of $10^{-6}$ throughout $\hatP \in [0, P_0]$ under the default settings, where both the DF integrand and $\hatvarrho[\hatP]$ develop steep gradients near both boundaries. These are described in detail in the extensive documentation available at
\url{https://pianistpy.com}.

\begin{figure*}
    \includegraphics[width=0.98\linewidth]{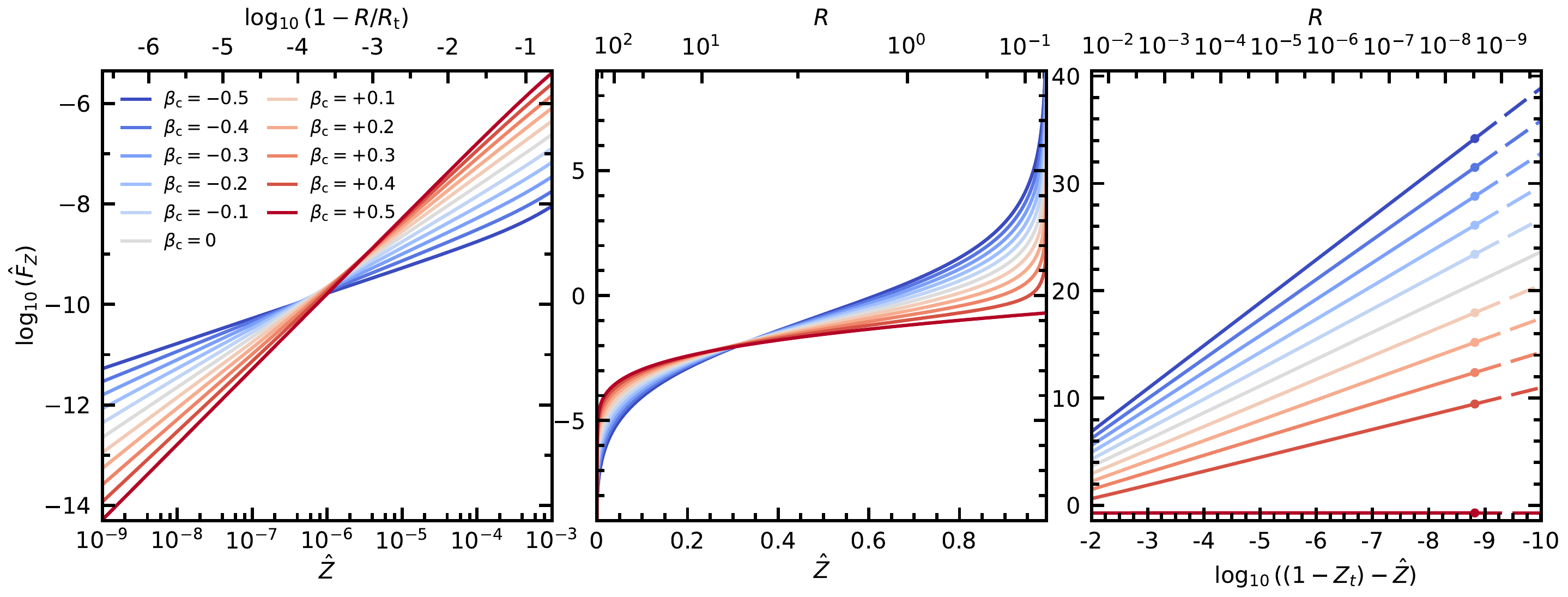}
    \caption{\PIANISTpyNS-generated constant-anisotropy DF solutions over the full range of relative energy (bottom $x$-axis) and corresponding radius (top $x$-axis, numerically mapped by setting $\hatZ = \hatP(R)$), demonstrating the numerical accuracy achieved by \PIANISTpyNS. The middle panel shows the entire solutions in linear scale in $\hatZ$; the left (right) panel zooms in on the low-$\hatP$ (high-$\hatP$) tail, respectively. This example is constructed by truncating an NFW profile with $c_\rho=30$ so that its total mass $\Mt=2\Mvir$. The isotropic DF is computed using Method I, while the anisotropic DFs are Method-II-inverted from this isotropic $\hatvarrho$--$\hatP$ pair. The dashed line style marks the \PIANISTpyNS-patched asymptotic solutions of the Method-II DFs (\aref{app:num_DF_high_tail}) above the anchor points (filled circles).}
    \label{fig:DF_Comparison}
\end{figure*}

\subsubsection{The $\hatP \rightarrow P_0$ tail}\label{app:num_DF_high_tail}

The Abel kernel requires the evaluation of $\rmd^{m+1}\varrho_\mathrm{aux}/\rmd\hatP^{m+1}$ at each integration point, where $m$ is set by the anisotropy model and $\betac$ as described in \aref{app:DF_Truncation}. For Method~I, whose input is an analytical double-power-law profile, these derivatives are evaluated exactly. Namely, the operator $\rmd/\rmd P = (\rmd P/\rmd R)^{-1}\,\rmd/\rmd R$, with $\rmd P/\rmd R = -\bigM(R)/R^{2}$ and the closed-form enclosed mass $\bigM(R)$ of \eref{eqn:double_power_law_mass}, is applied symbolically to the closed-form $\varrho_\mathrm{aux}(R)$ as many times as \eref{eqn:F_bigE_constant_beta_gen} requires, and the result is evaluated at $R = R(\hatP)$ by inverting the monotonic $P(R)$. For Method~II, whose input is a truncated potential--density pair, \PIANISTpy instead evaluates these derivatives as integrals of the input DF. From the velocity integral \eref{eqn:Poissonr_Truncation_rho}, each derivative of $\hatvarrho$ with respect to $\hatP$ is analytically equivalent to an integral of $\hat F$ and of its derivatives weighted by the kernel $(\hatP - \hat\bigK)^{-1/2}$, with $\rmd\hat F/\rmd\hat\bigK$ evaluated exactly at every grid point of the energy grid when the DF is inverted (by applying the inversion operator one order higher) rather than by directly differentiating the input table. This exact analytical recasting thus avoids any explicit (spline-based) differentiation of the tabulated input and the associated exponential rounding-error amplification of repeated finite differencing on a non-uniform grid \citep[e.g.,][]{Fornberg1988}. In Method~I, the DF is evaluated on the exact analytical derivatives up to a fixed ceiling at $P_0 - \hat\bigK = 10^{-9}$, where the composite map of \aref{app:num_DF} resolves the steepening integrand. In Method~II, the DF is continued above the last point of the energy grid, the anchor at $P_0 - \hat\bigK = 10^{-9}$ (or deeper, where the exact derivatives leave the double-precision range), by its asymptotic form (\aref{app:DF_inner_asymp}) with its closed-form amplitude and an additive constant fixed by continuity at the anchor.

Separately, for the output of the numerical ODE solver, we apply the analytical inner expansions given by Eqs.~(\ref{eqn:inner_density_expansion},~\ref{eqn:inner_potential_expansion},~\ref{eqn:inner_mass_expansion}) to patch the radial arrays of potential, density, and enclosed mass in the $R \to 0$ limit.

\subsubsection{The $\hatP \rightarrow 0$ tail}\label{app:num_DF_low_tail}

For untruncated models, the outward integration of \eref{eqn:Poissonr_Truncation} eventually tapers to a constant error floor in $\hatP$ of the order of the solver tolerance, as opposed to actually asymptoting $\hatP \to 0$ at large radii. \PIANISTpy then replaces these numerically affected outer grid values, beyond the radius at which $\hatP$ has fallen to half its central value $P_0$, with the closed-form potential, density and enclosed mass of the analytical profile, on a logarithmic grid extending to the radius where $\hatP = \hat\bigK_{\rm min}$; the ODE solution is kept only inside that join radius. The inversion grid has a fixed floor at $\hat\bigK_{\rm min}$, and untruncated DF tables are written from $\hat\bigK \geq 2\,\hat\bigK_{\rm min}$; below the floor the DF follows its low-$\hat\bigK$ asymptotic form (\aref{app:DF_lowK_asymp}).

For (C)OM Type~I models, whose DFs diverge as $F(\bigQ_- \to 0) \propto \bigQ_-^{\kappa_{\rm (C)OM}}$ (Eq.~[\ref{eqn:F_lowK_asymp}]), the contribution of $[0, \hat\bigK_{\rm min}]$ to the density integral \eref{eqn:Poissonr_Truncation_rho} is added in closed form for $\kappa_{\rm (C)OM} > -1$, with its amplitude set by the surface term at $\hat\bigK_{\rm min}$. For $\kappa_{\rm (C)OM} \leq -1$ the untruncated density integral diverges and \PIANISTpy refuses the model; any $\bigKt > 0$ removes the divergence.

\begin{figure*}
	\includegraphics[width=0.93\linewidth]{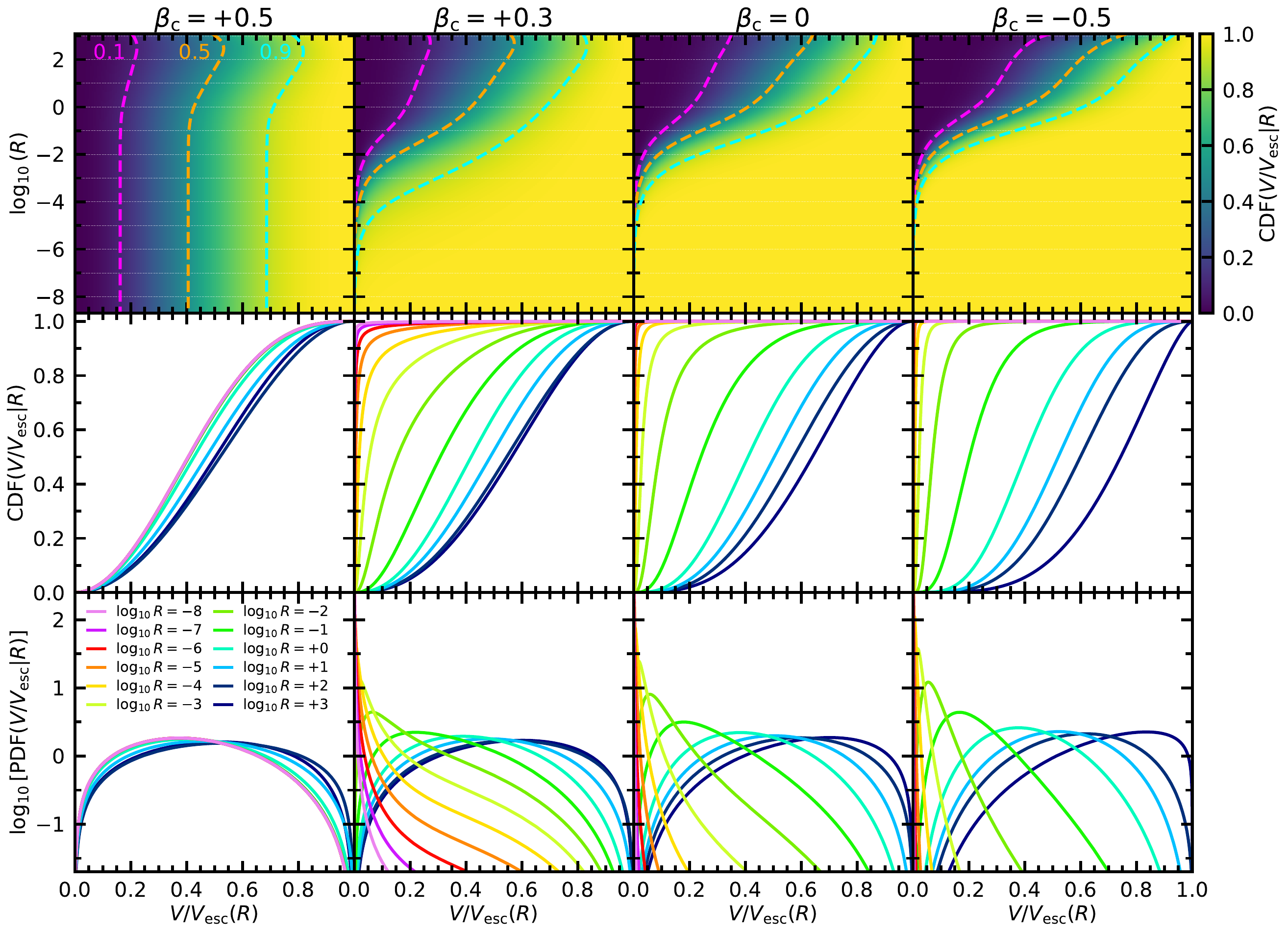}
	\caption{\PIANISTpyNS-generated velocity distributions from the truncated DFs with $\betac = +0.5, +0.3, 0, -0.5$ (left to right columns) shown in \fref{fig:DF_Comparison}. We show the full radial profile of the CDF over the entire range of evaluation in $R \leq \Rt$ (top row; 0.1, 0.5, and 0.9 values at fixed $R$ are marked by dashed curves), and selected CDF (middle row) and corresponding PDF (bottom row) profiles at $\log_{10}R=-8, -7, ..., +3$ (colour-coded as indicated). The well-resolved and smoothly varying velocity PDF profiles spanning eleven orders of magnitude in radius demonstrate the high numerical accuracy achieved by \PIANISTpy in DF evaluation (\aref{app:num_DF}) and CDF-based particle sampling (\aref{app:num_par}) across all anisotropy models shown.}
	\label{fig:CDF_PDF_Comparison}
\end{figure*}

As an example, \fref{fig:DF_Comparison} shows the \PIANISTpyNS-computed DFs for constant-anisotropy NFW haloes with $c_\rho=30$. The isotropic version ($\beta_\rmc=0$) is Method-I truncated with $Z_\rmt = 0.00796$, resulting in a total mass $\Mt = 2\Mvir$, while all other DFs are Method-II inverted from the isotropic $\hatvarrho$--$\hatP$ pair. Blue and red curves indicate systems with tangential and radial anisotropy, respectively, as indicated. The middle panel shows the DFs covering the entire range in $\hatZ$, while left and right panels zoom in on the low- and high-$\hatP$ regions, respectively. For the Method-II DFs, the dashed lines above the filled circles mark the DF entries patched by the corresponding asymptotic forms.

\subsection{Particle sampling pipeline}\label{app:num_par}

The particle sampling pipeline in \PIANISTpy consists of two sequential components: the construction of a table of the speed quantile function, the inverse of the cumulative velocity distribution function (CDF) $P(<v|\hatP)$, and particle generation by evaluating this pre-computed table. First, \PIANISTpy integrates the CDF at each $\hatP$ on a sine-clustered $\xi$ grid, $\xi \equiv V/V_\text{esc} \in [0,1]$ with $\xi(t) = \sin^{3/2}(\pi t/2)$ for $t$ uniform in $[0,1]$, that concentrates nodes near both endpoints $\xi = 0$ and $1$ where the DF integrand varies most steeply, removing the need for a separate singularity substitution at the upper-limit boundary. Its inverse is tabulated on a two-dimensional grid uniform in the logit variables $\ln[u/(1-u)]$ of the cumulative probability $u$ and $\ln[\hatP/(P_0-\hatP)]$ of the potential, which resolve both ends of either range; every node stores the quantile and its exact slope. Next, discrete particle realisations are drawn from the pre-tabulated quantile by cubic Hermite interpolation in both variables at the particle's $(\hatP, u)$. This eliminates the rejection overhead that becomes prohibitive for highly anisotropic models. Each particle's radius $R$ is drawn likewise, by cubic Hermite interpolation of the tabulated quantile of the enclosed-mass fraction in the same logit variables, making the position draw branch-free and \texttt{vmap}-safe. For anisotropic models, the velocity vector is then assigned in three steps: (1) draw the velocity angle $\varphi$ between $\mathbf{V}$ and $\hat{R}$ from $p_\varphi(\varphi|R)$ (Eq.~[\ref{eqn:varphi_prob}]) via inverse-CDF sampling, exploiting the exact symmetry $p_\varphi(\pi-\varphi|R) = p_\varphi(\varphi|R)$ to draw $\varphi$ over $[0,\pi/2]$ only, with the sign of $\Vr$ drawn independently as $\pm1$ with equal probability, exactly recovering the full $[0,\pi]$ distribution; a change of variable in $\sin^2\varphi$ removes the radius dependence of \eref{eqn:varphi_prob}, so that $\varphi$ is drawn from one fixed distribution at every radius, which eliminates the need of angle tables in $R$; (2) draw the speed $V$ from $p_V(V|R,\varphi)$ (Eq.~[\ref{eqn:V_prob}]) over $[0, V_\text{esc}(R)]$ via the $\varphi$-independent factorisation of the (C)OM conditional $p_x(x|R,\varphi) \propto x^{2-2\betac}F[P(1-x^2)]$ in terms of $x \eee V_\text{aux}/V_\text{esc}$; a single CDF in $x$ at each $\hatP$ therefore replaces the three-dimensional $(R, \varphi, V)$ conditional table, with the velocity directly recovered from $V_\text{aux}(V; R, \varphi)$, giving $\Vr = V\cos\varphi$ and $\Vt = V\sin\varphi$ (Eq.~[\ref{eqn:velocity_angle}]); and (3) obtain the direction of the tangential velocity component by Gram--Schmidt projection of an independent draw from a three-dimensional isotropic Gaussian onto the plane perpendicular to $\hat{R}$, giving exact uniform sampling on the tangent plane with no pole singularity. For the isotropic model, step (1) is bypassed and the full velocity direction is instead drawn uniformly on the unit sphere.

\label{lastpage}
\end{document}